\documentclass[twocolumn]{svjour3}
\pdfoutput=1
\usepackage{color,graphicx}
\usepackage{colortbl}
\usepackage{balance}
\usepackage{subfigure}
\usepackage{epsfig}
\usepackage{epstopdf}
\usepackage{multirow}
\usepackage{times}
\usepackage{alltt}
\usepackage{wrapfig}
\usepackage{algorithmic}
\usepackage[ruled]{algorithm2e}

\newcommand{\topic}[1]{\smallskip\smallskip \noindent {\bf #1}}
\newcommand{\eat}[1]{}
\newcommand{\squishlist}{
 \begin{list}{$\bullet$}
  { \setlength{\itemsep}{0pt}
     \setlength{\parsep}{3pt}
     \setlength{\topsep}{3pt}
     \setlength{\partopsep}{0pt}
     \setlength{\leftmargin}{1.5em}
     \setlength{\labelwidth}{1em}
     \setlength{\labelsep}{0.5em} } }

\newcommand{\squishlisttwo}{
 \begin{list}{$\bullet$}
  { \setlength{\itemsep}{0pt}
     \setlength{\parsep}{0pt}
    \setlength{\topsep}{0pt}
    \setlength{\partopsep}{0pt}
    \setlength{\leftmargin}{2em}
    \setlength{\labelwidth}{1.5em}
    \setlength{\labelsep}{0.5em} } }

\newcommand{\squishend}{
  \end{list}  }
\newcommand{\nframe}{\textsc{NScale}}

\newcommand{\Paragraph}[1]{\vspace{+1mm}\noindent \textbf{#1}\ \ }
\newcommand{\Paragraphul}[1]{\vspace{+1mm}\noindent \underline{\textbf{#1}}\ \ }

\newcommand{\red}[1]{\textcolor{black}{#1}}
\newcommand{\ignore}[1]{}

\begin{document}
\title{NScale: Neighborhood-centric Large-Scale Graph Analytics in the Cloud}
\author{Abdul Quamar   \and Amol Deshpande  \and Jimmy Lin }
\maketitle

\begin{abstract}

There is an increasing interest in executing complex analyses over large graphs, many of which 
require processing a large number of multi-hop neighborhoods or subgraphs.
Examples include ego network analysis, motif counting,
finding social circles, 
personalized recommendations, 
link prediction, anomaly detection, analyzing influence cascades, and others. 
These tasks are not well served by existing {\em vertex-centric} graph processing frameworks,
where user programs are only able to directly access the state of a single vertex at a time, 
resulting in high communication, scheduling, and memory overheads in executing such tasks.
Further, most existing graph processing frameworks ignore the challenges in extracting 
the relevant portions of the graph that an analysis task is interested in, and loading those onto distributed memory.
 
This paper introduces \nframe, a novel end-to-end graph processing framework that
enables the distributed execution of complex {\em subgraph-centric} analytics over large-scale graphs in the cloud.
\nframe~enables users to write programs at the level of
subgraphs rather than at the level of vertices. 
Unlike most previous graph processing frameworks,
which apply the user program to the entire graph, \nframe~allows users to declaratively specify subgraphs
of interest. Our framework includes a novel graph extraction and packing (GEP) module that
utilizes a cost-based optimizer to partition and pack the subgraphs of interest into
memory on as few machines as possible. \red{The distributed execution engine 
then takes over and
runs the user program in parallel on those subgraphs, restricting the scope of 
the execution appropriately, and utilizes novel techniques to minimize memory consumption by exploiting
overlaps among the subgraphs. 
We present a comprehensive empirical evaluation comparing against three state-of-the-art systems, namely, Giraph, GraphLab, and GraphX, on
several real-world datasets and a variety of analysis tasks.
Our experimental results show
orders-of-magnitude improvements in performance and drastic reductions in the cost of analytics compared to vertex-centric approaches.}
\end{abstract}

\section{Introduction}

Over the past several years, we have witnessed unprecedented growth in the size and availability of graph-structured data.
Examples includes social networks, citation networks, biological networks, IP traffic
networks, just a name a few.
There is a growing need to execute complex analytics over graph data to extract insights,
support scientific discovery, detect anomalies, etc.
A large number of these tasks can be viewed as
operations on local neighborhoods of vertices
in the graph (i.e., subgraphs). For example, there is much interest in analyzing ego networks, i.e., $1$- or
$2$-hop neighborhoods, for
identifying structural holes~\cite{burt2009structural}, brokerage analysis~\cite{burt2007secondhand}, counting motifs~\cite{MilEtAl02}, identifying
social circles~\cite{NIPS2012_0272}, social recommendations~\cite{Backstrom:2011:SRW:1935826.1935914}, 
computing statistics like local clustering coefficients or ego betweenness~\cite{everett2005ego}, and anomaly detection~\cite{oddball}. 
In other cases, we might be interested in analyzing induced subgraphs satisfying certain properties, for example, 
users who tweet a particular hashtag in the Twitter network or groups of users who
have exhibited significant communication activity in recent past.
More complex subgraphs can be specified as unions or intersections of neighborhoods of pairs
of vertices; this may be required for graph cleaning tasks like
entity resolution~\cite{DBLP:conf/icde/MoustafaNDG11}. 

In this paper, we propose a novel distributed graph processing
framework called \nframe, aimed at supporting complex graph analytics over very large graphs. Although there has been no shortage of new
distributed graph processing frameworks in recent years (see
Section~\ref{sec:relWork} for a detailed discussion), our work has three
distinguishing features:

\begin{list}{$\bullet$}{\leftmargin 0.15in \topsep 1pt \itemsep 1pt}

\item {\bf Subgraph-centric programming model.} Unlike vertex-centric
  frameworks, \nframe~allows users to write custom
  programs that access the state of entire {\em subgraphs}
  of the complete graph. This model
  is more natural and intuitive for many complex graph analysis tasks compared to 
  the popular vertex-centric model.

\item {\bf Extraction of query subgraphs.} Unlike existing graph
  processing frameworks, most of which apply user programs to the
  entire graph, \nframe~efficiently supports tasks that involve only a
  select set of subgraphs (and of course, \nframe~can execute programs
  on the entire graph if desired).

\item {\bf Efficient packing of query subgraphs.} To enable efficient
  execution, subgraphs of interest are packed into as few containers
  (i.e., memory) as possible by taking advantage of overlaps between
  subgraphs. The user is able to control resource allocation (for
  example, by specifying the container size), which makes our
  framework highly amenable to execution in cloud environments.

\end{list}

\noindent \nframe~is an end-to-end graph processing framework that enables scalable distributed execution
of subgraph-centric analytics over large-scale graphs in the cloud. 
In our framework, the user specifies:\ (a) the subgraphs of interest (for example, $k$-hop neighborhoods
around vertices that satisfy a set of predicates) and (b) a user program
to be executed on those subgraphs (which may itself be iterative). The user program is written against a general
graph API (specifically, BluePrints), and has access to the entire state of the subgraph against
which it is being executed. \nframe~execution engine is in charge of ensuring that the user program only 
has access to that state and nothing more; this guarantee allows existing graph algorithms to be used without modification.
\red{Thus a program written to compute, say, connected components 
in a graph, can be used as is to compute the connected components within each subgraph of interest. } 
Our current subgraph specification format allows users to specify subgraphs of interest as $k$-hop
neighborhoods around a set of query vertices, followed by a filter on the nodes and the edges in the
neighborhood. It also allows selecting subgraphs induced by certain attributes of the nodes; e.g.,
the user may choose an attribute like tweeted hashtags, and ask for induced subgraphs, one for
each hashtag, over users that tweeted that particular hashtag.

User programs corresponding to complex analytics may make arbitrary and random
accesses to the graph they are operating upon. Hence, one of our key design decisions was
to ensure that each of the subgraphs of interest would reside entirely in memory on a single machine while the user program ran against it.
\nframe~consists of two major components.
First, the graph extraction and packing (GEP) module extracts relevant subgraphs of interest and uses a cost-based
optimizer for data replication and placement that minimizes the number
of machines needed, while attempting to balance load across machines to guard
against the straggler effect.
Second, the distributed execution engine executes user-specified computation on the
subgraphs in memory.
It employs several optimizations that reduce the total memory footprint by exploiting overlap between subgraphs
loaded on a machine, without compromising correctness.

\red{Although we primarily focus on one-pass complex analysis tasks described above, \nframe~also supports the 
Bulk Synchronous Protocol (BSP) model for executing iterative analysis tasks like computation of PageRank or 
global connected components. \nframe's BSP implementation is most similar to that of GraphLab, and the information
exchange is achieved through shared state updates between subgraphs on the same partition and through use of 
``ghost'' vertices (i.e., replicas) and message passing between subgraphs across different partitions.}



We present a comprehensive experimental evaluation that illustrates that
extraction of relevant portions of data from the underlying graph and optimized data replication and placement helps 
improve scalability and performance with significantly fewer resources reducing the cost of data analytics substantially. 
\red{The  graph computation and execution model employed by \nframe~affects a drastic reduction in communication (message passing) overheads (with no message passing within
        subgraphs), and significantly reduces the memory footprint (up to 2.6X for applications over 1-hop neighborhoods and up to 25X for applications such as personalized page rank over 2-hop neighborhoods); the overall
performance improvements range from 3X to 30X for graphs of different sizes for applications over 1-hop neighborhoods and 20X to 400X for 2-hop neighborhood analytics. Further, our experiments show that 
GEP is a small fraction of the total time taken to complete the task, and is thus the crucial component that enables the efficient execution of the graph 
computation on the materialized subgraphs in distributed memory using minimal resources.} This enables \nframe\ to scale neighborhood-centric graph
analytics to very large graphs for which the existing vertex-centric approaches fail completely.

\section{Related Work}
\label{sec:relWork}
Here we focus on the large-scale graph processing frameworks and programming models; motivating
applications are discussed in the next section.

\noindent {\bf Vertex-centric approaches.} Most existing graph
processing frameworks such as
Pregel~\cite{Malewicz:2010:PSL:1807167.1807184}, Apache Giraph, 
GraphLab~\cite{DBLP:journals/corr/abs-1204-6078},
Kineograph~\cite{Cheng:2012:KTP:2168836.2168846},
GPS~\cite{Salihoglu:2013:GGP:2484838.2484843},
Grace~\cite{conf/cidr/WangXDG13}, etc.,
are vertex-centric. Users write
{\em vertex-level programs}, which are then executed by the framework
in either a bulk synchronous fashion (Pregel, Giraph) or asynchronous
fashion (GraphLab) using message passing or shared memory. These
frameworks fundamentally limit the user program's access to a single
vertex's state -- in most cases to the local state of the vertex and
its edges. 
This is a serious limitation for many complex analytics
tasks that require access to subgraphs.

For example, to analyze a 2-hop neighborhood around a vertex to find
social circles~\cite{NIPS2012_0272}, one would first need to gather all the information
from the 2-hop neighbors through message-passing, and reconstruct those
neighborhoods locally (i.e., in the vertex program local state). Even something 
as simple as computing the number of triangles for a node requires gathering information from
1-hop neighbors (since we need to reason about the edges between the neighbors, cf. Figure \ref{fig:graphCompute}).
This requires significant network communication and an enormous amount of
memory. Consider some back-of-the-envelope calculations for 
estimating the message passing and memory overhead for constructing neighborhoods of
various sizes at each vertex for 
the Orkut social network graph with
approx 3M nodes, 234M edges and an average degree of 77. The original
graph occupies 14GB of memory for a data structure that stores the
graph as a bag of vertices in adjacency list format. 
Table~\ref{table:overheads} provides an
estimate of the number of messages that would need to be exchanged and
the memory footprints required in order to construct 1- and 2-hop
neighborhoods at each vertex for ego network analysis. It is
clear that a vertex-centric approach requires inordinate amounts of
network traffic, beyond what can be addressed by ``combiners'' in
Pregel~\cite{Malewicz:2010:PSL:1807167.1807184} or
GPS~\cite{Salihoglu:2013:GGP:2484838.2484843}, and impractical amount of cluster memory. 
Although GraphLab is based on a shared memory model, it too would require two 
phases of GAS (Gather, Apply, Scatter) to construct a 2-hop neighborhood at each vertex
and suffers from duplication of state and high memory overhead.

We also see that even
for a modest graph, the memory requirements are quite high for most
clusters today. Furthermore, because most existing graph processing
frameworks hash-partition vertices by default, this approach will
create much duplication of neighorhood data structures. 
In recent work, 
Seo et al.~\cite{DBLP:journals/pvldb/SeoPSL13} also observe
that these frameworks quickly run out of memory and do not scale
for ego-centric analysis tasks.

\begin{table}[t]
\small
    \centering
    \begin{tabular}{| p{4.85cm} | c | c | }
    \hline\hline
    {\bf Neighborhood size} & {\bf 1-Hop} & {\bf 2-Hop}\\ \hline
    Messages required to construct neighborhoods & 231 M& $\approx$ 18 B\\ \hline
    Avg. Memory required per neighborhood  & 83 KB & 6 MB\\  \hline
    Total Cluster Memory required & 233 GB & $\approx$ 18 TB \\ 
         \hline\hline
     \end{tabular}    
     \caption{Message passing and memory overheads of an vertex-centric approach, for constructing neighborhoods of different sizes at each vertex for executing an ego-centric analysis 
     task (the input Orkut graph has 3M nodes and 234M edges).}
    \label{table:overheads}
\vspace{-15pt}
\end{table}

The other weakness of existing vertex-centric approaches is that they
almost always process the entire graph. In many cases, the user may
only want to analyze a subset of the subgraphs in a large graph (for
example, focusing in only on the neighborhoods surrounding ``persons
of interest'' in a social network, or only the subgraphs induced by a set of ``hashtags''
depicting current events in the Twitter network). Naively loading each partition of
the graph onto a separate machine may lead to unnecessary network
communication, especially since the number of messages exchanged increases
non-linearly with the number of machines.

 
\smallskip \noindent {\bf Existing subgraph-centric approaches.}
While researchers have proposed a few subgraph-centric
frameworks such as Giraph++~\cite{DBLP:journals/pvldb/TianBCTM13} and
GoFFish~\cite{DBLP:journals/corr/SimmhanKWNRRP13}, there are
significant limitations associated with both.
These approaches primarily target the message passing overheads and scalability issues in the vertex-centric, BSP model of computation.
Giraph++~partitions the graph onto multiple machines, and runs a sequential algorithm on the entire subgraph in a partition in each superstep.
GoFFish is very similar and partitions the graph using {\sc metis} (another scalability issue) and runs a connected components algorithm in each partition. 
An important distinction is that in both cases, the subgraphs are determined by the system, in contrast to our framework, which explicitly allows users to specify the subgraphs of interest.
Furthermore, these previous
frameworks use serial execution within a partition and the onus of parallelization is left to the
user. It would be extremely difficult for the end user to incorporate tools and libraries to parallelize these sequential algorithms to exploit powerful multicore architectures available today.

\smallskip \noindent {\bf Other graph processing frameworks.}
There are several other graph programming frameworks that have been recently proposed.
SociaLite~\cite{DBLP:conf/icde/SeoGL13} describes an extension of a Datalog-based query language to
express graph computations such as PageRank, connected components, shortest path, etc. The system
uses an underlying relational database with tail-nested tables and enables users to hint at the
execution order. Galois~\cite{nguyen13}, LFGraph~\cite{lfgraph}, 
are among highly scalable general-purpose graph processing frameworks that target systems- or
hardware-level optimization issues, but support only low-level or vertex-centric programming frameworks. 
\red{
Facebook's Unicorn system~\cite{Unicorn} constructs a distributed inverted index and supports online graph-based
searches using a programming API that allows users to compose queries using set operations like AND, OR, etc.;
thus Unicorn is similar to an online SPARQL query processing system and can be used to identify nodes or entities
that satisfy certain conditions, but it is not a general-purpose complex graph analytics system.
}

\red{X-Stream~\cite{xstream} provides an edge-centric graph processing model using streamed partitions on a 
single shared memory machine. The programming API is based on scatter and gather functions that are 
executed on the edges and that update the states maintained in the vertices. 
Any multi-hop traversal in X-Stream would be expensive as it requires multiple iterations of the scatter, shuffle and 
gather phases. Since the stream partitioning used by the framework does not take the neighborhood structure 
into account, such operations would necessitate a large amount of data to be shuffled to the gather phase across 
different stream partitions. X-Stream also fundamentally relies on the vertex state remaining constant in size, and it
would negate the key benefits of X-Stream if variable-sized neighborhoods were constructed in the vertex state. 
Finally, X-Stream provides a restricted edge-centric API that would make it hard to 
encode neighborhood-centric computations such as those supported by \nframe.
}

\red{GraphX, built on top of Apache Spark, supports a flexible set of operations on large graphs~\cite{GraphX}; however, GraphX stores
the vertex information and edge information as separate RDDs, which necessitates a join operation for each 
edge traversal. Further, the only way to support subgraph-centric operations in GraphX is
through its emulation of the vertex-centric programming framework, and our experimental comparisons with GraphX 
show that it suffers from the same limitations of the vertex-centric frameworks as discussed above. }

\section{Application Scenarios}
\label{sec:appScenarios}

\begin{figure}[t]
\centering
\includegraphics[width=\linewidth]{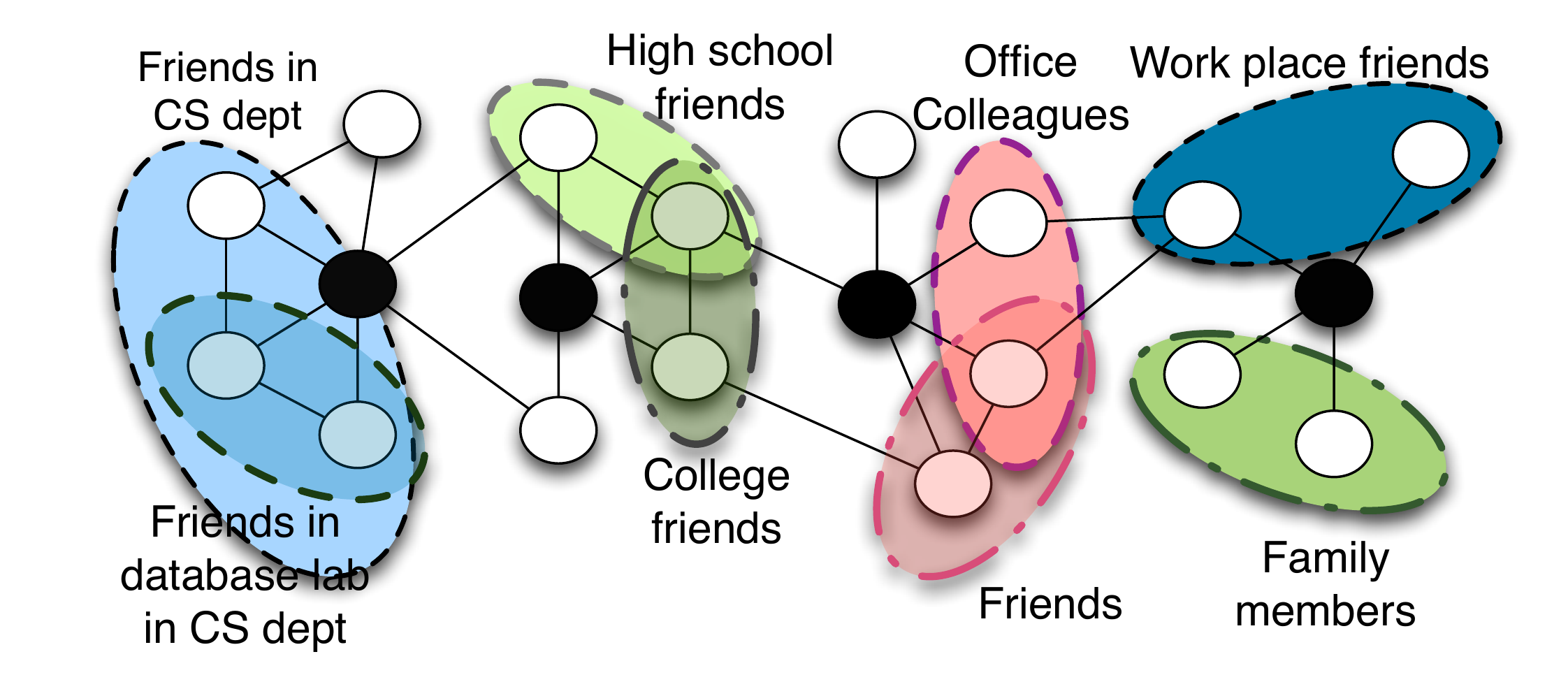}
\vspace{-10pt}
\caption{An example of neighborhood-centric analysis: identify users' {\em social circles} in a social network.}
\label{fig:socialCircles}
\vspace{-5pt}
\end{figure}

This section discusses several representative graph analytics tasks that are ill-suited
for vertex-centric frameworks, but fit well with \nframe's subgraph-centric computation model. 

\topic{Local clustering coefficient (LCC).}
In a social network, the LCC quantifies, for a user, the fraction of his or her friends
who are also friends---this is an important starting point for many graph analytics tasks.
Computing the LCC for a vertex requires
constructing its ego network, which includes the vertex, its 1-hop neighbors, and all the edges between the neighbors.
Even for this simple task, the limitations of vertex-centric approaches are apparent, since they 
require multiple iterations to collect the ego-network before performing the LCC
computation (such approaches quickly run out of memory as we increase the number of vertices we are interested in).

\topic{Identifying social circles.} Given a user's social network ($k$-hop neighborhood), the goal is to identify 
the social circles (subsets of the user's friends), which provide the basis for information dissemination and other tasks.
Current social networks either do this manually, which 
is time consuming, or group friends based on common attributes, which fails to capture the individual aspects of the user's communities.
Figure~\ref{fig:socialCircles} shows examples of different social circles in the ego networks
of a subset of the vertices (i.e., shaded vertices). Automatic identification of social circles
can be formulated as a clustering problem in the user's $k$-hop neighborhood, for example, based on
a set of densely connected alters~\cite{NIPS2012_0272}. 
Once again, vertex-centric approaches are not amenable to algorithms that consider subgraphs as primitives,
both from the point of view of performance and ease of programming.



\begin{figure}[t]
\centering
\includegraphics[scale=0.50]{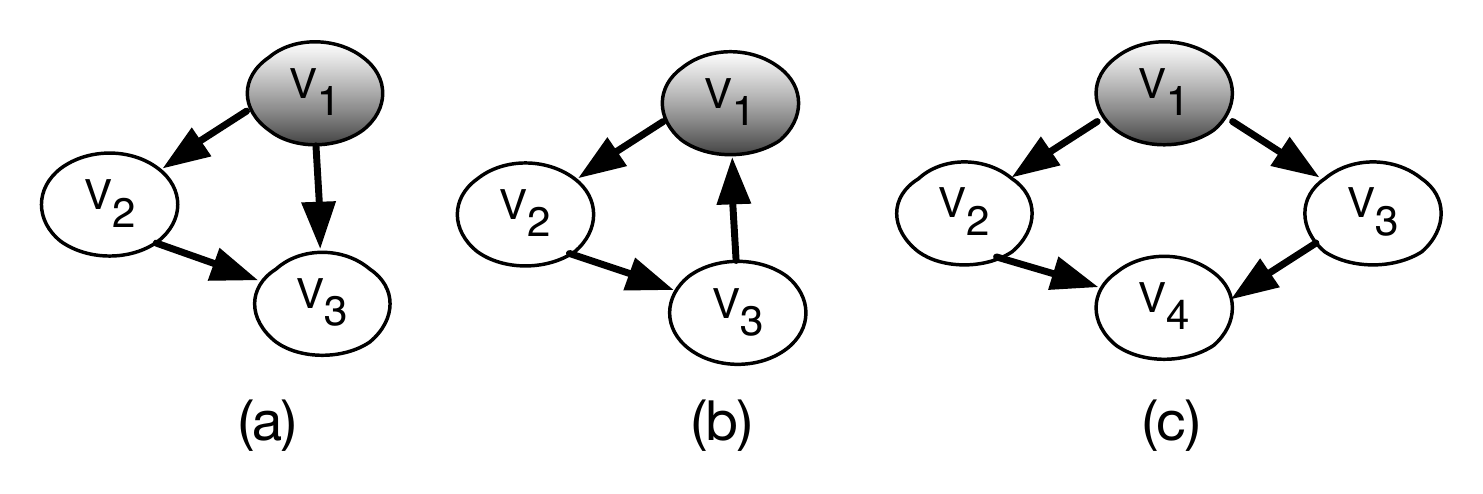}
\vspace{-10pt}
\caption{Counting different types of network motifs:\ (a) Feed-fwd Loop, (b) Feedback Loop, (c) Bi-parallel Motif.}
\label{fig:NW_motif}
\vspace{-5pt}
\end{figure}

\topic{Counting network motifs.} Network motifs are subgraphs that appear in
complex networks (Figure~\ref{fig:NW_motif}), which have important applications in biological networks and other domains.
However, counting network motifs over large graphs is quite challenging~\cite{kashtan2004efficient} as it involves identifying and counting subgraph patterns in the neighborhood of every query vertex that the user is interested in.
Once again, in a vertex-centric framework, this would entail
message passing to gather neighborhood data at each vertex,
incurring huge messaging and memory overheads.

\topic{Social recommendations.} Random walks with restarts (such as personalized PageRank~\cite{Backstrom:2011:SRW:1935826.1935914}) lie at the core of several social
recommendation algorithms. These algorithms can be
implemented using Monte-Carlo methods~\cite{Gupta:2013:WFS:2488388.2488433} where the random walk starts at a vertex $v$, and
repeatedly chooses a random outgoing edge and updates a visit counter with the restriction that the walk jumps back only to
$v$ with a certain probability. The stationary distribution of such a walk assigns a PageRank score to each vertex in the
neighborhood of $v$; these provide the basis for link prediction and recommendation algorithms. Implementing random walks in a vertex-centric framework would involve one iteration with message
passing for each step of the random walk.
In contrast, with \nframe~the complete state of the $k$-hop neighborhood around a vertex is available to the user's program, which can then directly execute personalized PageRank or any existing algorithm of choice.

\topic{Subgraph Pattern Matching and Isomorphism.}
\red{
Subgraph pattern matching or subgraph isomorphism have important applications in a 
variety of application domains including biological networks, chemical interaction networks,
social networks, and many others; and a wide variety of techniques have been developed for exact
or approximate subgraph pattern matching~\cite{Shasha:2002:AAT:543613.543620,Yan:2004:GIF:1007568.1007607,Cheng:fg:index:towards,Zhao:2007:GIT:1325851.1325957,Zou:2008:NSC:1353343.1353369,Ullmann:1976:ASI:321921.321925,P.Cordella:2004:GIA:1018035.1018377,Shang:2008:TVH:1453856.1453899,He:2008:GQL:1376616.1376660,Tian:2008:TTA:1546682.1547209,journals/jbcb/MongioviNGPFS10} (see Lee et al.~\cite{Lee:2012:ICS:2448936.2448946} for a recent comparison of the state-of-the-art techniques). 
Many of those techinques work by identifying potential matches for a central node in the pattern, and 
then exploring the neighborhood around those nodes to look for matches. This second step can often 
involve fairly sophisticated algorithms, especially if the patterns are large or contain
sophisticated constructs, or if the goal is to find approximate matches, or if the data is
uncertain. Most of those algorithms are not easily parallelizable, and hence it would not be easy
to execute them in a distributed fashion using the vertex-centric programming frameworks. On the other hand, 
\nframe~could be used to construct the relevant neighborhoods in memory in many of those cases, 
and those search algorithms could be used as is on those neighborhoods.
}

%
%


\section{NScale Overview}
\label{sec:overview}

\subsection{Programming Model}

We assume a standard definition of a graph $G(V,E)$ where $V=\{v_1,v_2,...,v_n\}$ denotes the set of vertices and $E=\{e_1,e_2,. .. , e_m\}$ denotes the set of edges in $G$. Let $A= \{ a_{1}, a_{2}, ... ,a_{k}\}$ denote the union of the sets of attributes associated with the vertices and edges in $G$.
In contrast to vertex-centric programming models, \nframe~allows users to
specify subgraphs or neighborhoods as the scope of computation. 
More specifically, users need to specify: (a) subgraphs of interest on which to run the computations
through a {\em subgraph extraction query}, and (b) a user program.

\begin{figure*}[t]
{
    \hspace{.6in}
\includegraphics[scale=0.55]{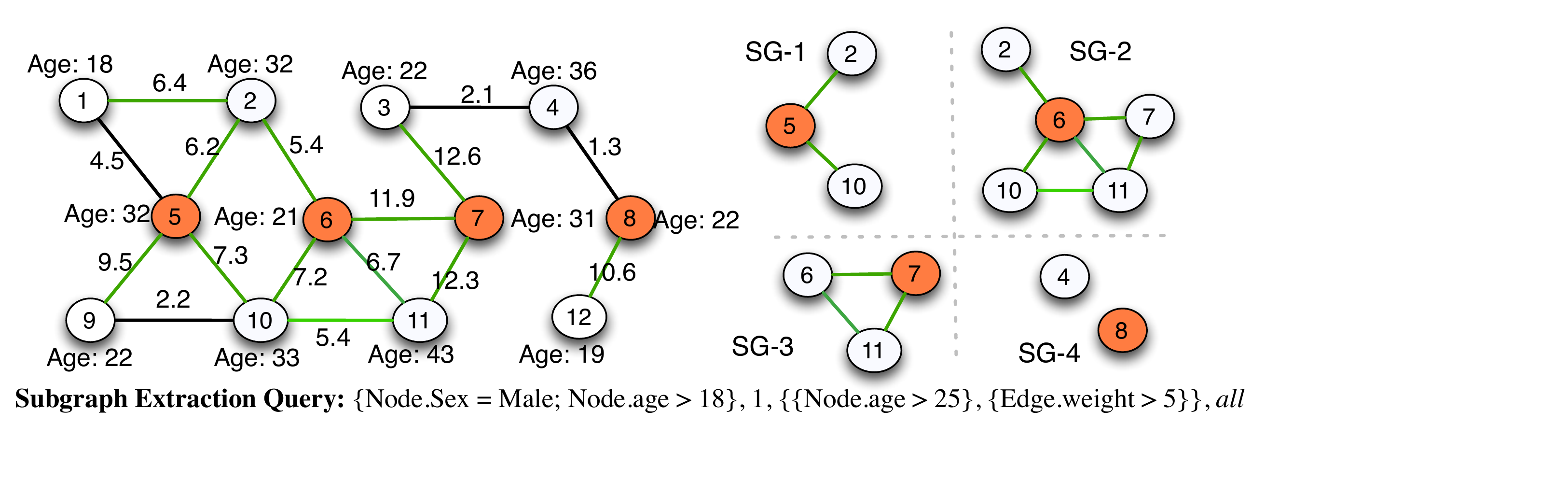}
}
\vspace{-30pt}
\caption{A subgraph extraction query on a social network}
\label{fig:socNW}
\end{figure*}

\topic{Specifying subgraphs of interest.} 
We envision that \nframe\ will support a wide range of subgraph extraction queries, including
pre-defined parameterized queries, and declaratively specified queries using a Datalog-based
language that we are currently developing. Currently, we support 
extraction queries that are specified in terms of four parameters:\ (1) a predicate on vertex attributes that
identifies a set of {\em query vertices} ($P_{QV}$), (2) $k$ -- the radius of the subgraphs of interest, (3) edge and vertex predicates
to select a subset of vertices and edges from those $k$-hop neighborhoods ($P_E, P_V$), and (4) a list of edge and vertex attributes that are of
interest ($A_E, A_V$). 
This captures a large number of subgraph-centric graph analysis tasks, including all of the tasks discussed earlier. 
For a given subgraph extraction query $q$, we denote the subgraphs of interest by $SG_1(V_1, E_1), ..., SG_q(V_q, E_q)$.

Figure~\ref{fig:socNW} shows an example subgraph extraction query, where the query vertices are
selected to be vertices with $age > 18$, radius is set to 1, and the user is interested in extracting
induced subgraphs containing vertices with $age > 25$ and edges with $weight > 5$. The four extracted
subgraphs, $SG_1, ..., SG_4$ are also shown.

\begin{figure}[t]
\begin{verbatim}
ArrayList<RVertex> n_arr = new ArrayList<RVertex>();
for(Edge e: this.getQueryVertex().getOutEdges)
     n_arr.add(e.getVertex(Direction.IN));

int possibleLinks = n_arr.size()* (n_arr.size()-1)/2;
		
// compute #actual edges among the neighbors
for(int i=0; i < n_arr.size()-1; i++)
	   for(int j=i+1; j < n_arr.size(); j++)
	      if(edgeExists(n_arr.get(i), n_arr.get(j)))
	          numEdges++;
double lcc = (double) numEdges/possibleLinks;
\end{verbatim}
\vspace{-2pt}
\caption{Example user program to compute {\em local clustering coefficient} written using the BluePrints API. The {\em edgeExists()}
call requires access to neighbors' states, and thus this program cannot be executed as is in a
    vertex-centric framework. 
}
\label{fig:graphCompute}
\vspace{-2pt}
\end{figure}

\topic{Specifying subgraph computation user program.} 
The user computation to be run against the subgraphs is specified as a Java program against the
BluePrints API~\cite{Blueprints:Online}, a collection of interfaces analogous to
JDBC but for graph data. Blueprints is a generic graph Java API used by many graph processing and programming
frameworks (e.g., Gremlin, a graph traversal language~\cite{Gremlin:Online}; Furnace, a graph algorithms package~\cite{Furnace:Online}; etc.). By supporting the Blueprints API,
we immediately enable use of many of these already existing toolkits over large graphs. 
Figure~\ref{fig:graphCompute} shows a sample code snippet
of how a user can write a simple local clustering coefficient computation using the BluePrints API.
The subgraphs of interest here are the 1-hop neighborhoods of all vertices (by definition, a 1-hop neighborhood
includes the edges between the neighbors of the node).

\red{\nframe~supports the Bulk Synchronous Protocol (BSP) for iterative execution, where the analysis task is executed 
using a number of iterations (also called {\em supersteps}). In each iteration, the user program is independently executed in parallel on all the subgraphs
(in a distributed fashion). The user program may then change the state of the query vertex on which it is operating 
(for consistent and deterministic semantics, we only allow the user program to change state of the query vertex that it
owns; otherwise we would need a mechanism to arbitrate conflicting changes to a vertex state and we are not aware of any 
clean and easy model for achieving that). The state changes are made visible across all the subgraphs during the 
synchronization barrier, through use of shared state for subgraphs on the same partition and through message passing
for subgraphs on different partitions. 
We provide a more detailed description of the provision of support for iterative computation in \nframe, including
the consistency and ownership model used, in Section~\ref{sec:discussion}.}

\red{Certain user applications might require customized aggregation of the values produced
as a result of executing the user-specified program on the subgraphs of interest. 
Our mechanism to handle state updates for iterative tasks can also be used for aggregating information 
across all the nodes in the graph in the synchronization step. To briefly summarize, the nodes can 
send messages to the coordinator that it can use to make various decisions (e.g., when to stop). 
The messages can be first locally aggregated, and the final aggregation is done by the 
coordinator (depending on the aggregation function).
}

\subsection{System Architecture}
\begin{figure}[t]
\centering
\includegraphics[width=\linewidth]{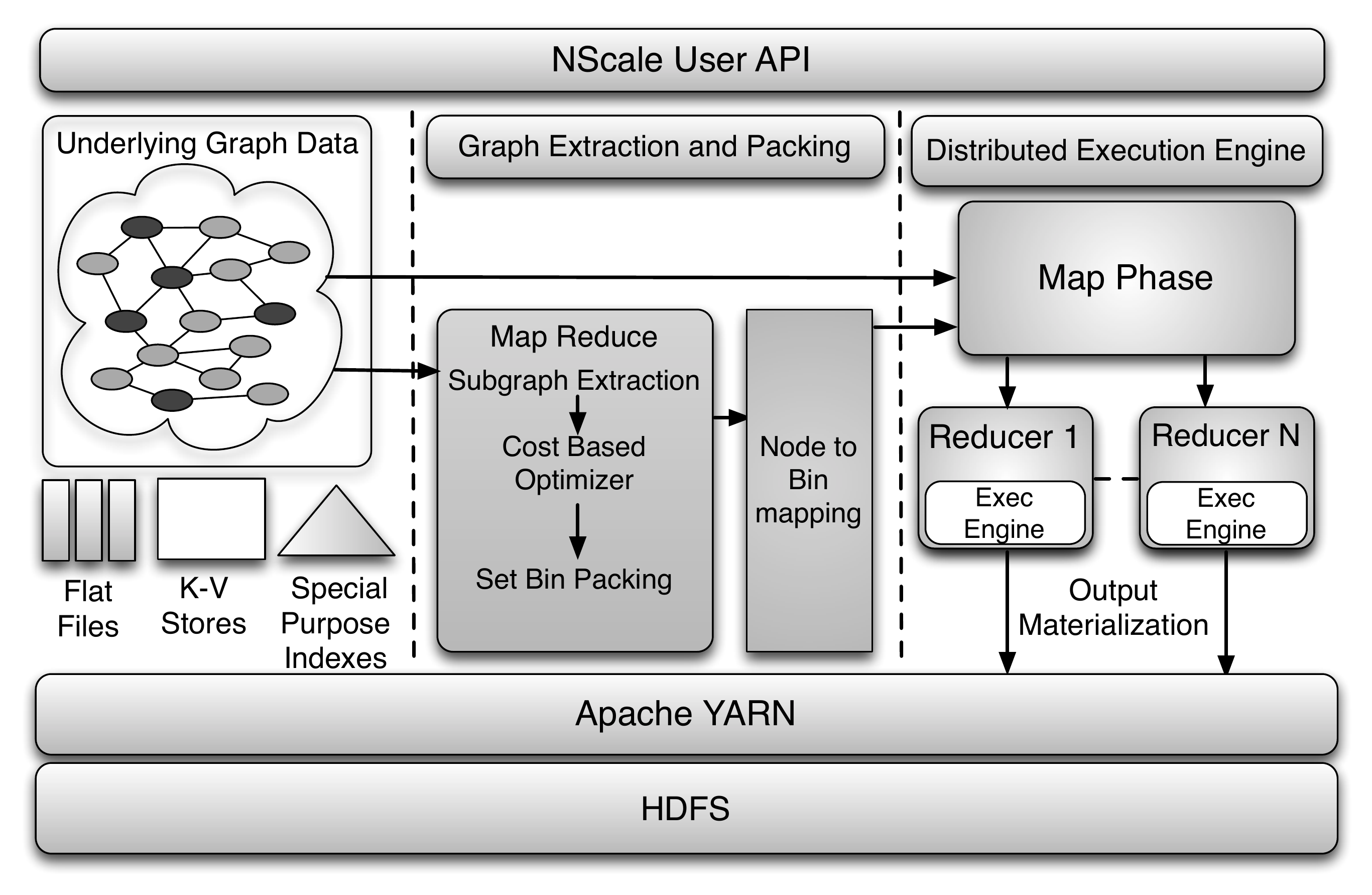}
\caption{\nframe~architecture. The GEP module is responsible for
  extracting and packing subgraphs of interest and then handing off
  the partitions to the distributed execution engine.}
\label{fig:arch}
\vspace{-2pt}
\end{figure}

Figure~\ref{fig:arch} shows the overall system architecture of
\nframe, which is implemented as a Hadoop YARN application.
The framework supports ingestion of the underlying graph in a variety of
different formats including edge lists, adjacency lists, and in a
variety of different types of persistent storage engines including
key--value pairs, specialized indexes stored in flat files, relational
databases, etc. The two major components of \nframe~are the graph
extraction and packing (GEP) module and the distributed execution
engine. We briefly discuss the key functionalities of these two
components here, and present details in the following sections.


\topic{Graph Extraction and Packing (GEP) Module.}
The user specifies the subgraphs of interest and the graph computation to be executed on them using the \nframe~user API. 
Unlike prior graph processing frameworks, the GEP module forms a major component of the 
overall \nframe~framework. From a usability perspective, it is important to provide the ability to
read the underlying graph from the persistent storage engines that are not naturally graph-oriented.
However, more importantly, partitioning and replication of the graph data are more critical for graph
analytics than for analytics on, say, relational or text data.

Graph analytics tasks, by their very
nature, tend to traverse graphs in an arbitrary and unpredictable manner. If the graph is
partitioned across a set of machines, then many of these traversals are made over the network, incurring
significant performance penalties. Further, as the number of partitions of a graph grows, the number of {\em cut} edges (with endpoints in different
partitions), and hence the number of distributed traversals, grows in a non-linear fashion. This is in contrast
to relational or text analytics where the number of machines used has a minor impact on the execution
cost. 

This is especially an issue in \nframe,
where user programs are treated as black-boxes. 
Hence, we have made a design decision to avoid distributed traversals altogether by replicating vertices and edges
sufficiently so that every subgraph of interest is fully present in at least one partition. Similar
approach has been taken by some of the prior work on efficiently executing ``fetch neighbors'' 
queries~\cite{journals/ton/PujolESYLCR12} and SPARQL queries~\cite{scalable-subgraphs} in distributed settings. 
The GEP module is used to ensure this property, and is responsible for extracting the subgraphs of interest and packing them
onto a small set of partitions such that every subgraph of interest is fully contained within at least one partition.
GEP is implemented as multiple MapReduce jobs (described in detail later).
The output is a {\em vertex-to-partition mapping}, which consists of a mapping from the graph vertices to 
partitions to be created.
This data is either written to HDFS or directly fed to the execution engine.


\topic{Distributed Execution Engine.}
The distributed execution phase in \nframe~is implemented as a MapReduce job, which
reads the original graph and the mappings generated by GEP, shuffles graph data onto 
a set of reducers, each of which constructs one of the partitions. 
Inside each reducer, the execution engine is instantiated along with the user program,
which then receives and processes the graph partition.

The execution engine supports both serial and parallel execution modes for executing user programs
on the extracted subgraphs. For serial execution, the execution engine uses a single thread and
loops across all the subgraphs in a partition, whereas for parallel execution, it uses a pool of
threads to execute the user computation in parallel on multiple subgraphs in the partition. 
However, this is not straightforward because the different subgraphs of interest in a partition are 
stored in an overlapping fashion in memory to reduce the total memory requirements. 
The execution engine employs several bitmap-based techniques to ensure correctness 
in that scenario.

\vspace{-15pt}
\section{Graph Extraction and Packing}
\vspace{-5pt}

\vspace{-2pt}
\subsection{Subgraph Extraction}
\vspace{-5pt}
\label{sec:gelPhase1}
Subgraph extraction in the GEP module has been implemented as a set of MapReduce (MR) jobs. 
The number of  MR stages needed depends on the size of the graph, how the graph is laid out,
size(s) of the machine(s) available to do the extraction, and the complexity of the subgraph
extraction query itself. The first stage of GEP is always a map stage that reads in the
underlying graph data, and identifies the {\em query vertices}. It also applies the filtering
predicates ($P_E, P_V$) to remove the vertices and edges that do not pass the predicates. It also
computes a {\em size} or {\em weight} for each vertex, that indicates how much memory is needed
to hold the vertex, its edges, and their attributes in a partition. This allows us to estimate the 
memory required by a subgraph as the sum of the weights of its constituent vertices. (Only the attributes identified 
in the extraction query are used to compute these weights.) The rest of 
the GEP process only operates upon the network structure (the vertices and the edges), and the vertex weights.

\topic{Case 1: Filtered graph structure is small enough to fit in a single machine.} In that case,
the vertices, their weights, and their edges are sent to a single reducer. That reducer constructs the subgraphs
of interest and represents them as subsets of vertices, i.e., each subgraph is represented as a list of 
vertices along with their weights (no edge information is retained further); this is sufficient for the subgraph packing 
purposes.
The subgraph packing algorithm takes as input these subsets of vertices
and the vertex weights, and produces a vertex-to-partition mapping. 

\topic{Case 2: Filtered graph structure does not fit on a single machine.} In that case, the subgraph extraction
and packing both are done in a distributed fashion, with the number of stages dependent on the radius ($k$) of
subgraphs of interest.
\begin{figure}[t]
\centering
\includegraphics[width=\linewidth]{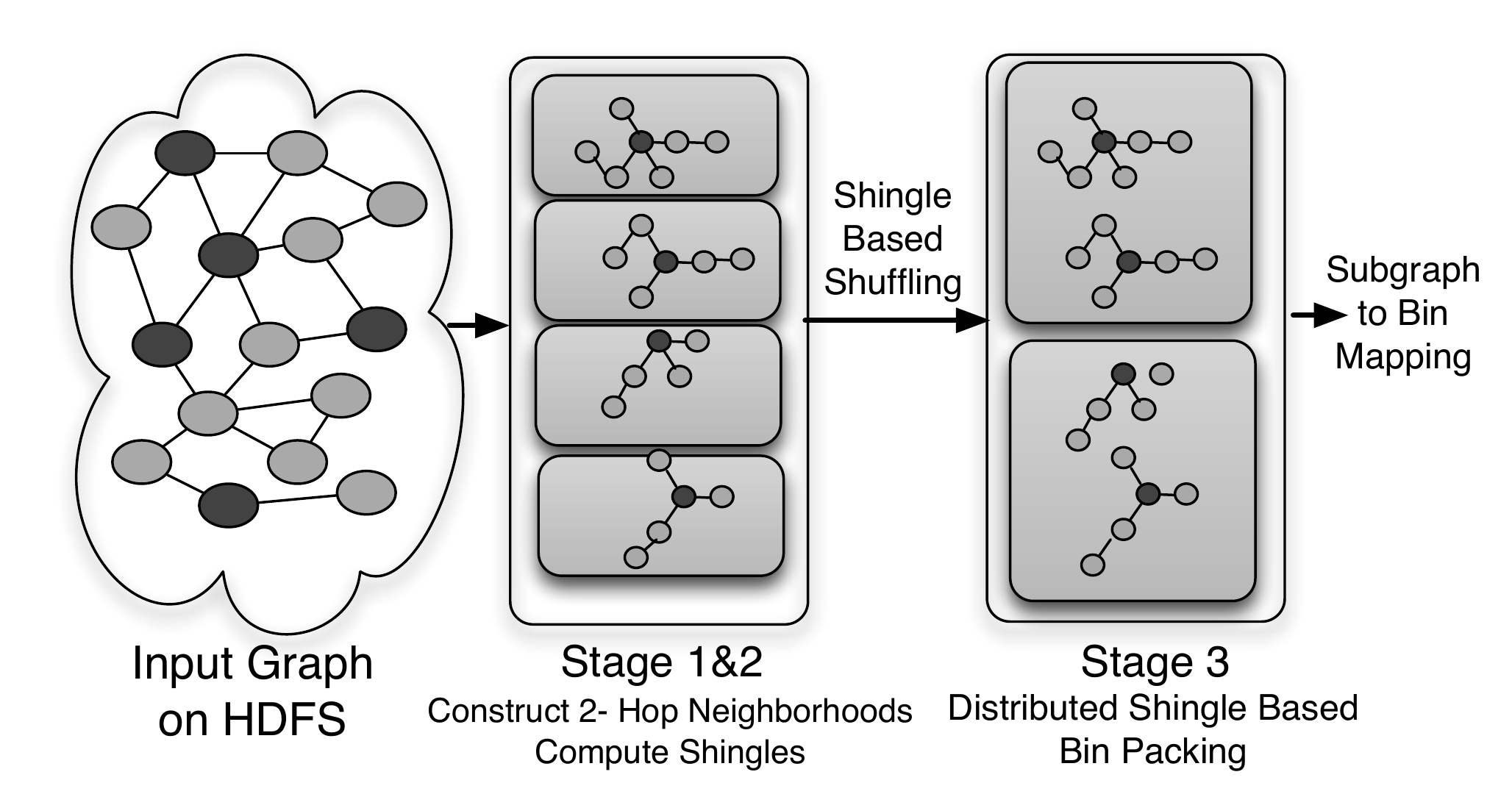}
\vspace{-15pt}
\caption{Distributed GEP Architecture: Stages 1 and 2  construct the 2-hop neighborhoods; Stage 3 
    does the distributed shingle based bin packing producing the final subgraph to bin mapping.}
\label{fig:DistrGel}
\vspace{-12pt}
\end{figure}
We explain the process assuming $k = 2$, i.e., assuming our subgraphs of interest are 2-hop neighborhoods around
a set of query vertices. We also assume an adjacency list representation of the data\footnote{\red{For input graphs represented as an edge list with the vertex attributes available as a separate mapping, we have a 
minor modification to the first stage that uses a MapReduce job to join the edge and vertex data and produce a 
distributed adjacency list in the required format.}} (i.e., the IDs of the neighbors of a vertex
are stored along with rest of its attributes); 

\red{Figure~\ref{fig:DistrGel} shows the 3-stage distributed architecture of GEP. We begin with providing a brief sketch of
the process. Given an input graph and a user query, the first two 
stages essentially are responsible for gathering for each query-vertex, its 2 hop neighborhood along with the weight 
attributes associated with each vertex in the 2-hop  neighborhood. This is done iteratively, wherein the first stage 
constructs the 1-hop neighborhood of the query-vertices specified by the query with all the required information on a set of reducers. Subsequently,  
the second stage takes the output of the first stage as input, constructs the 2-hop neighborhoods of the query-vertices  and computes
their shingle values in a distributed fashion, and outputs them as keys associated with these query-vertex neighborhoods. The final stage 
shuffles the neighborhoods based on these keys to multiple reducers in an attempt to group together neighborhoods with 
high overlap on a single reducer. The reducers in stage 3  run the bin packing in parallel which is followed by a post-processing step to produce
the final neighborhood-to-bin mapping. 
}

Next, we provide an in-depth description of the process. For a node $u$, let $N(u) = u_1, ..., u_{N(u)}$ denote its neighbors.
The following steps are taken:\\[2pt]
\underline{\bf MapReduce Stage 1:} For each vertex $u$ that passes the filtering predicates ($P_V$), the map stage 
emits $N(u)+1$ records:\\ $\langle key, (u, weight(u), isQueryVertex, N(u)) \rangle$, \\ where $key = u, u_1, ..., u_{N(u)}$.
Thus, given a vertex $u$, we have $N(u)'+1$ records that were emitted with $u$ as the key, one for its own information,
and one for each of its $N(u)'$ neighbors that satisfies $P_V$ (emitted while those neighbors are processed). 
In the reduce stage, the reducer
responsible for vertex $u$ now has all the information for its 1-hop neighbors, and IDs of all its 2-hop neighbors (obtained from its neighbors' neighborhoods), but it 
does not have the weights of 
its 2-hop neighbors or whether they satisfied the filtering predicates $P_V$. 
For each query vertex $u$, the reducer creates a list of 
the nodes in its 2-hop neighborhood, and outputs that information with key $u$. For each vertex $v$ and for each of its 2-hop
neighbors $w$, it also emits a record $\langle key = w, (v, weight(v)) \rangle$. \\[2pt]
\underline{\bf MapReduce Stage 2:} The second MapReduce stage groups the outputs of the first MapReduce stage by the vertex ID.
Each reducer processes a subset of the vertices. There are two types of records that a reducer might process for a vertex $u$: (a) a record
containing a list of $u$'s 1- and 2-hop neighbors and the weights of its 1-hop neighbors, and (b) several records each containing 
the weight of a 2-hop neighbor of $u$. 
If a reducer only sees the records of the second type, then $u$ is not a query vertex, and those records are discarded. Otherwise, the 
reducer adds the weight information for 2-hop neighbors, and completes the subgraph corresponding to $u$.
For each of the subgraphs, the reducer then computes a min-hash signature, i.e., a set of {\em shingles}, over the vertex set of the subgraph, 
and emits a record with the set of shingles as the key and the subgraph as the value (we use 4 shingles in our experiments).
A shingle is computed by applying a hash function to each of the vertex IDs in the subgraph, and taking the minimum of the hash values; it is well
known that if two
sets share a large fraction of the shingles, then they are likely to have a high overlap~\cite{Rajaraman:2011:MMD:2124405}. 
\\[2pt]
\underline{\bf MapReduce Stage 3:} The third MapReduce phase uses the shingle value of the subgraphs to shuffle the subgraphs to appropriate
reducers. As a result of this shuffling, the subgraphs that are assigned to a reducer are likely to have high overlap and the subgraph packing
algorithm is executed on each reducer separately. Finally, a post-processing step combines the results of all the
reducers by merging any partitions that might be underutilized in the solutions produced by the individual reducers.


Intuitively, the above sequence of MapReduce stages constructs the required subgraphs, and then does a shuffle using the shingles technique 
in an attempt to create groups that contain overlapping subgraphs. Those groups are then processed independently and the resulting
vertex-to-partition mappings are concatenated together.

\vspace{-15pt}
 \subsection{Subgraph Packing}
\vspace{-5pt}
 \label{sec:binPacking}
\Paragraph{Problem Definition.} We now formally define the problem of packing the extracted subgraphs into a minimum number of
partitions (or bins)\footnote{We use the terms {\em partitions} and {\em bins} interchangeably in this paper.}, such that each subgraph is contained within a partition and the computation
load across the partitions is balanced. 
Let $SG = \{SG_1, SG_2,..,SG_q \}$ be the set of subgraphs extracted from the underlying graph data 
(at a reducer).
As discussed earlier, we assume that the memory required to hold a subgraph $SG_i$ can be 
estimated as the sum of weights
of the nodes in it. Let $BC$ denote the bin capacity. This is set based on the maximum 
container capability of a YARN cluster node, a configuration parameter that needs to be set
for the YARN cluster keeping in mind the maximum allocation of resources to individual tasks on the
cluster

Without considering overlaps between subgraphs and the load balancing objective, this problem
reduces to the standard {\em bin packing} problem, where the goal is to minimize the number of
bins required to pack a given set of objects. The variation of the problem where the objects are
{\em sets}, and when packing multiple such objects into a bin, a {\em set union} is taken (i.e.,
overlaps are exploited), has been called {\em set bin packing}; that problem is considered much
harder and we have found very little prior work on that problem~\cite{izumi1998computational}. 

Further, we note that we have a dual-objective optimization problem; we reduce it to a single-objective 
optimization problem by putting a constraint on the number of subgraphs that can be assigned to a bin.
Let $MAX$ denote the constraint, i.e., the maximum number of subgraphs that can be assigned to a bin.

\Paragraph{Subgraph Bin Packing Algorithms.} The subgraph bin packing problem is NP-Hard and appears to be much harder to solve than the 
standard {\em bin packing} problem, as it also exhibits some of the features of the 
{\em set cover} and the {\em graph partitioning} problems. Next, we develop several 
scalable heuristics to solve this problem. We also developed and implemented an optimal
algorithm for this problem (OPT), where we construct an Integer Program for the given problem instance
and use the Gurobi Optimizer to solve the Integer Program. We were, however, able to run OPT successfully
only for a very few small graphs; we present those results in Section~\ref{sec:GELEval}.

\vspace{-8pt}
\subsubsection{Bin Packing-based Algorithms}
The first set of heuristics that we develop exploit the similarity between subgraph packing problem and 
the bin packing problem. All of these heuristics use the standard greedy bin packing algorithm, where the items
are considered in a particular order and placed in the first bin where they fit. More specifically, 
the algorithm (Algorithm~\ref{alg:vbp}) 
    takes as input an ordered list of subgraphs, as determined 
by the heuristic, processes them in order, and packs each subgraph into the first available
bin that has the available residual capacity, without violating the constraint on the maximum
number of subgraphs in a bin. The addition of a subgraph to a bin is a set union operation that takes care of the overlap
between the subgraphs. 
Each bin represents a partition onto which the actual graph data, associated with the nodes mapped to the bin
using this algorithm, would be distributed for final execution step. 

\red{The complexity of this algorithm in the worst case in terms of the number of comparison operations required is  
$\mathcal{O}(nm)$ where $n$ is the number of subgraphs and $m$ is the number of bins required ($= n$ in the worst case). 
Each comparison operation compares the estimated size of the union (accounting for the overlap) and the bin capacity. In addition to these 
comparisons, there would be $n$ set union operations for inserting the subgraphs into bins. The complexity of the 
comparison and the set union operations is implementation dependent. For a hashtable-based approach, those operations would be
linear in the number of set elements, giving us an overall complexity of $\mathcal{O}(nmC)$, where $C$ is the bin capacity.
However this worst-case complexity is quite pessimistic, and in practice, the algorithms run very fast.}

We now describe three different heuristics to provide the input ordering of the subgraphs to be packed into bins. 


\begin{algorithm}
\SetKwInOut{Input}{Input}\SetKwInOut{Output}{Output}
\Input{Ordered list of subgraphs $SG_1, ..., SG_q$, each represented as a list of vertices and edges}
\Input{Bin capacity $BC$; Maximum number of subgraphs per bin $MAX$}
\Output{Partitions}
\For {$i = 1, 2, ..., q$}
{
     \For{$j = 1, 2,..., B$}
      {
          \If{number of subgraphs in Bin j $<$ MAX} 
          {
          \If{$SG_i$ fits in Bin $j$ (accounting for overlap)}
          {
              Add $SG_i$ to Bin $j$\;
              break\;
          }
          }
      }
      \If{$SG_i$ not yet placed in a bin}
      {
        Create a new bin and add $SG_i$ to it\;
      }
}
\label{alg:vbp}
\caption{Bin Packing Algorithm.}
\end{algorithm}

\Paragraph{1. First Fit bin packing algorithm.}  The first fit algorithm is a standard greedy
2-approximation algorithm for bin packing, and processes the subgraphs in the order in which they were received
(i.e., in arbitrary order). 

\Paragraph{2. First Fit Decreasing bin packing algorithm.} The first fit decreasing algorithm is a variant of the first fit algorithm wherein
the subgraphs are considered in the decreasing order of their sizes. 

\Paragraph{3. Shingle-based bin packing algorithm.} The key idea behind this heuristic is to
order the subgraphs with respect to the similarity of their vertex sets. The ordering so
produced will maximize the probability that subgraphs with high overlap are processed
together, potentially resulting in a better overall packing.

\red{The shingle-based ordering is based on the min-hashing
technique~\cite{mmds} which produces signatures for large sets that can be used
to estimate the similarity of the sets. For computing the {\em min-hash}
signatures (or shingles) of the subgraphs of interest over their vertex
set, we choose a set of $k$ different random hash functions to simulate the
effect of choosing $k$ random permutations of the characteristic matrix that
represents the subgraphs. For each query vertex and each hash function, we apply
the hash function to the set of nodes in the subgraph of the
query vertex and find the minimum among the hash values. 
}

\red{Thus the output of the shingle computation algorithm (Ref
Algorithm~\ref{alg:shinglesAlgorithm}) is a list of $k$ shingles (min-hash values) 
for each subgraph of interest, where the order of the hash functions within the list is effectively
arbitrary\footnote{The higher the value of $k$, the better the quality of the result. We have chosen $k = 6$ for our
implementation which was determined experimentally to strike a fine balance between
the quality of shingle-based similarity and computation time.}. To compute the shingle
ordering, we sort-order the subgraphs of interest based on this list
of shingle values associated with the subgraphs in a lexicographical fashion.
The sorted order so obtained using this
technique places subgraphs with high Jaccard similarity (i.e., overlap) in
close proximity to each other. This shingle-based order is then used to pack
the neighborhoods into bins using the greedy algorithm.}

\vspace{2pt}

\noindent
\red{{\bf Handling skew.} A high variance in the sizes of subgraphs could lead
to a bin packing where some partitions have only a few large subgraphs and few
partitions have a very large number of small subgraphs. This might lead  to
load imbalance and skewed execution times across partitions. To handle this
skew in the sizes of the subgraphs, the bin packing algorithm (Algorithm 1)
accepts a constraint on the maximum number of subgraphs (MAX) in a bin in addition
to the bin capacity. This limits the number of small subgraphs that
can be binned together in a partition and mitigates the potential of load imbalance between
partitions to some degree. The trade-off here is that, we may need to use a higher number
of bins to satisfy the constraints while some of the bins are not fully utilized. 
The MAX parameter can be set  empirically depending on the nature
of user computation and the underlying graph keeping in view the above
mentioned trade-off.}


\begin{algorithm}[t]
\SetKwInOut{Input}{Input}\SetKwInOut{Output}{Output}
\Input{Subgraph $SG(V, E)$; A family of pairwise-independent hash functions $H$}
$shingles[SG_i] \leftarrow \{\}$\;
\For{$h \in H$}
{
    $shingles[SG] \leftarrow \{ shingles[SG],  min_{v \in V} h(v) \}$\;
}
\Return shingles\;
\label{alg:shinglesAlgorithm}
\caption{Computing shingles for a subgraph}
\end{algorithm}

\vspace{-5pt}
\subsubsection{Graph Partitioning-based Algorithms}
The subgraph packing problem has some 
similarities to the {\em graph partitioning} problem, with the key difference
being that: standard graph partitioning problem asks for disjoint balanced partitions, whereas the partitions that we need 
to create typically have overlap in order to satisfy the requirement that each subgraph be completely contained within 
at least one partition. Graph partitioning is very well-studied and a number of packages are available that can partition 
large graphs efficiently, METIS perhaps being the most widely used~\cite{Metis:Online}. 

Despite the similarities, graph partitioning algorithms turn out to be a bad fit for the subgraph packing problem, because
it is not easy to enforce the constraint that each subgraph of interest be completely contained in a partition. One option is
to start with a disjoint partitioning returned by a graph partitioning algorithm, and then ``grow'' each of the partitions to ensure
that constraint. However, we also need to ensure that the enlarged partitions obey the bin capacity constraint, which is hard to achieve since
different partitions may get enlarged by different amounts. 

We instead take the following approach (Algorithm~\ref{alg:graphPkgAlgorithm}). We {\em overpartition} the graph using a standard graph partitioning algorithm (we use METIS in our
implementation) into a large number of fine-grained partitions. We then grow each of those partitions as needed. This requires that for each query vertex
in the fine grained partition, we check is its $k$-hop neighborhood lies within the partition. If not, we replicate the required nodes in the partition. This
ensures that each subgraph of interest is fully contained in one of the partitions, and finally use the
shingle-based bin packing heuristic to pack those partitions into bins. While packing, we also keep track of the nodes that are owned by the bin (or partition)
and the ones that are replicated (ghosts) from other bins, to maintain the invariant of keeping each subgraph of interest fully in the memory of one of the 
partitions.

\begin{algorithm}[t]
\SetKwInOut{Input}{Input}\SetKwInOut{Output}{Output}
\Input{Graph $G(V, E)$; Num of over partitions $k$}
\Output{Bins $\mathcal{B}$}
//Over partition $G$ into $k$ partitions.\;
$\mathcal{P} \leftarrow$ Metis(G); where $|\mathcal{P}|=k$\;
\For{$p \in P$}
{
    \For{$qv \in p$}
    {
      \If{ ! $(k-hop$  $neighborhood)  \in p$}
      {
      	Grow: Replicate the required nodes\
	adding them to $p$\;
      }
    }
    
}
//Compute Shingles for each grown partition\;
\For{i = 1 to $|P|$}
{
	$s_i = ComputeShingles(p_i)$\;
}
//Sort the partitions based on shingle values ($s_i$) \;
Sort($P$)\;
$\mathcal{B} = BinPackingAlgo(P)$\;
\Return $\mathcal{B}$\;
\label{alg:graphPkgAlgorithm}
\caption{Graph Partitioning-based algorithm.}
\end{algorithm}

\subsubsection{Clustering-based Algorithms}
The subgraph packing problem also has similarities to {\em clustering}, since our goal can be seen
as identifying similar (i.e., overlapping) subgraphs and grouping them together into bins. We
developed two heuristics based on the two commonly used clustering techniques.

\topic{Agglomerative Clustering-based Algorithm.} Agglomerative clustering refers to a class of
bottom-up algorithms that start with
each item being in its own cluster, and recursively merge the closest clusters till the
requisite number of clusters is reached. For handling large volumes of data, a threshold-based
approach is typically used where in each step, pairs of clusters that are sufficiently close to
each other are merged, and the threshold is slowly increased. 
Next we sketch our adaptation of this technique to subgraph packing. 

We start with computing a set of shingles for each subgraph and ordering the subgraphs in the
shingle order. This is done in order to reduce the number of pairs of clusters that we
consider for merging; in other words, we only consider those pairs for merging that are sufficiently
close to each other in the shingle order. The function $createAggClusters()$ in Algorithm~\ref{alg:agglomerativeAlgorithm} 
does the actual scanning of sets and merges close by sets together.
The algorithm uses two parameters, both of which are adapted
during the execution: (1) $\tau$, a threshold that controls when we merge clusters, and (2) $l$, that
controls how many pairs of clusters we consider for merging. In other words, we only merge a pair of
clusters if they are less than $l$ apart in the shingle order, and the Jaccard distance between them is less than
$\tau$. The set of merged clusters are available as $AC$.

\begin{algorithm}[t]
\SetKwInOut{Input}{Input}\SetKwInOut{Output}{Output}
\Input{Set of subgraphs $SG=\{SG_1, ..., SG_q\}$}
\Input{Merge size $l$ (Number of pairs to be considered for merging.)}
\Output{Agglomerative Clusters (Bins) $\mathcal{AC}$}
//Compute Shingles of each subgraph\;
\For{$i=1$ to $q$}
{
	$s_i=ComputeShingles(SG_i)$\;
	
}
/*Sort the subgraphs based on their \
shingle values ($S=\{s_1, s_2,..s_q\}$)*/\;
Sort($SG$) \;
Done=false\;

//Create an empty set of agglomerative clusters\;
 $\mathcal{AC}\leftarrow \phi$\;
\While{!Done}
{
	$\mathcal{\tau}$ = setThreshold()\;
	$numMerges = createAggCluster(SG,\mathcal{AC}, \mathcal{\tau},I )$\;
	\If{$numMerges = 0$}
	{
		Done=True\;
		break;
		
	}
	//adjust the merge size if required\;
	$I$ = adjustMergeSize()\;
	//Re-Compute Shingles of each merged cluster\;
	$m=|\mathcal{AC}|$\;
	\For{$i=1$ to $m$} 
	{
		$s_i=ComputeShingles(AC_i)$\;
	
	}
	//Sort clusters based on their shingle values ($s_i$).\
	Sort($AC$) \;
	$SG=\mathcal{AC}$;
}

\Return $\mathcal{AC}$\;
\label{alg:agglomerativeAlgorithm}
\caption{Agglomerative Clustering-based algorithm.}
\end{algorithm}

To reduce the number of parameters, we use a sampling-based approach in the function $setThreshold()$ in 
Algorithm~\ref{alg:agglomerativeAlgorithm},  to set $\tau$ at the beginning
of each iteration. We choose a random sample of the eligible pairs (we use 1\%
sample), compute the Jaccard distance for each pair, and set
$\tau$ such that 10\% of those pairs of clusters would have distances below $\tau$. We experimented
with different percentage thresholds, and we observed that 10\% gave us the best mix of quality and
running time.  

After computing $\tau$, we make a linear scan over the clusters that have been constructed so far.
For each cluster, we compute its actual Jaccard distance with the $l$ clusters 
that follow it. If the smallest of those distances is less than $\tau$, then we merge the two
clusters and re-compute shingles for the merged cluster (this is done by simply picking the minimum
of the two values for each shingle position).  This is only done if the merged cluster does not
exceed the bin capacity (pairs of clusters whose union exceeds bin capacity are also excluded from
the computation of $\tau$).  

During computation of $\tau$, we also keep track of the number of pairs excluded because the size of
their union is larger than the bin capacity. If those pairs form more 50\% of sampled pairs, then we
increase $l$ ($adjustMergeSize()$) to increase the pool of eligible pairs. Since this usually happens towards the end
when the number of clusters is small, we do this aggressively by increasing $l$ by 50\% each time.
The algorithm halts when it cannot merge any pair of clusters without violating the bin capacity 
constraint.

\begin{algorithm}[t]
\SetKwInOut{Input}{Input}\SetKwInOut{Output}{Output}
\Input{Set of subgraphs $SG=\{SG_1, ..., SG_q\}$; Bin Capacity $BC$}
\Input{$k$: The number of K-Means Clusters; MAX: maximum iterations}
\Output{Bins $\mathcal{B}$}

//Create an empty centroid set\
$\mathcal{KC}\leftarrow \phi$\;

//Randomly pick k subgraphs and assign them as the k-centroids\;
\While{(Sizeof($\mathcal{KC}) < k$)}
{	//Generate a random number from 1 to k
	i=GenerateRandom(k)\;
	$\mathcal{KC}=\mathcal{KC} \bigcup SG_i$
}

//Scan over the set of subgraphs and assign them to nearest centroid\;
$AssignmentMap\leftarrow \phi$\;
\For{i = 1 to q}
{	\If{!($SG_i \in \mathcal{KC}$)}
	{
		Max = $-\infty$\;
		CentroidAssigned =0\;
		\For{j=1 to k}
		{
			dist = $computeDistance(SG_i, KC_j, BC)$\;
			\If{($Max  <  dist$)}
			{	
				Max = dist\;
				CentroidAssigned = j\;
			}

		}
		UpdateCentroid($SG_i, KC_{CentroidAssigned}$)\;
		AssignmentMap.Put(i,CentroidAssigned)\;
	}
}

//Update assignments iteratively to improve clustering\;
numIterations=0\;
\While{$numIterations<MAX$} 
{
	\For{i = to q}
	{
		CurrentAssignment = AssignmentMap.Get(i)\;
		\For{j = 1 to k}
		{
			SwapGain = ComputeGain(i, CurrentAssignment, j)\;
			\If{($SwapGain>0$)}
			{
				Swap(i, CurrentAssignment, j)\;
			}
		}
	}
	numIterations++\;
}

$\mathcal{B} = BinPackingAlgo(\mathcal{KC})$\;
\Return $\mathcal{B}$\;
\label{alg:KMeansAlgorithm}
\caption{KMeans Clustering-based algorithm.}
\end{algorithm}

\begin{algorithm}
\SetKwInOut{Input}{Input}\SetKwInOut{Output}{Output}
\Input{Subgraph $SG$; Centroid $C$; Bin Capacity $BC$}
\Output{Distance between $SG$ and $C$}
\eIf{$|SG \cup C|> BC$}{
\Return{-$\infty$}
}{
\Return{$|SG \cap C|$}
}
\caption{ComputeDistance()}
\end{algorithm}

\topic{K-Means-based Algorithm.} K-Means is perhaps the most commonly used algorithm for clustering,
and is known for its scalability and for constructing good quality clusters. Our adaptation of
K-means (Ref Algorithm~\ref{alg:KMeansAlgorithm}) is sketched next. 

We start by picking $k$ of the subgraphs randomly as {\em centroids}. We then make a linear scan over the
subgraphs and for each subgraph, we compute the distance to each centroid using the function $computeDistance()$.  
We assign the subgraph to the centroid with which it has the highest
intersection (in other words, we assign it to the centroid whose size needs to increase the least
to include the subgraph). This is only done if the total size of the vertices in the cluster
does not exceed $BC$.
After assigning the subgraph to the centroid, we recompute the
centroid ($UpdateCentroid()$) as the union of the old centroid and the subgraph. The function also keeps track of multiplicities of
the vertices in the centroid at all times (i.e., for each vertex in a centroid, we keep track of how many of the
assigned subgraphs contain it). 

As with K-Means, we make repeated passes over the list of subgraphs in order to improve the
clustering. In the subsequent iterations, for each subgraph, we check if it may improve the solution 
using the function $ComputeGain()$. If the swap gain is positive, i.e. there is a net decrease in the sum of the size of the centroids
involved in the swap, we reassign the subgraph
to a different centroid, using the
multiplicities to remove it from one centroid and assign it to the other centroid ($Swap()$). Finally the k cluster obtained are
packed into bins (or partitions).

Having to choose a value of $k$ a priori is one of the key disadvantages of K-Means. We estimate a
value of $k$ based on the subgraph sizes and the bin capacity. If at the end of first iteration, we
discover that we are left with too many unassigned subgraphs, we increase the value of $k$ and
repeat the process till we are able to find a good clustering.

\subsection{Handling Very Large Subgraphs} 
Most machines today, even commodity machines, have large amounts of RAM available, and can easily handle 
very large subgraphs, including 2-hop neighborhoods of high-degree nodes in large-scale networks. However, in the rare
case of a subgraph extraction query where one of the subgraphs extracted is too large to fit into
the memory of a single machine, we have two options. The first option is to use disk-resident
processing, by storing the subgraph on the disk and loading it into memory as needed. The user
program may need to be modified so that it does not thrash in such a scenario. We note here that our
flexible programming model makes it difficult to process the subgraph in a distributed fashion (i.e., by partitioning
        the subgraph across a set of distributed machines); if 
this scenario is common, we may wish to enforce a vertex-centric programming model within \nframe,
and that is something we plan to consider in future work.

The other option, that we currently support in \nframe\ and is arguably better suited for handling large subgraphs, 
is to use {\em sampling} to reduce the size of the subgraph. We currently assume that the subgraph skeleton
(i.e., the network structure of a subgraph) can be held in the memory of a single machine during
GEP; this is needed to support many of the effective random sampling techniques like forest fire or
random walks (independent random sampling can be used without making this assumption)~\cite{Leskovec:2006:SLG:1150402.1150479},~\cite{Popescu:2013:PTP:2556549.2556553}. The key idea 
here is to construct a random
sample of a subgraph during GEP, if the size of the subgraph is estimated to be larger than the
bin capacity. We provide built-in support for two random
sampling techniques: {\em random node selection}, and {\em random walk-based sampling}. The former technique
chooses an independent random sample of the nodes to be part of the subgraph, whereas the latter
technique does random walks starting with the query vertex and including all visited nodes in the
sample (till a desired sample size is reached). \nframe~also provides a
flexible API for users to implement and provide their own graph sampling/compression technique. 
The random sampling is performed at the reduce stage in GEP where the subgraph skeleton is first
constructed.

 \begin{figure}[t]
\centering
\includegraphics[scale=0.60]{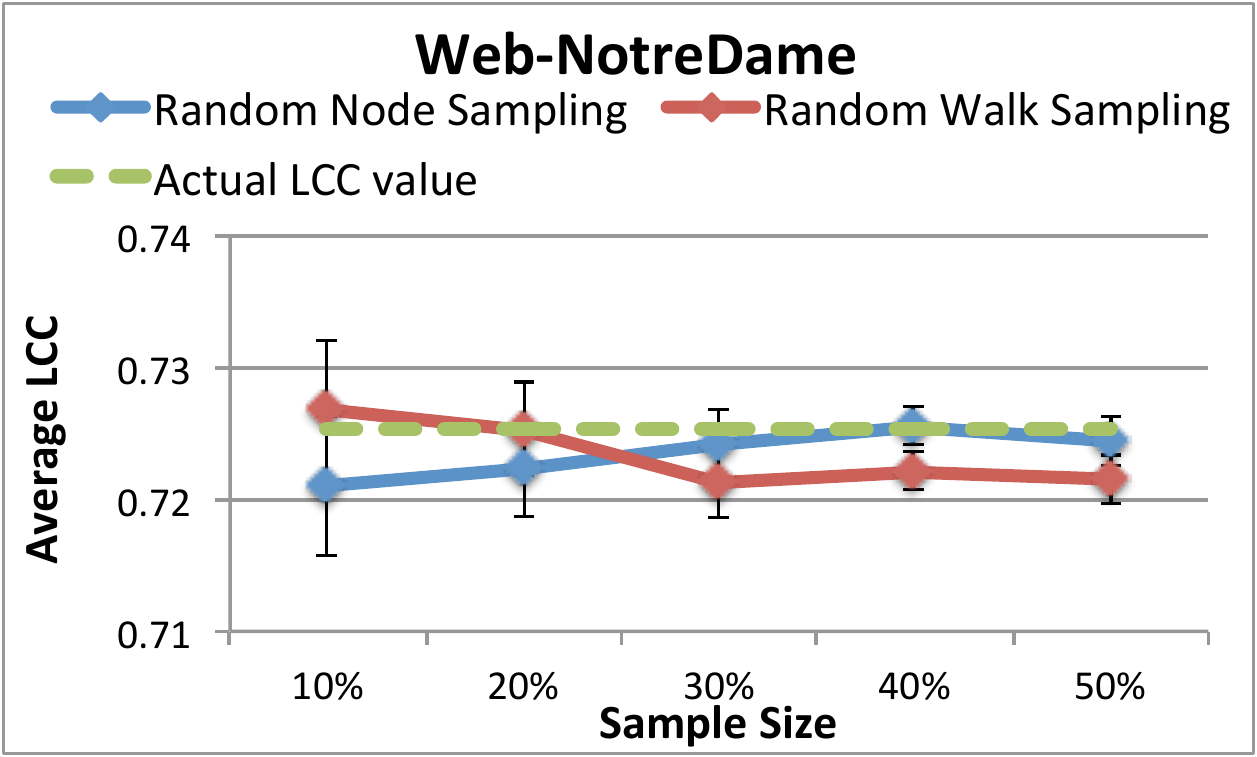}\\
\vskip 1ex
\includegraphics[scale=0.60]{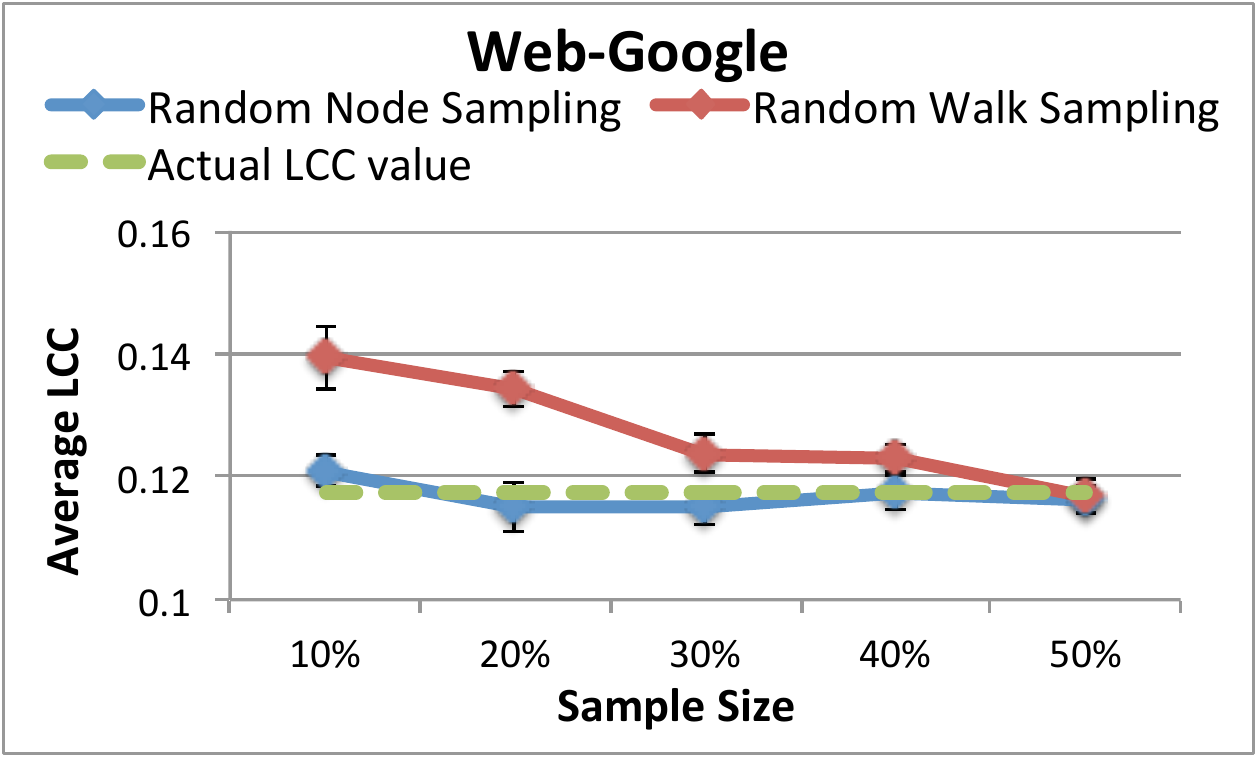}\\
\vskip -1ex
\caption{Effect of Graph Sampling}
\label{fig:GraphSampling}
\vskip -2ex
\end{figure}

\red{
Figure~\ref{fig:GraphSampling} shows the effect of using our random node and random
walk-based sampling algorithms on the accuracy of the
local clustering coefficient (LCC)
computation. We plot the
 average LCC computed on samples of different sizes for two
different data sets, and compare them to the actual result.
Each data point is an average of 10 runs. We also show the standard deviation error bars.
For the random node-based sampling techniques, the standard deviation across multiple random runs decreases and the accuracy increases
as the sampling ratio increases (as seen in that figure). 
This is not surprising since the estimated LCC through this technique is an unbiased estimator for the true average LCC (although it has a very high 
variance).
For the random walk-based sampling, the numbers do not show any consistent trend since the set of sampled nodes does not have any uniformity guarantees
and in fact, the set of sampled nodes would be biased towards the high degree nodes (and the effect on the estimated LCC would be arbitrary since the degree
of a node is not directly correlated with the LCC for that node).}


\vspace{-15pt}
\section{Distributed Execution Engine}
\label{sec:DistrExecEng}

The \nframe~distributed execution engine runs inside the reduce stage of a MapReduce job (Figure
        \ref{fig:arch}). The map stage takes as input the original graph and the vertex-to-partition
mappings that are computed by the GEP module, and it replicates and shuffles the graph data so
that each of the reducers gets the data corresponding to one of the partitions. Each reducer
constructs the graph in memory from the data that it receives, and identifies the subgraphs owned by
it (the vertex-to-partition mappings contain this information as well).
It then uses a 
worker thread pool to execute the user computation on those subgraphs.
The output of the graph computation is written to HDFS. 
 
%

\subsection{Execution modes} The execution engine provides several different execution modes.
The {\em vector bitmap mode} associates a bit-vector with
each vertex and edge in the partition graph, and enables parallel execution of user computation on
different subgraphs.
The {\em batched bitmap mode} is an optimization that uses smaller bitmaps to reduce
memory consumption, at the expense of increased execution time.
The {\em single bit bitmap mode} associates a single bit with each vertex and edge,
consuming less memory but allowing for only serial execution of the computation on the subgraphs in a partition.

\Paragraphul{Vector Bitmap Mode.} Here each vertex and edge 
is associated with a bitmap, whose size is equal to the number of subgraphs in the partition. Each vector bit position is associated with one subgraph
and is set to 1 if the vertex or the edge participates in the subgraph computation. A master process on each partition schedules a set of
worker threads in parallel, one per subgraph. 
Each worker thread executes the user computation on its subgraph, using the corresponding bit to
control what data the user computation sees. 
     Specifically, our BluePrints API implementation interprets the bitmaps to only return the 
     elements (vertices or edges or attributes) that the callee should see. The
     use of bitmaps thus obviates the need for state duplication and enables efficient parallel execution
     of user computation on subgraphs. \red{For consistent and deterministic execution of the user computation,
     each worker thread can only update the state of the query-vertex contained in its subgraph. We 
     discuss the details of this consistency mechanism in greater detail in Section~\ref{sec:discussion}.}

\begin{figure}[t]
 \vspace{-10pt}
\centering
\includegraphics[scale=0.42]{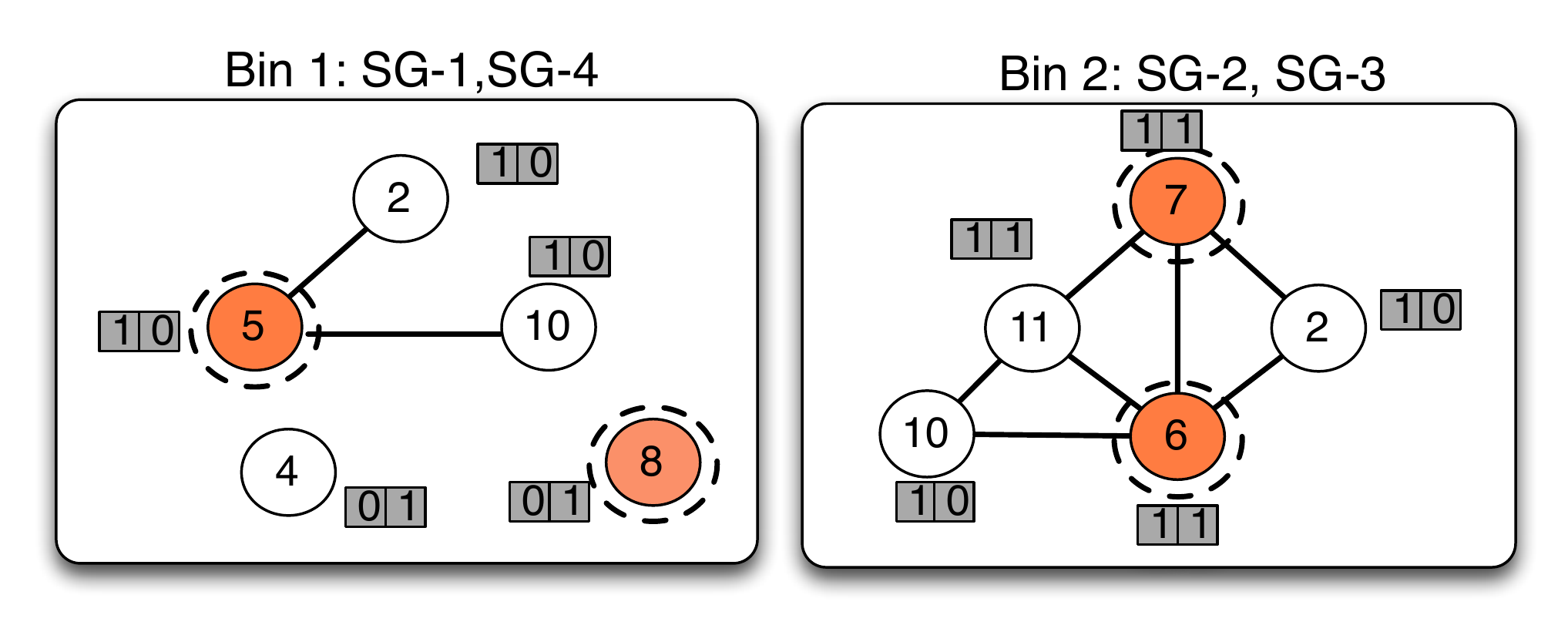}
\vspace{-10pt}
\caption{Bitmap based parallel execution}
\label{fig:execModel}
\vspace{-5pt}
\end{figure}

 Figure~\ref{fig:execModel} shows an example bitmap setting for the subgraphs extracted in
 Figure~\ref{fig:socNW}. In Bin 2, subgraphs 2 and 3 share nodes 6 and 7 which have both the bits in the vector bitmap set to 1 indicating that they belong to both the subgraphs. 
 All other nodes in the bins have only one of their bits set, indicating appropriate subgraph membership. 

\Paragraphul{Batching Bitmap Mode.}  As the system scales to a very large number of subgraphs per
reducer, 
the memory consumed by the bitmaps can grow rapidly. At the same time, the maximum parallelism that
can be achieved is constrained by the hardware configuration, and it is likely that only a small
number of subgraphs can actually be processed in parallel. The batching bitmap mode exploits this by limiting
up front the number of subgraphs that may be processed in parallel.
Specifically, we batch the subgraphs into batches of a fixed size (called {\em batch-size}), and
process the subgraphs one batch at a time. A bitmap of length {\em batch-size} is sufficient now to
indicate to which subgraphs in the batch a vertex or a node contributes. After a batch is finished, the 
bitmaps are re-initialized and the next batch commences.

\begin{figure}[t]
  \centering
   \subfigure[]{\label{fig:batching_1}\includegraphics[scale=0.60]{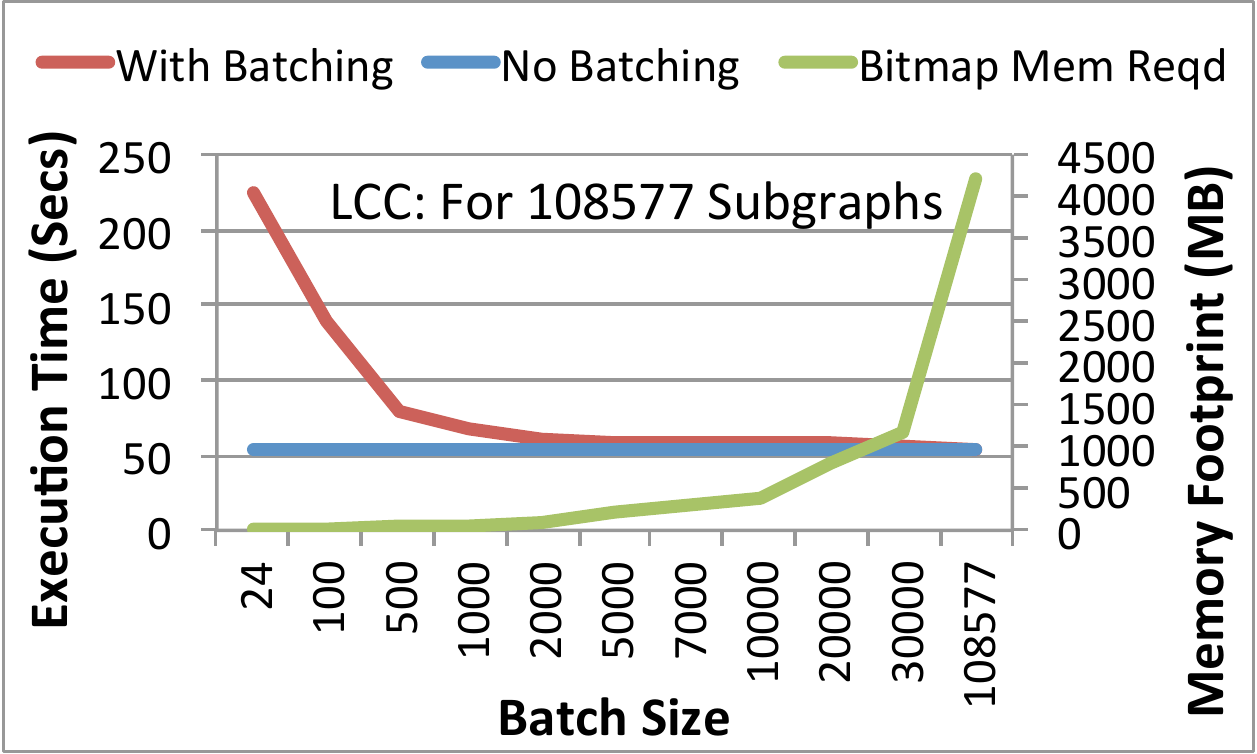}}\\
\subfigure[]{\label{fig:batching_2}\includegraphics[scale=0.60]{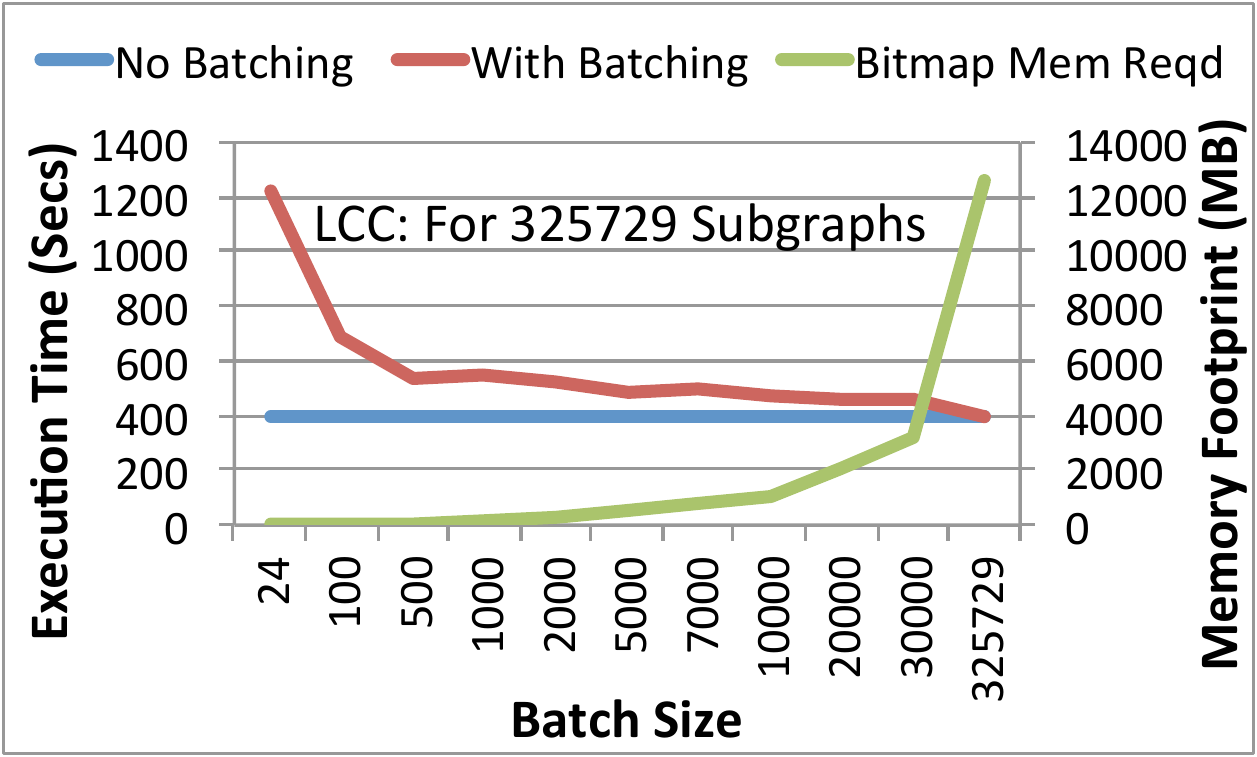}}\\
 \vspace{-12pt} 
\caption{Effect of batching on execution time and memory footprints on two different graph datasets.}
 
  \label{Batching}
\vspace{-10pt} 
  \end{figure}

The key question is how to set the batch size. A small batch size may impact the parallelism and may
lead to an increased total execution time. A small batch size is also susceptible to the {\em
straggler effect}, where the entire batch completion is held up for one or a few subgraphs (leading
to wasted resources and low utilization). A very large batch size, on the other hand, can lead to
high memory overheads for negligible reductions in total execution time. 

Figures~\ref{fig:batching_1} and \ref{fig:batching_2} show the results of a
set of experiments that we ran to understand the effect of batch size on total execution time and
the amount of memory consumed. As we can see, a small batch size
indeed leads to underutilization of the available parallelism and consequently higher execution times. 
However, we also observe that beyond a certain value, increasing the batch size further did not lead
to significant reduction in the execution time. We do a small penalty for batching that can be
attributed to the overhead of reinitializing bitmaps across batched execution and to minor straggler
effects. However, there is a wide range of parameter values where the execution time penalty is
acceptable, and the total memory consumed by the bitmaps is low. Based on our evaluation, we set the
batch size to be 3000 for most of our experiments; a lower number should be used if the hardware
parallelism is lower (these experiments were done on a 24-core machine), and a higher number is
warranted for machines with more cores.

\Paragraphul{Single-Bit Mode.} To further reduce the memory overhead associated with bit vectors, we
provide a single bit execution mode wherein each node and edge is associated with a single bit which
is set if the node participates in the current subgraph computation. The subgraphs are processed in
a serial order, one at a time, with the bits re-initialized after each computation is finished.
This mode is supported to cater to distributed computation on low end commodity machines, but it is not
expected to scale to large graphs.

\subsection{Bitmap Implementation}
\label{sec:bitmapImpl}
Given the central role played by bitmaps in our execution engine, we carefully analyzed and compared
different bitmap implementations that are available for use in \nframe.

\topic{Java BitSet.} Java provides a standard BitSet class that implements a vector of bits that
grows as needed. The Java BitSet class provides generic functionality implementing additional interfaces
and maintains some additional state to support this 
functionality. As a consequence, as the bitmap size grows, the 
Java BitSet object can take up a significant amount of memory, resulting in a relatively high memory overhead.

\topic{LBitSet.} To reduce the memory overhead of the Java BitSet class, we implemented the
LBitSet class as a bare bones implementation; LBitSet uses an array of Java primitive type 'long' (64 bits). Depending on the bitmap
size, an appropriate size of the array is chosen. To set a bit, the long array is considered as a
contiguous set of bits and the appropriate bit position is set to 1 using binary bit operations.
To unset a bit the corresponding bit index position is set to 0. LBitSet incurs less memory overhead
than native Java BitSet, which also uses an array of longs underneath, for the
reasons described above.

\topic{CBitSet.} The CBitSet Java class has been implemented using hash buckets. Each bit index in
the bitmap hashes (maps) to a unique bucket which contains all the bitmap indexes that are set to
1. To set a bit, the bit index is added to the corresponding hash bucket. To unset a bit, the bit
index is removed from the corresponding hash bucket if it is present. This bitmap construction works
on the lines of set association, wherein we can hash onto the set and do a linear search within it,
thereby avoiding allocation of space of all bits explicitly. 

\begin{table}[t]
    \small	
    \centering
    \begin{tabular}{| p{0.85cm} | p{0.73cm} | p{0.73cm} | p{0.73cm} | p{0.73cm} | p{0.73cm} | p{0.73cm} |}
    \hline\hline
{\bf Bitmap size} & Java BitSet & L BitSet & C BitSet (Init) & C BitSet (1) & C BitSet (2) & C BitSet (25\%)\\ \hline
    70  &  54  & 39  & 134  & 138 & 142 & 204\\ \hline
    144  & 63 & 39  & 134  & 138 & 142 & 278\\ \hline
    3252  & 484  & 254  & 134 & 138 & 142 & 3386 \\ \hline
    5000  & 632  & 321  & 134  & 138 & 142 & 5134\\
     \hline\hline
     \end{tabular}    
    \vspace{-2pt}
     \caption{Memory footprints in Bytes for different bitmap constructions and bitmap sizes in bits. For CBitSet, the
             table shows the initial memory footprint and how it increases when 1 bit is set, 2 bits are set and 25\% bits are set (\#bits set indicate 
             the \#subgraphs the vertex is part of).}
    \label{table:bitmaps}
\vspace{-12pt}
\end{table}

We conducted a micro-benchmark comparing these bitmap implementations to get an estimate of the memory
overhead for each bitmap, using a memory mapping utility. Table~\ref{table:bitmaps} gives an 
estimate of the memory requirements per node for each of these bitmaps. Memory footprints for CBitSet
    shown in the table include a column for the initial allotment when the bitmaps are initialized.
At run time, when bits are set, this would increase (by about 4 bytes per bit set). The table shows the increase in CBitSet memory as 1, 2, and 25\% bits are set. 
The number of bits set in each bitmap is indicative of the overlap among them. As we can see,
CBitSet would have a lesser memory footprint if the overlap is less. In other cases LBitSet has the least memory footprint.  
A more detailed performance evaluation of the
different bitmap implementations can be found in Section~\ref{sec:execEngEval}.

\subsection{Support for Iterative computation.}
\label{sec:discussion}
\begin{figure*}[t]
\centering
\includegraphics[scale=0.50]{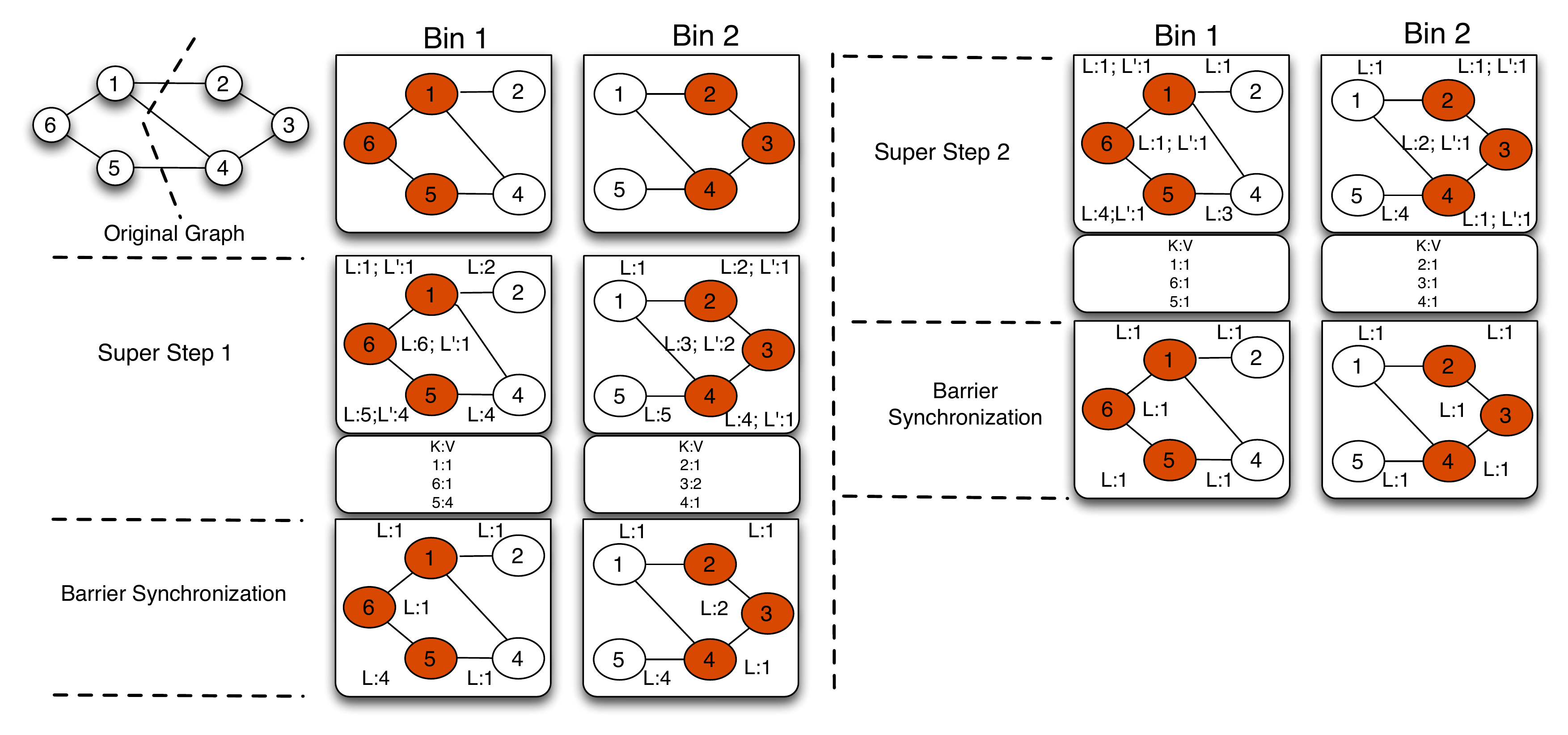}
\vspace{-20pt}
\caption{Iterative execution of global connected components algorithm on an example graph on \nframe.}
\label{fig:iterSupport}
\end{figure*}

\red{\nframe~can naturally handle iterative tasks as well where information must be exchanged across subgraphs between iterations. 
Below we briefly sketch a description of \nframe's iterative execution model.}

\vspace{3pt}\noindent
\red{{\bf Execution model.} \nframe~uses the Bulk Synchronous Protocol (BSP), 
used by Pregel, Giraph, GraphX, and several other distributed graph processing systems.
The analysis task is executed in a number of iterations (also called {\em supersteps})
with barrier synchronization steps in between the iterations. Since subgraphs of interest
typically overlap, the main job of the barrier synchronization step is to ensure that
all the updates made by the user program locally to the query vertices are propagated to other subgraphs
containing those vertices.
During barrier synchronization, after each
superstep, the information exchange between subgraphs co-located on the same
physical partition is done through shared state updates (saving the overhead of
message passing). Information exchange between subgraphs on different physical
partitions is done using message passing which is amenable to optimizations
such as batching of all updates for a particular partition together, to reduce
the overhead. 
}

\vspace{3pt}\noindent
\red{{\bf Consistency model.} To provide deterministic execution of iterative
computation, the updating of state is closely linked to the query-vertex
ownership in \nframe. Each partition in \nframe~owns a disjoint set of
query-vertices and each worker thread is responsible for one query-vertex and
its neighborhood. We only allow updating the state of the query-vertex in
each subgraph by the worker thread that owns (or is currently associated with)
the query vertex.  The state of the query-vertex updated in the current
superstep is available for consumption by other subgraphs in the next
superstep. This BSP-based consistency model thus does away with the requirement
of any explicit locking-based synchronization and its associated overheads
making the system easy to parallelize and scalable for large graphs. 
}

\red{We note that, this restriction on the consistency model is equivalent to the
restrictions imposed by the other vertex-centric graph processing frameworks, and 
does not preclude any 
iterative execution task that we are aware of.}

\vspace{3pt}\noindent
\red{{\bf Implementation details.} The barrier synchronization required by the BSP execution model can be achieved using any mechanism for reliably maintaining centralized state that can be accessed by different partitions (e.g., one option on YARN is Zookeeper). Further, the message passing model for information exchange between partitions can be built using an in-memory distributed and fault tolerant key-value store like Cassandra~\cite{Cassandra}  or a distributed in-memory key-value cache such as Redis~\cite{Redis:Online}, as we do not envision the messages to be very large. 
The number of components (or partitions) of the distributed  key-value store
(or cache) can be set equal to the number of partitions in \nframe~with one
component co-located with each partition to minimize the network overhead. Each
query vertex would mark its updated state in the key-value store that is
co-located with the partition to which the query vertex belongs, keyed by the
query-vertex ID. In our current implementation, we use Redis for both barrier
synchronization using a counter and for message passing.
We explain the step-by-process with an example for computing global connected
components. Note that, for this application, each vertex in the graph is a query
vertex and the set of its 1-hop neighbors constitutes a subgraph of interest.}

\vspace{3pt}\noindent
\red{{\bf Example.} Figure~\ref{fig:iterSupport} shows an example execution of the global connected components algorithm using multiple supersteps. The figure shows an input graph with vertex IDs as labels of vertices. The GEP phase in \nframe~ extracts the subgraphs for each query vertex and instantiates them in two bins (Bin 1 and 2) in an overlapped fashion. Each partition is associated with a disjoint set of query-vertices that it owns. The colored vertices are the query vertices and the other vertices are copies created to enforce the 1-hop neighborhood guarantee.   A key-value store shard is also co-located with each partition. Every vertex has an initial label value $L$ (its vertex ID).}

\red{In superstep 1, each query vertex accesses the labels of its one-hop
neighbors and computes the minimum label and assigns a new value to its own
label; the new label is stored in a temporary copy denoted $L'$. Also each query vertex
inserts an entry in the local shard of the distributed K-V store with its ID as the key 
and its new state ($L'$) as the value.
Superstep 1 is followed by barrier synchronization during which the updated values in $L'$
are copied into $L$ for each query vertex, and all non query-vertices in the
partition are updated with the values in the distributed key-value store. This
is where the message passing takes place between partitions, which is handled by
the distributed key-value store under the hood. For improved performance, we use
multiple threads to read and write to the Redis key-value cache. In superstep 2,
each query vertex repeats the same procedure and updates its $L'$ values and the
key-value store entries. In the subsequent barrier synchronization phase, all
the vertices converge to the same label hence terminating the iterations.}

\begin{table*}[t]
\small
    \centering
    \begin{tabular}{| l | c | c | c | c | c | c |}
    \hline\hline
    Dataset & \#~Nodes & \#~Edges & Avg Degree & Avg Clust. Coeff. & \# Triangles & Diameter  \\ \hline
    EU Email Comn Network & 265214  & 840090 & 3.16 & 0.0671 & 267313 & 14 \\ \hline
    Notre Dame Web Graph & 325729 & 2,994,268 & 9.19 & 0.2346 & 8910005 & 46 \\ \hline 
    Google Web Graph & 875713 & 10,210,078 & 11.66 & 0.5143 & 13391903 & 21 \\ \hline
    Wikipedia Talk Network &  2,394,385 & 10,042,820 & 4.2 & 0.0526 & 9203519 & 9\\ \hline
    LiveJournal Social Network  & 4,847,571& 137,987,546 & 28.5 & 0.2741 & 285730264 & 16 \\ \hline
    Orkut Social Network & 3,072,441 & 234,370,166 & 76.3 & 0.1666 & 627584181 & 9 \\ \hline
    ClueWeb Graph & 428,136,613 & 1,448,223,018 & 3.38 & 0.2655 & 4372668765 & 11 \\
     \hline\hline
     \end{tabular}
     \caption{Dataset Statistics}
    \label{table:datasets}
\end{table*}

\section{Experimental Evaluation}
\label{sec:exps}
We performed an extensive experimental evaluation of different design facets of \nframe~and also compared it with three
popular distributed graph programming platforms. We briefly discuss some additional 
implementations details of \nframe~here, and describe the experimental setup.

\Paragraphul{Implementation Details.}
\nframe~has been written in Java (version ``1.7.0\_45'') and deployed on a YARN
cluster. The framework implements and exports the generic BluePrints API to write graph computations.  
The GEP module takes the subgraph extraction query, the bin packing heuristic to be
used, the bin capacity, and an optional parameter for graph compression/sampling (if required). 
The YARN platform distributes the user computation and the
execution engine library using the distributed cache mechanism to the appropriate machines on the cluster. 
The execution engine has been parametrized to vary its execution modes, and use different batch sizes and bitmap construction techniques.  
Although \nframe~has been designed for the cloud, its deployability and design 
features are not tied to any cloud-specific features; it could be deployed on any cluster of machines or a private cloud that
supports YARN or Hadoop as the underlying data-computation framework.

\Paragraphul{Data Sets.}
\label{sec:expEval}
We conducted experiments using several different datasets, majority of which have been taken from the Stanford SNAP dataset
repository~\cite{snap:online} (see Table~\ref{table:datasets} for details and some statistics).

\begin{itemize}
\item {\bf Web graphs:} We have used three different web graph datasets: {\em Notre Dame Web Graph},
    {\em Google Web Graph}, and {\em ClueWeb09 Dataset}; in all of these, the nodes represent
web pages and directed edges represent hyperlinks between them. 

\item {\bf Communication/Interaction networks:} We use: (1)  {\em EU Email Communication Network},
    generated using email data from a European research institution for a period from October 2003
    to May 2005; and (2) The {\em Wikipedia Talk network}, created from the talk pages of registered users on Wikipedia until Jan 2008. 

\item {\bf Social networks:} 
We also use two social network datasets: the {\em Live Journal social network} and {\em
    Orkut social network}. 

\item {\bf Small-scale synthetic graphs.} For comparing against the optimal algorithm, we
generated a set of small-scale synthetic graphs (100-1000 nodes, 500-20000 edges) using the 
Barabasi-Albert preferential attachment model. 
\end{itemize}

\vspace{2pt}
\Paragraphul{Graph Applications.}
\label{sec:grApps}
We evaluate \nframe~over \red{6} different applications. Three of them, namely, Local Clustering Coefficient (LCC),
Motif Counting: Feed-Forward Loop (MC), and Link Prediction using Personalized Page Rank (PPR), are 
described in Section~\ref{sec:appScenarios}. In addition, we used:

\begin{itemize}
\item{\bf Triangle Counting (TC):} 
Here the goal is to count the number of triangles  each vertex is
part of. These statistics are very useful for complex network
analysis~\cite{DBLP:journals/im/KolountzakisMPT12} and real world applications such as spam
detection, link recommendation, etc.
\item{\bf Counting Weak Ties (WT):} A weak tie is defined to be a pattern where the center node is
connected to two nodes that are not connected to each other. The goal with this task is to find the
number of weak ties that each vertex is part of. 
Number of weak ties is considered an important metric in social science~\cite{granovetter2010strentgh}. 
\end{itemize}

\red{In addition to the above graph applications that involve single-pass analytics, we also evaluated 
\nframe\ using a global iterative graph application, computing the connected components, as described in Section~\ref{sec:discussion}. }

\Paragraphul{Comparison platforms.}
\label{sec:baselineEM}
We compare \nframe~with \red{three} widely used graph programming frameworks.

\squishlist
\item{\bf Apache Giraph~\cite{Giraph:Online}. } The
open source version of Pregel, written in Java, is a vertex-centric graph
programming framework and widely used in many production systems (e.g., at Facebook). We deploy
Apache Giraph \red{(Version 1.0.0)} on Apache YARN with Zookeeper for synchronization for the BSP model of computation.
Deploying Apache Giraph on YARN with HDFS as the underlying
storage layer enables us to provide a fair comparison using the same datasets and graph
applications.    

\item{\bf GraphLab~\cite{DBLP:journals/pvldb/LowGKBGH12}.} GraphLab, a distributed graph-parallel API written in C++, is an open source
vertex-centric programming model that supports both synchronous and asynchronous execution.
GraphLab uses the GAS model of execution wherein each vertex program is decomposed into {\em
    gather,}
{\em apply}, and {\em scatter} phases; the framework uses MPI for message passing across machines.
We deployed GraphLab v2.2 which supports OpenMPI 1.3.2 and MPICH2 1.5, on our cluster.

\item\red{{\bf GraphX~\cite{GraphX}.} GraphX is a graph programming library that sits on top of Apache Spark. We used the GraphX library version
2.10 over Spark version 1.3.0 which was deployed on Apache YARN with HDFS as the underlying storage layer.}
\squishend

\Paragraphul{Evaluation metrics.} We use the following evaluation metrics to evaluate the performance of \nframe.
\begin{itemize}
\item {\bf Computational Effort ($\mathcal{CE}$}). $\mathcal{CE}$ captures the total cost of doing
analytics on a cluster of nodes deployed in the cloud. Let $T=\{T_1,T_2,...,T_N\}$ be the set of
tasks (or processes) deployed by the framework on the cluster during execution of the analytics task. 
Also, let $t_i$ be the time taken by the task $T_i$ to be executed on node $i$. We
define $\mathcal{CE}= \sum_{i=1}^{N} t_{i}$. 
The metric captures the cost of doing data analytics in terms of {\em node-secs} which is appropriate for the cloud environment.  
\item {\bf Execution Time.}  This is the measure of the wall clock time or elapsed time for executing an end-to-end graph computation on a cluster of machines. It includes the time taken by the GEP phase for extracting the subgraphs as well as the time taken by the distributed execution engine  to execute the user computation on all subgraphs of interest.  
\item {\bf Cluster Memory.}  Here we measure the maximum total physical memory used across all nodes in
the cluster. 
\end{itemize}

\Paragraphul{Experimental Setup.}
We use two 16 node clusters wherein each data node has 2 4-core Intel Xeon
E5520 processors, 24GB RAM and 3 2 TB disks. The first cluster runs Apache
YARN (MRv2 on \red{Cloudera's CDH version 5.1.2)} and Apache Zookeeper for coordination. Each process on this cluster 
runs in a container with a max memory capacity restricted to 15GB with a maximum 
of 6 processes per physical machine. We run \nframe,
Giraph \red{and GraphX} experiments on this cluster.
The second cluster supports MPI for message passing and uses a TORQUE (Terascale Open-Source Resource and 
QUEue) Manager. We run GraphLab in this cluster and restrict the
 max memory per process on each machine to 15GB for a fair comparison.
 
 \red{For all our baseline comparisons and scalability experiments, we have used the shingle-based bin packing heuristic 
 as the GEP algorithm for packing subgraphs into bins. We have chosen shingle-based bin packing as it finds good quality solutions 
 efficiently, while consuming fewer resources as compared to the other heuristics. Also, for smaller graphs such as NotreDame 
 web graph, Google web graph, etc., where the filtered structure can fit onto a single machine, we used the centralized 
 GEP solution (Ref Case 1, Section~\ref{sec:gelPhase1}). On the other hand, for larger graphs such as the Clue Web graph, we use the distributed 
 GEP solution (Ref Case 2 Section~\ref{sec:gelPhase1}).  }
%

\section{Experimental Results}
\begin{table*}[!ht]
\small
    \centering
   
    \begin{tabular}{|p{1.80cm} |>{\columncolor[gray]{0.9}}p{1.45cm} |  p{1.45cm} |>{\columncolor[gray]{0.9}}p{1.45cm} |  p{1.45cm} | >{\columncolor[gray]{0.9}}p{1.45cm} |  p{1.45cm} | >{\columncolor[gray]{0.9}}p{1.45cm} |  p{1.45cm} | }
    \hline\hline
   \multirow{3}{*}{Dataset} & \multicolumn{8}{|c|}{\bfseries Local Clustering Coefficient} \\
  \hline
  & \multicolumn{2}{|c|}{\nframe} & \multicolumn{2}{|c|}{Giraph} & \multicolumn{2}{|c|}{GraphLab} & \multicolumn{2}{|c|}{GraphX} \\
  \hline
   & $\mathcal{CE}$ (Node-Secs) &  Cluster Mem(GB)& $\mathcal{CE}$ (Node-Secs) &  Cluster Mem(GB)& $\mathcal{CE}$ (Node-Secs) &  Cluster Mem(GB) & $\mathcal{CE}$ (Node-Secs) &  Cluster Mem(GB) \\ \hline
    EU Email & 377  & 9.00 & 1150 & 26.17 & 365 & 20.1 & 225 & 4.95  \\ \hline
    NotreDame & 620 & 19.07 & 1564 & 30.14 & 550 & 21.4 &340 & 9.75   \\ \hline 
    GoogleWeb & 658 &  25.82 & 2024 & 35.35 & 600 & 33.5 &1485 &21.92   \\ \hline
    WikiTalk &  726 &  24.16 & DNC & OOM & 1125 & 37.22 & 1860 & 32 \\  \hline
    LiveJournal &  1800 &  50 & DNC & OOM & 5500 & 128.62 & 4515 & 84   \\ \hline
    Orkut &  2000 &  62 & DNC & OOM & DNC & OOM &20175 & 125  \\ 
   \hline
     \end{tabular}
     
      \vspace{14pt}
         \begin{tabular}{|p{1.80cm} |>{\columncolor[gray]{0.9}}p{1.45cm} |p{1.45cm} |>{\columncolor[gray]{0.9}}p{1.45cm} |p{1.45cm} |>{\columncolor[gray]{0.9}}p{1.45cm} |p{1.45cm} |>{\columncolor[gray]{0.9}}p{1.45cm} |p{1.45cm} |}
    \hline
   \multirow{3}{*}{Dataset}  & \multicolumn{8}{|c|}{\bf Motif Counting: Feed-Forward Loop}\\
  \hline
  &  \multicolumn{2}{|c|}{\nframe} & \multicolumn{2}{|c|}{Giraph} & \multicolumn{2}{|c|}{GraphLab} & \multicolumn{2}{|c|}{GraphX} \\
  \hline
   & $\mathcal{CE}$ (Node-Secs) &  Cluster Mem(GB) &  $\mathcal{CE}$ (Node-Secs) &  Cluster Mem(GB) & $\mathcal{CE}$ (Node-Secs) &  Cluster Mem(GB)& $\mathcal{CE}$ (Node-Secs) &  Cluster Mem(GB) \\ \hline
    EU Email & 279 & 8.76& 1371 & 24.43 & 285 & 20.8 &4125 &7.2\\ \hline
    NotreDame & 524 & 18.02 & 1923 &  28.98 & 575 & 21.6 &10875 &15.6 \\ \hline 
    GoogleWeb &  812 & 23.64 & 2164 & 37.27 & 625 & 31.9 &DNC&- \\ \hline
    WikiTalk &  991 &  29.34 & DNC & OOM & 1150 & 36.81 &DNC&- \\  \hline
    LiveJournal & 1886 &  51 & DNC & OOM & 4750 & 130.74 &DNC &-  \\ \hline
    Orkut & 2024 &  63 & DNC & OOM & DNC & OOM &DNC &- \\ 
   \hline\hline
     \end{tabular}

  \vspace{14pt}
         \begin{tabular}{|p{1.80cm} |>{\columncolor[gray]{0.9}}p{1.45cm} |p{1.45cm} |>{\columncolor[gray]{0.9}}p{1.45cm} |p{1.45cm} |>{\columncolor[gray]{0.9}}p{1.45cm} |p{1.45cm} |>{\columncolor[gray]{0.9}}p{1.45cm} |p{1.45cm} |}
    \hline
   \multirow{3}{*}{Dataset}  & \multicolumn{8}{|c|}{\bf Per-Vertex Triangle Counting}\\
  \hline
  &  \multicolumn{2}{|c|}{\nframe} & \multicolumn{2}{|c|}{Giraph} & \multicolumn{2}{|c|}{GraphLab} & \multicolumn{2}{|c|}{GraphX} \\
  \hline
   & $\mathcal{CE}$ (Node-Secs) &  Cluster Mem(GB) &  $\mathcal{CE}$ (Node-Secs) &  Cluster Mem(GB) & $\mathcal{CE}$ (Node-Secs) &  Cluster Mem(GB)& $\mathcal{CE}$ (Node-Secs) &  Cluster Mem(GB) \\ \hline
    EU Email & 264 & 15.36 & 1012 &  26.10 & 250 & 21.1 & 240 &4.5\\ \hline
    NotreDame & 477 &  17.62 & 1518 &  30.16 & 425 & 22.7 & 270 &9 \\ \hline 
    GoogleWeb &  663 &  25.86 & 1978 &  35.39 & 550 & 31.3 &1230 &21 \\ \hline
    WikiTalk &  715 &  21.29 & DNC  & OOM & 975 & 32.22 &1590 &30.2 \\  \hline
    LiveJournal & 1792 &  49.34 & DNC & OOM & 4750 & 129.61 &4335 &74  \\ \hline
    Orkut &1986 &  61.32 & DNC & OOM & DNC & OOM &13875 & 115 \\ 
   \hline\hline
     \end{tabular}

    \vspace{14pt}
    
         \begin{tabular}{|p{1.80cm} |>{\columncolor[gray]{0.9}}p{1.45cm} |p{1.45cm} |>{\columncolor[gray]{0.9}}p{1.45cm} |p{1.45cm} |>{\columncolor[gray]{0.9}}p{1.45cm} |p{1.45cm} |>{\columncolor[gray]{0.9}}p{1.45cm} |p{1.45cm} |}
    \hline
   \multirow{3}{*}{Dataset}  & \multicolumn{8}{|c|}{\bf Identifying Weak Ties}\\
  \hline
  &  \multicolumn{2}{|c|}{\nframe} & \multicolumn{2}{|c|}{Giraph} & \multicolumn{2}{|c|}{GraphLab}& \multicolumn{2}{|c|}{GraphX} \\
  \hline
   & $\mathcal{CE}$ (Node-Secs) &  Cluster Mem (GB) &  $\mathcal{CE}$ (Node-Secs) &  Cluster Mem (GB) & $\mathcal{CE}$ (Node-Secs) &  Cluster Mem (GB) & $\mathcal{CE}$ (Node-Secs) &  Cluster Mem (GB) \\ \hline
    EU Email & 278  & 7.34 & 1472 & 25.49 & 281 & 20.4 &4215 & 7.3 \\ \hline
    NotreDame & 390 & 13.26 & 2024 & 29.99 & 400 & 20.6 & 11795 &16.6 \\ \hline 
    GoogleWeb &  555 & 21.60 & 2254 & 39.26 & 525 & 30.7 & DNC &- \\ \hline
    WikiTalk &  592 &  18.18 & DNC &  OOM & 925 & 31.71 & DNC &- \\  \hline
    LiveJournal & 1762 &  48.32 & DNC &  OOM & 4625 & 126.71 & DNC &- \\ \hline
    Orkut &  1972 &  60.45 & DNC &  OOM & DNC & OOM  &DNC&-\\ 
   \hline\hline
     \end{tabular}

      \vspace{14pt}
      
    \begin{tabular}{|p{1.70cm} | p{1.3cm} | p{1.2cm} | >{\columncolor[gray]{0.9}} p{1.3cm} | p{1.2cm} | >{\columncolor[gray]{0.9}}p{1.3cm} | p{1.2cm} |>{\columncolor[gray]{0.9}}p{1.3cm} |p{1.2cm} |>{\columncolor[gray]{0.9}}p{1.3cm} |}
    \hline
   \multirow{3}{*}{Dataset} & \multicolumn{9}{|c|}{\bf Personalized Page Rank on 2-hop Neighborhood}   \\
  \hline
  & &\multicolumn{2}{|c|}{\nframe} & \multicolumn{2}{|c|}{Giraph} & \multicolumn{2}{|c|}{GraphLab}& \multicolumn{2}{|c|}{GraphX} \\
  \hline
   & {\bf \#Source Vertices} & $\mathcal{CE}$ (Node-Secs)   & Cluster Mem(GB)&   $\mathcal{CE}$ (Node-Secs)  & Cluster Mem(GB) & $\mathcal{CE}$ (Node-Secs) &  Cluster Mem(GB) & $\mathcal{CE}$ (Node-Secs) &  Cluster Mem(GB) \\ \hline
    EU Email  &  {\bf 3200}& 52 & 3.35 & 782 & 17.10 & 710 & 28.87 & 9975 & 85.5\\ \hline
    NotreDame   &  {\bf 3500}& 119 & 9.56 & 1058 & 31.76 & 870 & 70.54 & 50595 & 95\\ \hline 
    GoogleWeb  &  {\bf 4150} & 464 & 21.52 & 10482 & 64.16 & 1080 & 108.28 & DNC &- \\ \hline
    WikiTalk & {\bf 12000} & 3343 & 79.43 & DNC & OOM & DNC & OOM & DNC &-\\  \hline
     LiveJournal &  {\bf 20000} &  4286 & 84.94 & DNC & OOM & DNC & OOM & DNC &-  \\ \hline
    Orkut &  {\bf 20000} &  4691 & 93.07 & DNC & OOM & DNC & OOM & DNC &- \\ 
   \hline
     \end{tabular}
    
     \caption{Comparing \nframe~with Giraph, GraphLab and GraphX}
   \label{table:blComp}
\end{table*}

\begin{table*}
 \centering
\begin{tabular}{|p{1.60cm} | >{\columncolor[gray]{0.9}}p{0.80cm} |  p{0.80cm} | >{\columncolor[gray]{0.9}}p{0.80cm} |  p{0.80cm} | >{\columncolor[gray]{0.9}}p{0.80cm} |  p{0.80cm} | >{\columncolor[gray]{0.9}}p{0.80cm} |  p{0.80cm} | >{\columncolor[gray]{0.9}}p{0.80cm} | p{0.80cm} |>{\columncolor[gray]{0.9}}p{0.80cm} | p{0.80cm} |}
    \hline\hline
   \multirow{3}{*}{Dataset} & \multicolumn{6}{|c|}{\bfseries Local Clustering Coefficient} &
   \multicolumn{6}{|c|}{\bf Motif Counting: Feed-Forward Loop}  \\
  \hline
  & \multicolumn{2}{|c|}{Giraph} & \multicolumn{2}{|c|}{GraphLab} & \multicolumn{2}{|c|}{GraphX} &  \multicolumn{2}{|c|}{Giraph} & \multicolumn{2}{|c|}{GraphLab} & \multicolumn{2}{|c|}{GraphX} \\
  \hline
   &  $\mathcal{CE}$ (Node-Secs) &  Cluster Mem (GB)&  $\mathcal{CE}$ (Node-Secs) &  Cluster Mem (GB) & $\mathcal{CE}$ (Node-Secs) &  Cluster Mem (GB) & $\mathcal{CE}$ (Node-Secs) &  Cluster Mem (GB) & $\mathcal{CE}$ (Node-Secs) &  Cluster Mem (GB) & $\mathcal{CE}$ (Node-Secs) &  Cluster Mem (GB) 
    \\ \hline
    EU Email &  3.05X & 2.9X & 0.96X & 2.23X & 0.66X & 0.55X &  4.91X & 2.78X & 1.02X & 2.37X &14.78 &0.82X \\ \hline
    NotreDame & 2.52X & 1.58X & 0.88X & 1.12X &0.54 & 0.51 &   3.66X &  1.60X & 1.09X & 1.19X & 20.75X & 0.86X \\ \hline 
    GoogleWeb &  3.07X & 1.36X & 0.91X & 1.29X & 2.25X &0.84X &  2.66X & 1.57X & 0.76X & 1.34X &- &- \\ \hline
    WikiTalk &   - & - & 1.54X & 1.54X & 2.56X & 1.32X &  - & - & 1.16X & 1.25X &- &- \\  \hline
    LiveJournal &   - & - & 3.05X & 2.57X & 2.50X & 1.68X &  - & - & 2.51X & 2.56X &- &- \\ \hline
    Orkut &  - & - & - & - &  10.08X & 2.01X & - & - &- &- &- &-  \\ 
   \hline
     \end{tabular}
     
\vspace{10pt}
\begin{tabular}{|p{1.60cm} | >{\columncolor[gray]{0.9}}p{0.80cm} |  p{0.80cm} | >{\columncolor[gray]{0.9}}p{0.80cm} |  p{0.80cm} | >{\columncolor[gray]{0.9}}p{0.80cm} |  p{0.80cm} | >{\columncolor[gray]{0.9}}p{0.80cm} |  p{0.80cm} | >{\columncolor[gray]{0.9}}p{0.80cm} | p{0.80cm} | >{\columncolor[gray]{0.9}}p{0.80cm} | p{0.80cm} | >{\columncolor[gray]{0.9}}p{0.80cm} | p{0.80cm} |}
    \hline\hline
   \multirow{3}{*}{Dataset} & \multicolumn{6}{|c|}{\bfseries Per-Vertex Triangle Counting} &
   \multicolumn{6}{|c|}{\bf Identifying Weak Ties}  \\
  \hline
  & \multicolumn{2}{|c|}{Giraph} & \multicolumn{2}{|c|}{GraphLab}& \multicolumn{2}{|c|}{GraphX} &  \multicolumn{2}{|c|}{Giraph} & \multicolumn{2}{|c|}{GraphLab} & \multicolumn{2}{|c|}{GraphX}\\
  \hline
   &  $\mathcal{CE}$ (Node-Secs) &  Cluster Mem (GB) & $\mathcal{CE}$ (Node-Secs) &  Cluster Mem (GB) & $\mathcal{CE}$ (Node-Secs) &  Cluster Mem (GB) & $\mathcal{CE}$ (Node-Secs) &  Cluster Mem (GB) &  $\mathcal{CE}$ (Node-Secs) &  Cluster Mem (GB) &  $\mathcal{CE}$ (Node-Secs) &  Cluster Mem (GB)  \\ \hline
    EU Email &  3.83X & 1.69X & 0.94X & 1.37X  & 0.90X & 0.29X &  5.29X & 3.47X & 1.01X & 2.77X &15.16X & 0.99X   \\ \hline
    NotreDame & 3.18X & 1.71X & 0.89X & 1.28X & 0.56X & 0.51X &   5.18X &  2.26X & 1.02X & 1.55X &30.24X & 1.25X  \\ \hline 
    GoogleWeb &2.98X & 1.36X & 0.82X  & 1.21X & 1.85X & 0.81X  &  4.06X & 1.81X  & 0.94X & 1.42X &- &- \\ \hline
    WikiTalk &    - & - & 1.36X & 1.51X & 2.22X & 1.41X &  - & - & 1.56X & 1.74X &- &- \\  \hline
    LiveJournal &  - & - & 2.65X & 2.62X & 2.41X & 1.49X &  - & - & 2.62X & 2.62X  &- &-\\ \hline
    Orkut & - & - &  - & - & 6.98X & 1.87X & - & - &- &-&- &- \\ 
   \hline
     \end{tabular}
     
     \vspace{10pt}
\begin{tabular}{|p{1.60cm} | >{\columncolor[gray]{0.9}}p{1.20cm} |  p{1.20cm} | >{\columncolor[gray]{0.9}}p{1.20cm} |  p{1.20cm} | >{\columncolor[gray]{0.9}}p{1.20cm} |  p{1.20cm} | >{\columncolor[gray]{0.9}}p{1.20cm} |  p{1.20cm} |}
    \hline\hline
   \multirow{3}{*}{Dataset} & \multicolumn{6}{|c|}{\bfseries Personalized Page Rank}  \\
  \hline
  & \multicolumn{2}{|c|}{Giraph} & \multicolumn{2}{|c|}{GraphLab} & \multicolumn{2}{|c|}{GraphX} \\
  \hline
   &  $\mathcal{CE}$ (Node-Secs) &  Cluster Mem (GB) & $\mathcal{CE}$ (Node-Secs) &  Cluster Mem (GB) & $\mathcal{CE}$ (Node-Secs) &  Cluster Mem (GB) \\ \hline
    EU Email &  15.03X & 5.10X & 13.65X & 8.61X &191.82X &25.52X   \\ \hline
    NotreDame & 8.89X & 3.32X & 7.31X & 7.37X &425.16X  & 9.93X \\ \hline 
    GoogleWeb &  22.59X  & 2.98X & 2.32X & 5.03X &- &-  \\ \hline
    WikiTalk &  - & -  & - & -&- &- \\  \hline
    LiveJournal &  - & - & - & - &- &-  \\ \hline
    Orkut & - & - & - & - & - & - \\ 
   \hline
     \end{tabular}

\caption{Performance (X) improvement of \nframe~over Giraph, GraphLab and GraphX; a ``-'' indicates that the other system ran out of memory or did not complete.}
\label{table:improvement}
\end{table*}

\vspace{-5pt}
\subsection{Baseline Comparisons}
We begin with comparing \nframe~with Apache Giraph and GraphLab for different datasets for the five different
applications. For four of the applications (LCC, MC, TC, WT), the subgraphs of interest are
specified as 1-hop neighborhoods of a set of query vertices which could be chosen randomly or 
specified using query-vertex predicates.
On the other hand, Personalized Page Rank (PPR) is computed on the 2-hop
neighborhood of a set of query vertices. \red{For a fair comparison with all the other baselines, we choose each 
vertex as a query-vertex for \nframe~and run the the first four applications (LCC, MC, TC, WT) on their 1-hop neighborhoods in a single pass. 
For the Personalized page rank application we choose different number of source
(or query) vertices for different datasets. The Personalized page rank is computed with respect to
these source vertices on their 2-hop neighborhoods in all frameworks.} 

Table~\ref{table:blComp} shows the results for the baseline comparisons. 
Since all of these applications require access to neighborhoods, Apache Giraph runs them using multiple iterations. 
In the first superstep it gathers neighbor information using message passing and in the second
superstep, it does the required graph computation (for PPR, Giraph needs two supersteps to gather
the 2-hop neighborhoods). 

As we can see, for most of the graph analytics tasks, Giraph does not scale to larger graphs. 
It runs out of memory (OOM)  a short while into the map phase, and does not complete (DNC) the computation. 
Hence these baseline comparisons have been shown on relatively smaller graphs. The cluster logs confirmed
that the poor scalability of Giraph for such applications is due to the high message passing
overhead between the vertices, characteristic of vertex-centric approaches like Giraph, and
high memory requirements due to duplication of state at each vertex. 
             
Compared to Giraph, GraphLab performs a little better. For smaller graphs such as NotreDame and Google Web, 
GraphLab's performance is comparable to \nframe\ and for some applications like Local Clustering Coefficient,  it is a little 
better than \nframe~in terms of $\mathcal{CE}$. However, in all cases, GraphLab consumes much more cluster memory depending on 
the graph partitioning mechanism and the replication factor it uses, the latter of which varies with
the number of machines on which the job is executed. Like Giraph, GraphLab too does not scale to larger graphs for neighborhood-centric applications. \\

\red{GraphX does well for 1-hop graph applications such as LCC and TC on smaller graphs both in terms of memory and $\mathcal{CE}$ (node-secs). However as the graph
size increases, $\mathcal{CE}$ grows rapidly and surpasses that of \nframe, quite significantly. For applications such as MC and WT, GraphX performs poorly as these
applications require explicit edge information between the 1-hop neighbors of the query-vertex which necessitates joins and triplet aggregations across the vertex and edge RDDs, leading 
to poor scalability for larger graphs for such applications. For similar reasons, the performance 
of GraphX further deteriorates for 2-hop neighborhood applications such as PPR and it does not complete for any of the larger graph datasets (Web-Google and beyond).}

\red{Table~\ref{table:improvement} shows the performance gain of \nframe, over Giraph, GraphLab and GraphX
both in terms of $\mathcal{CE}$ and cluster memory consumption. Even for the smaller graphs, depending on the type of application and the size of neighborhood, 
\nframe~performs 3X to 22X better in terms of $\mathcal{CE}$, and consumes a lot less (up to 5X less) total cluster memory as compared to Giraph. }

\red{GraphLab follows a similar trend.  As can be seen, for all the five applications,
as the graph size increases, both $\mathcal{CE}$ and required memory increase sharply, and GraphLab fails 
to complete, running out of memory, for real world graphs such as WikiTalk, Orkut and Live Journal. Even for relatively smaller graphs, 
the performance difference is significant, especially for 2-hop applications such as Personalized Page Rank where GraphLab
 is up to 13X slower and consumes up to 8X more memory.}

\red{GraphX performs better for smaller graphs for applications such as
LC and TC. However, for relatively larger graphs, \nframe~is up to 10X better in terms of $\mathcal{CE}$ and consumes up to 2X less memory.  For MC and WT applications, \nframe~is up to 30X better in terms of $\mathcal{CE}$ and consumes up to 1.25X less memory for smaller graphs. For larger graphs GraphX fails
to complete. The most significant difference is seen for PPR where \nframe~performs up to 425X better in terms of $\mathcal{CE}$  and consumes up to 25X lesser memory for smaller graphs. Again, for larger
graphs GraphX fails to complete. }
                 
The improved performance of \nframe~can be attributed to the
\nframe~computation and execution models which (1) allow a user computation to access the
entire subgraph state and hence do not require multiple iterations avoiding the message passing
overhead, and  (2) avoid duplication of state at each vertex reducing memory requirements
drastically. Further, the extraction and loading of required subgraphs by the GEP module helps
\nframe~to scale to larger graphs using minimal resources.  


\begin{figure*}[t]
  \centering
 \subfigure[]{\label{fig:BPM_1}\includegraphics[scale=0.44]{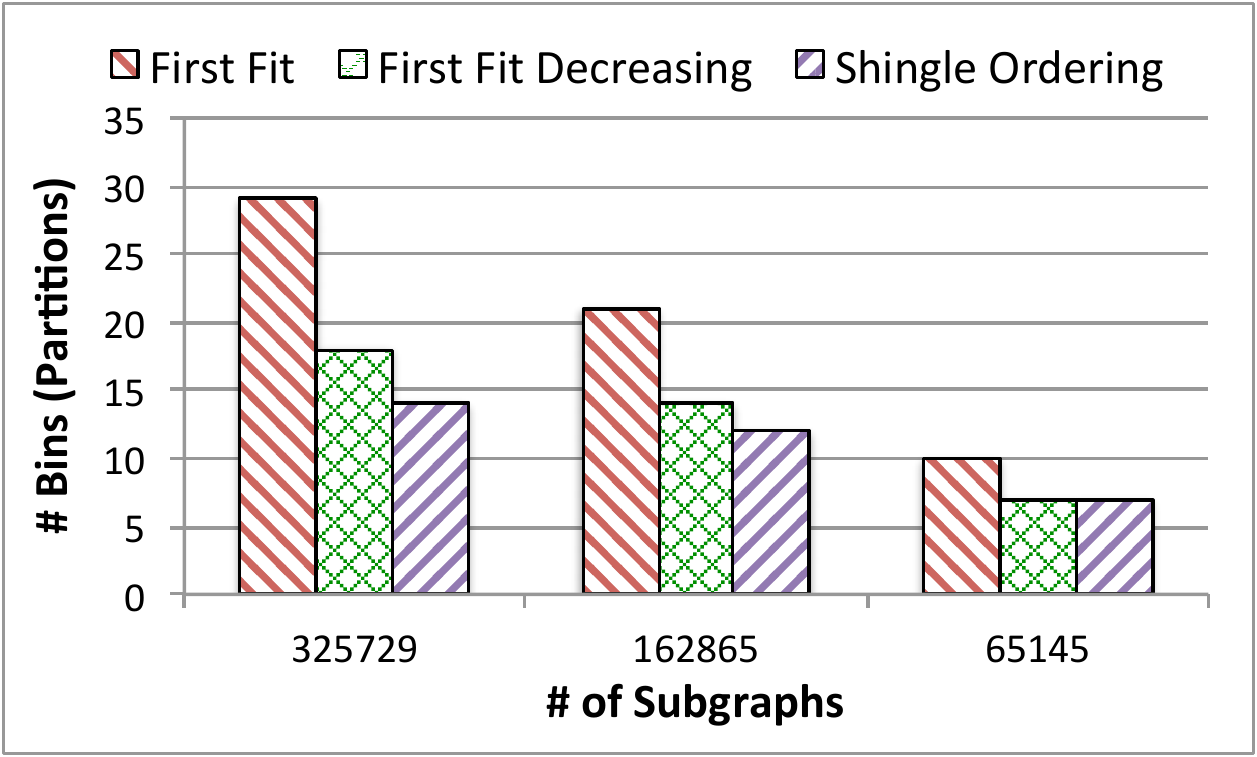}}
\subfigure[]{\label{fig:BPM_2}\includegraphics[scale=0.44]{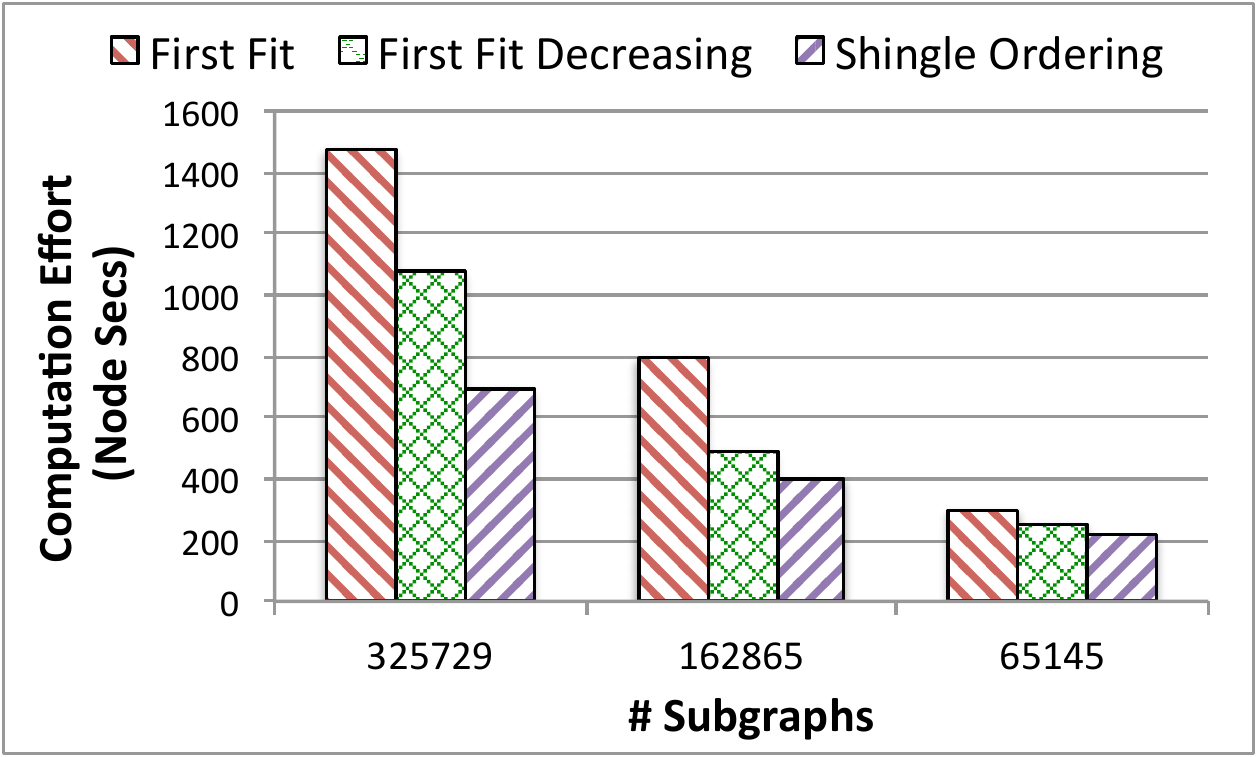}}
\subfigure[]{\label{fig:BPM_3}\includegraphics[scale=0.44]{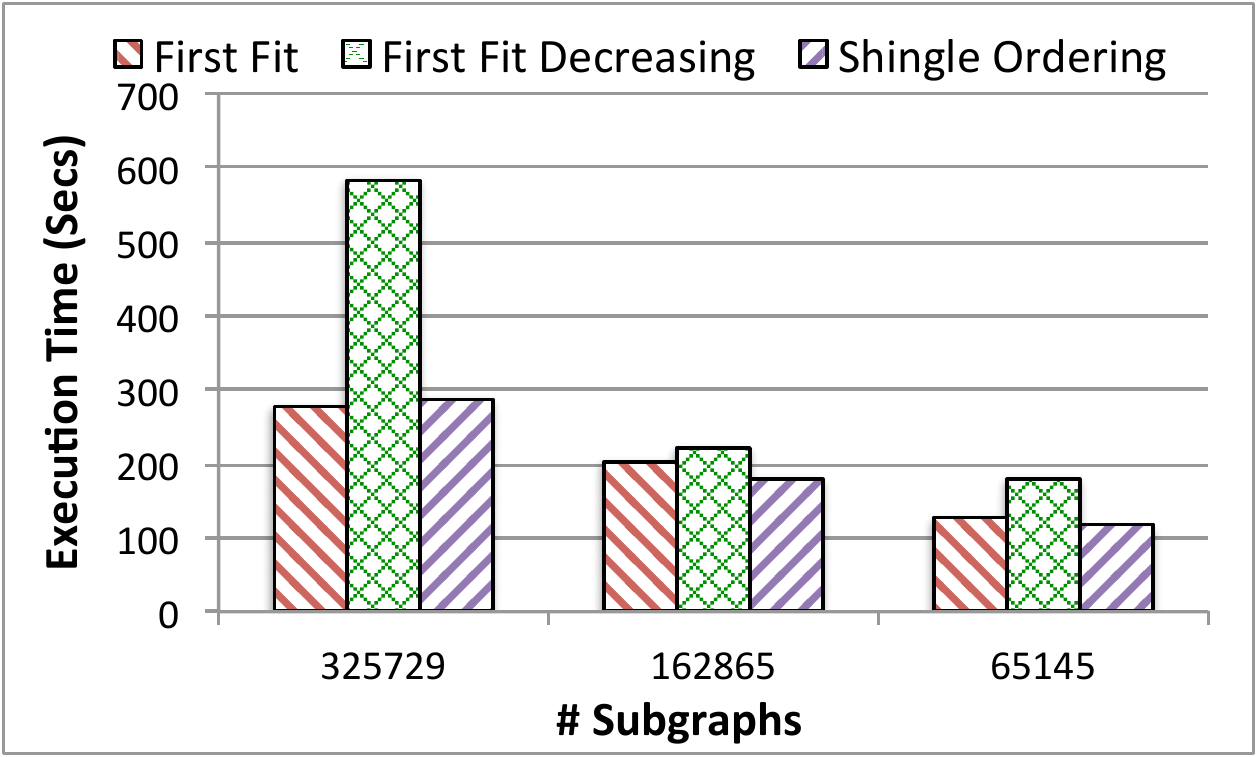}}\\
\vspace{-2pt}
\subfigure[]{\label{fig:BPM_4}\includegraphics[scale=0.44]{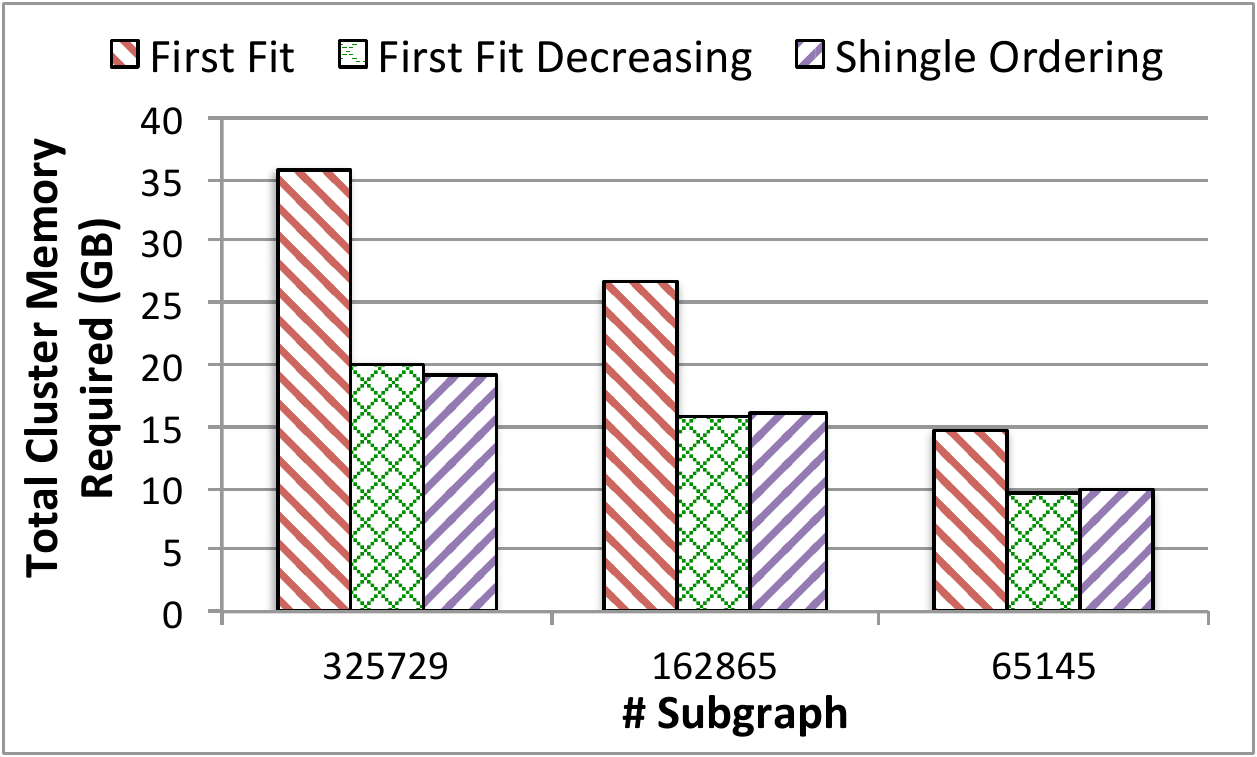}}
\subfigure[]{\label{fig:BPM_5}\includegraphics[scale=0.44]{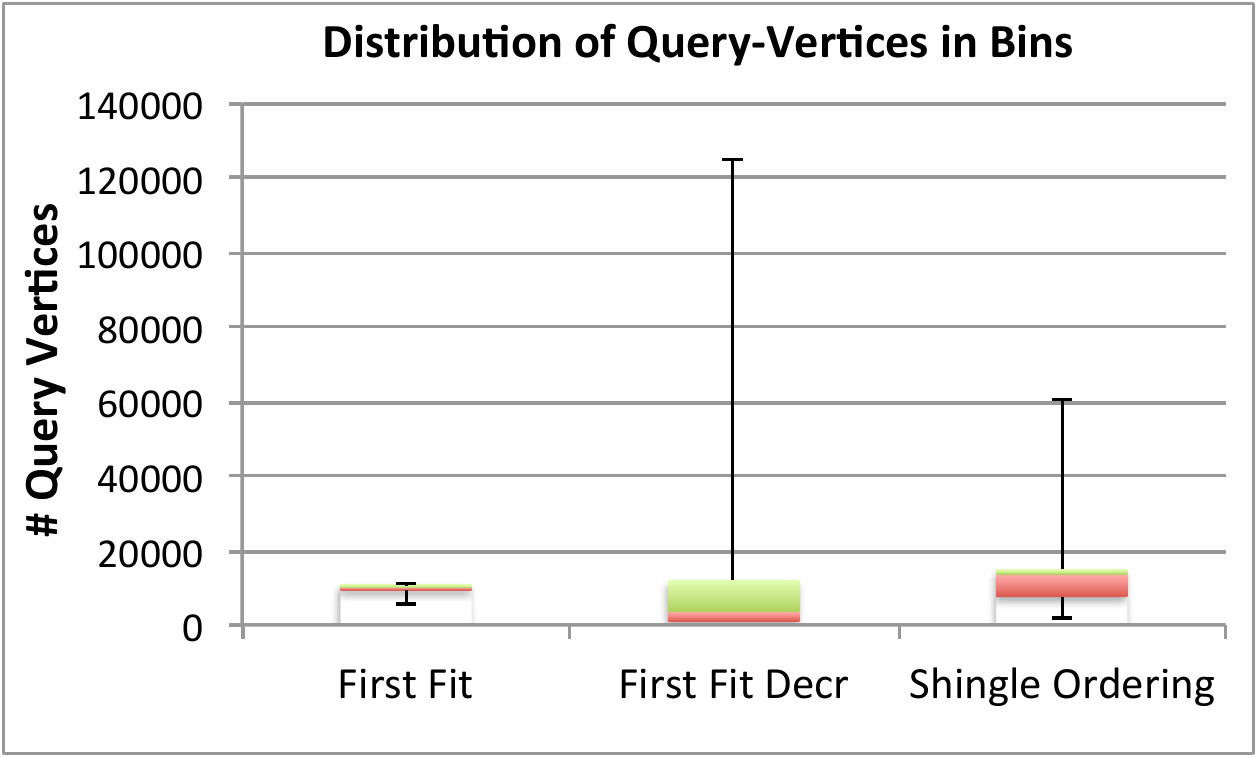}}
\subfigure[]{\label{fig:BPM_6}\includegraphics[scale=0.44]{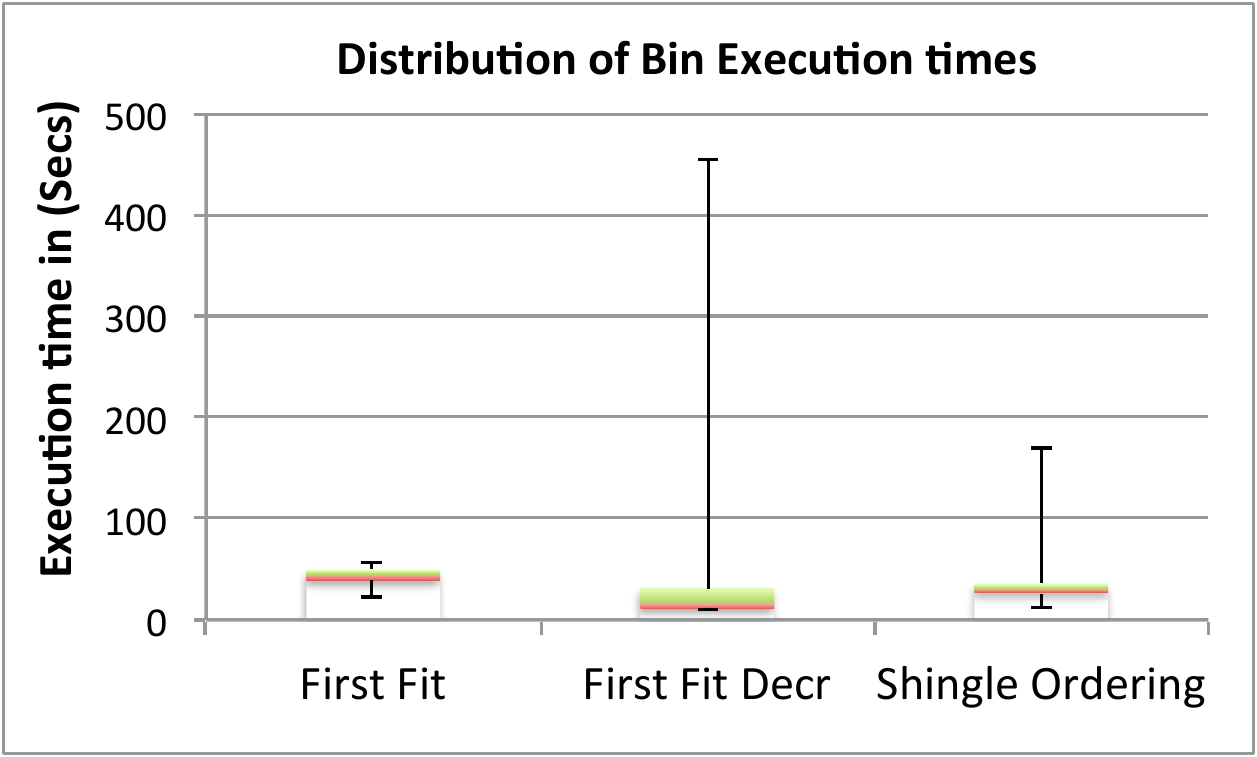}}\\  
\vspace{-2pt}
  \subfigure[]{\label{fig:BPM_7}\includegraphics[scale=0.44]{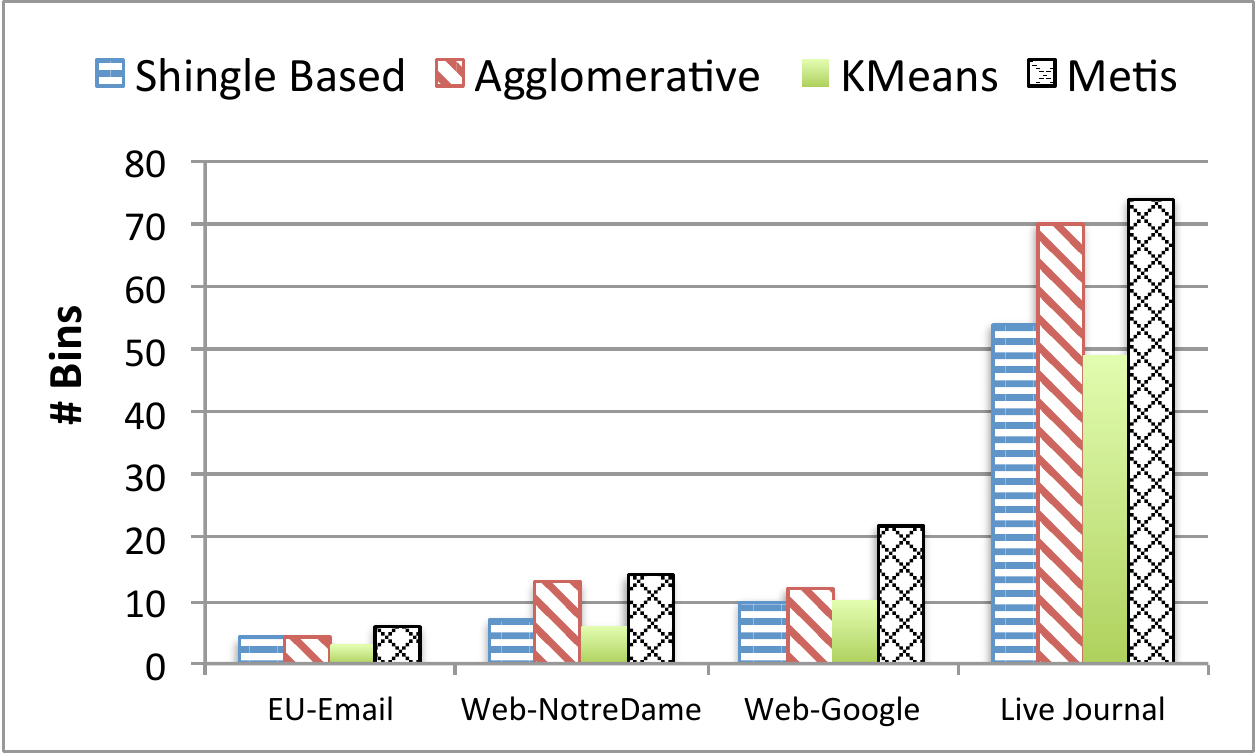}}
\subfigure[]{\label{fig:BPM_8}\includegraphics[scale=0.44]{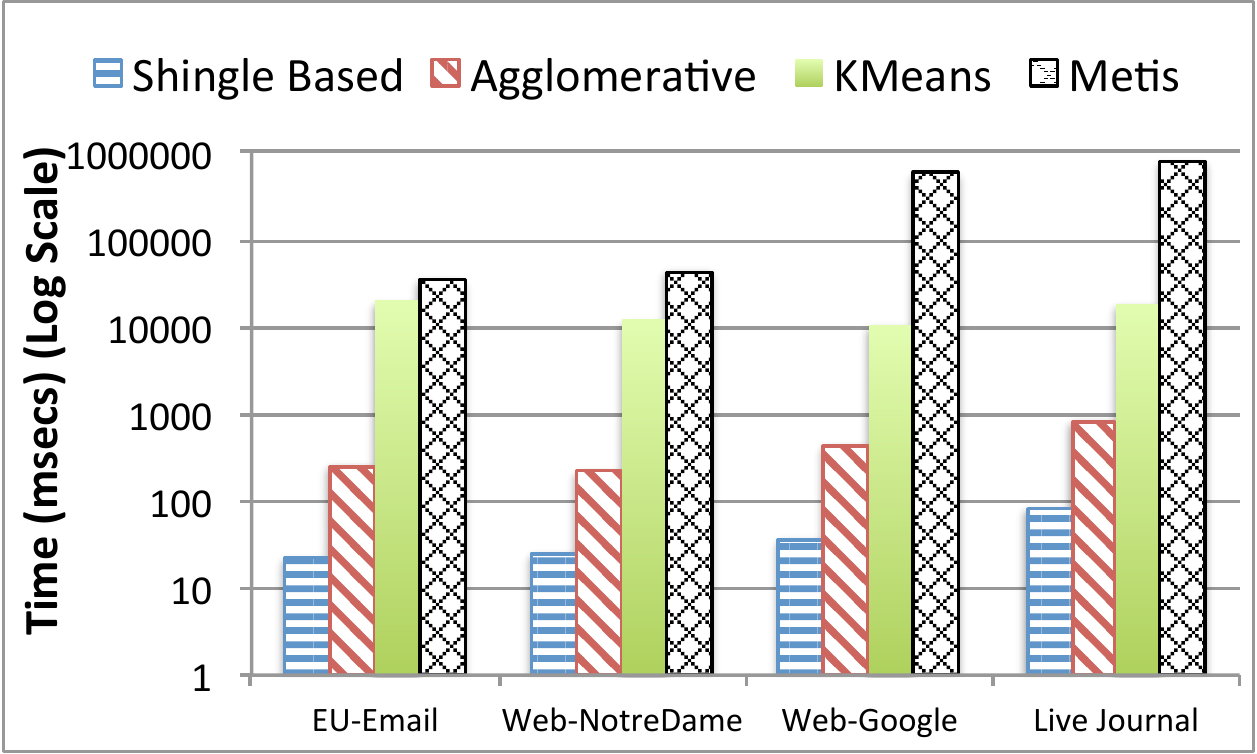}}
\subfigure[]{\label{fig:BPM_9}\includegraphics[scale=0.44]{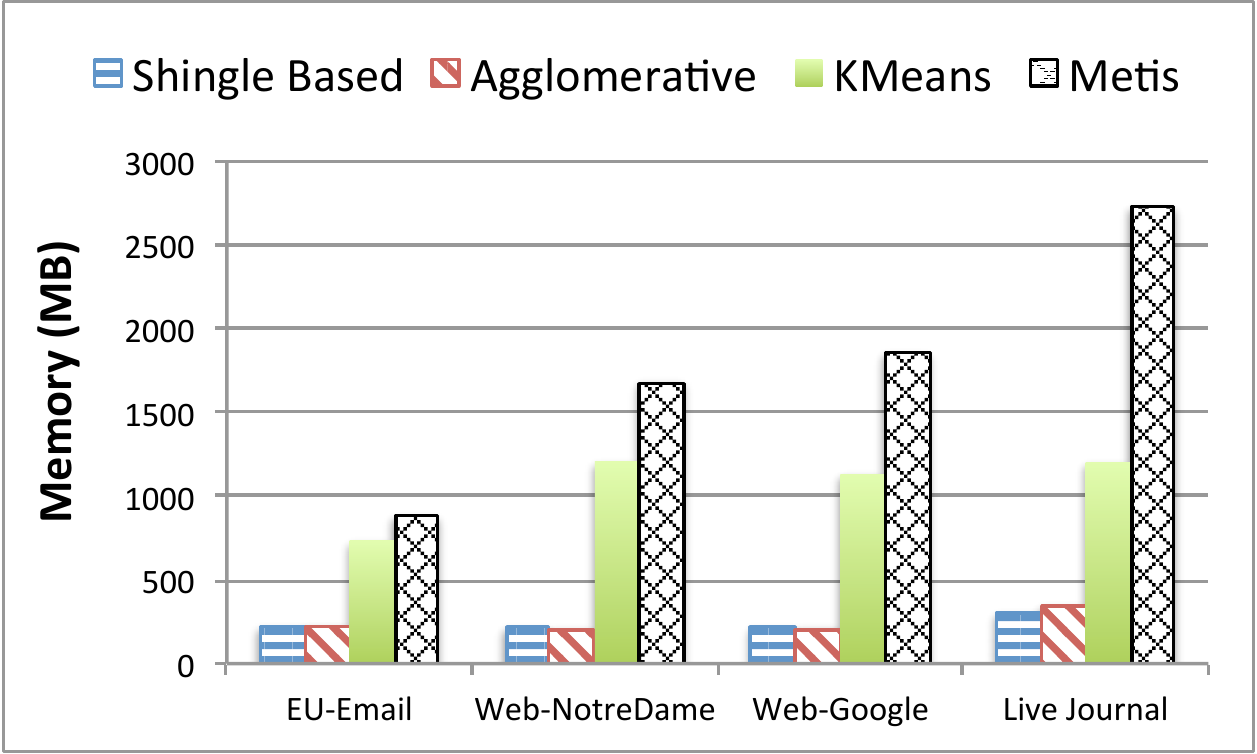}}\\
\vspace{-2pt}

\caption{For the different shingle based subgraph packing heuristics, we compare: (a) \#bins required; (b) total
    computational effort required; (c) total elapsed time (wall clock time) for running the LCC
        computation on the subgraphs; (d) total cluster memory required for GEP and execution of the
        LCC computation; (e)-(f) distribution of \# subgraphs and of execution engine running times over the bins; Comparison of shingle based subgraph packing heuristics with the other bin packing heuristics; 
        we compare: (g) \#bins required; (h) total time taken for bin packing; (i) memory required. }
 
  \label{fig:Graph1}
\vspace{-8pt} 
  \end{figure*}

\vspace{-5pt}
\subsection{GEP Evaluation}
\label{sec:GELEval}
\Paragraph{Comparing subgraph bin packing (SBP) algorithms.} 
We first evaluated the the performance and quality of the bin packing-based algorithms: First Fit, First 
Fit Decreasing and Shingle Based bin packing on the LiveJournal data set. 

Figure~\ref{fig:BPM_1} shows the number of bins required to
partition the subgraphs as we vary the number of subgraphs specified by the query
(using predicates on the query vertices). The number of bins increases as the number of subgraphs
increases for all the three heuristics. We see that the First Fit algorithm requires the maximum number
of bins as expected whereas the shingle-based packing algorithm performs the best in terms of packing
the subgraphs into a minimum number of bins. This is due to the fact that the shingle-based bin packing algorithm orders the
subgraphs based on neighborhood similarity thereby taking maximum advantage of the overlap amongst
them. 

To ascertain the cost of data analytics we study the effect of bin packing on the computation
effort $\mathcal{CE}$. 
Figure~\ref{fig:BPM_2} shows that the $\mathcal{CE}$ for the First Fit algorithm is the maximum
making it the most expensive, while the $\mathcal{CE}$ for shingle-based packing algorithm is the
minimum making it the most cost effective bin packing solution. Figures~\ref{fig:BPM_3},
        \ref{fig:BPM_4} show the execution (elapsed) time and the total cluster memory usage for
        binning and execution with respect to these three heuristics and different number of
        subgraphs. The First Fit has the best execution time and the shingle-based bin packing
        algorithm has an execution time which closely follows that of the First Fit algorithm.
        On the other hand, the First Fit Decreasing algorithm takes the largest execution time. This can be
        attributed to the fact the First Fit is expected to produce the most uniform distribution of
        subgraphs across the bins and the First Fit Decreasing is likely to produce a skewed
        distribution (packing a large number of smaller subgraphs in later bins) leading to larger
        execution times due to the straggler effect. 

We further study the distribution of the number of subgraphs (or query-vertices) packed per bin and
the distribution of running times of each instance of a execution engine on a bin (partition).
Figures~\ref{fig:BPM_5},\ref{fig:BPM_6} show the box plots with whiskers for both the distributions.
As expected the First Fit algorithm has the most uniform distribution across the bins in both cases. The
shingle-based packing algorithm also performs well and provides a distribution almost as good as the
First Fit algorithm, while First Fit Decreasing has the most skewed distribution in both cases, which also explains the highest end-to-end execution timings for the heuristic.  We thus see that the shingle-based packing algorithm performs the best in terms of minimizing the \# bins and $\mathcal{CE}$, having low execution times and almost uniform bin distributions thus minimizing the straggler effect. 

To summarize, our results showed that 
our proposed shingle-based packing algorithm performs much better than the other two algorithms in terms of minimizing the \# bins 
and $\mathcal{CE}$. It also has low execution times and almost uniform bin distributions thus minimizing the straggler effect. 

We next compare the shingle-based bin packing heuristic with the two clustering-based algorithms, 
and the METIS-based algorithm.
Figures~\ref{fig:BPM_7},~\ref{fig:BPM_8} and~\ref{fig:BPM_9}
show the performance of the four subgraph packing approaches for three different real-world datasets. 
We see that K-Means provides generally found solutions with minimum number of bins, but takes much 
longer and consumes significantly more memory.
The shingle-based solution finds almost as good solutions, but is much more efficient.
METIS-based partitioning does poorly both in terms of
binning quality and the efficiency (notice the log scale in Figure~\ref{fig:BPM_8}), and we did not
consider it for the rest of experimental evaluation.

\begin{figure}[t]
\centering
\subfigure[]{\label{fig:optimal_1}\includegraphics[scale=0.43,trim=10 10 10 10]{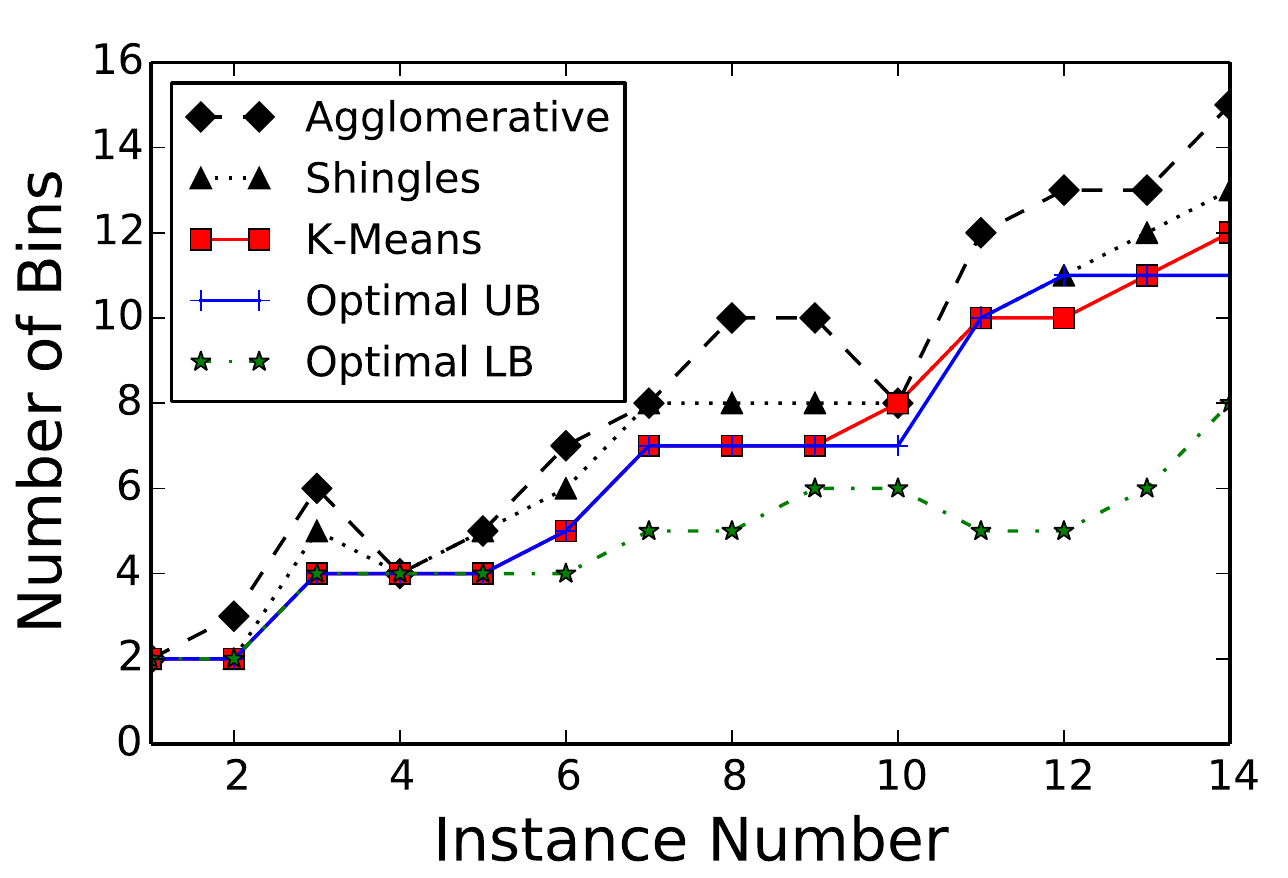}}\hspace{10pt}
\subfigure[]{\label{fig:optimal_2}\includegraphics[scale=0.43,trim=10 20 10 10]{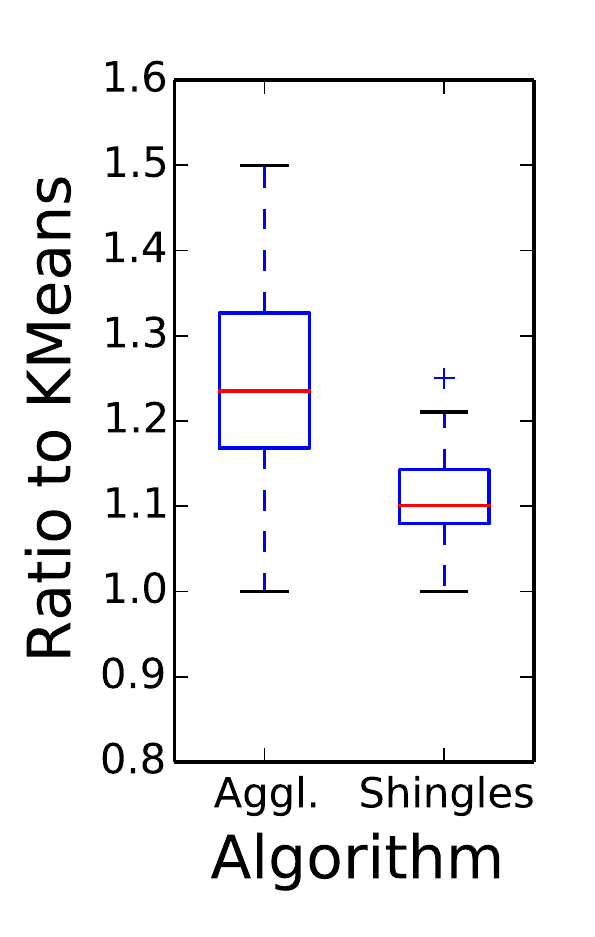}}\\
 \vspace{-12pt} 
\caption{Comparing subgraph packing heuristics to (a) the optimal solution and (b) each other, for
    synthetic graphs}
\label{fig:optimal}
\end{figure}

To better evaluate the performance of the heuristics, we also compared them with an optimal
algorithm (OPT), that constructs an Integer Program for the problem instance, and uses the Gurobi Optimizer
to find an optimal solution. Unfortunately, even after many hours on a powerful server per problem
instance, OPT was unable
to find a solution for most of our small-scale synthetically generated problem instances; 
for 14 of 64 synthetic datasets, it found either an optimal solution or 
reasonable bounds, and we have plotted those in Figure~\ref{fig:optimal_1} (the x-axis is sorted
by the value of the best solution found by OPT). We note that the only instances
where OPT found the optimal solution (i.e., where upper bound = lower bound) were solutions with 2
or 4 bins. As we can see, for almost all of these problem instances, our K-Means heuristics was able to
match the OPT solution.

\red{
Overall, the reason K-Means performs so well can be attributed to the fact that it explores the solution
space more extensively and in general, does more
pair-wise comparisons between the sets (corresponding to the subgraphs). The behavior was 
consistent across a wide range
of experiments that we did. 
The shingle-based heuristic, on other other hand, restricts the comparisons to subgraphs that are close
in the shingle order, and thus may miss out on pairs of sets that have high overlap.
At the same time, we want to note that KMeans takes much longer 
to run and consumes significantly more memory, whereas the shingle-based heuristic is much faster and finds
solutions with comparable quality.
%
%
}

Figure~\ref{fig:optimal_2} compares the K-Means heuristics against
the other two heuristics for all 64 datasets. The results are consistent with the results we
saw on the real-world datasets -- K-Means is consistently better than both of those heuristics, but
the shingle-based heuristic comes quite close to its performance.

\begin{figure*}{t}
\centering
\subfigure[]{\label{fig:Distr_GEL_1}\includegraphics[scale=0.44]{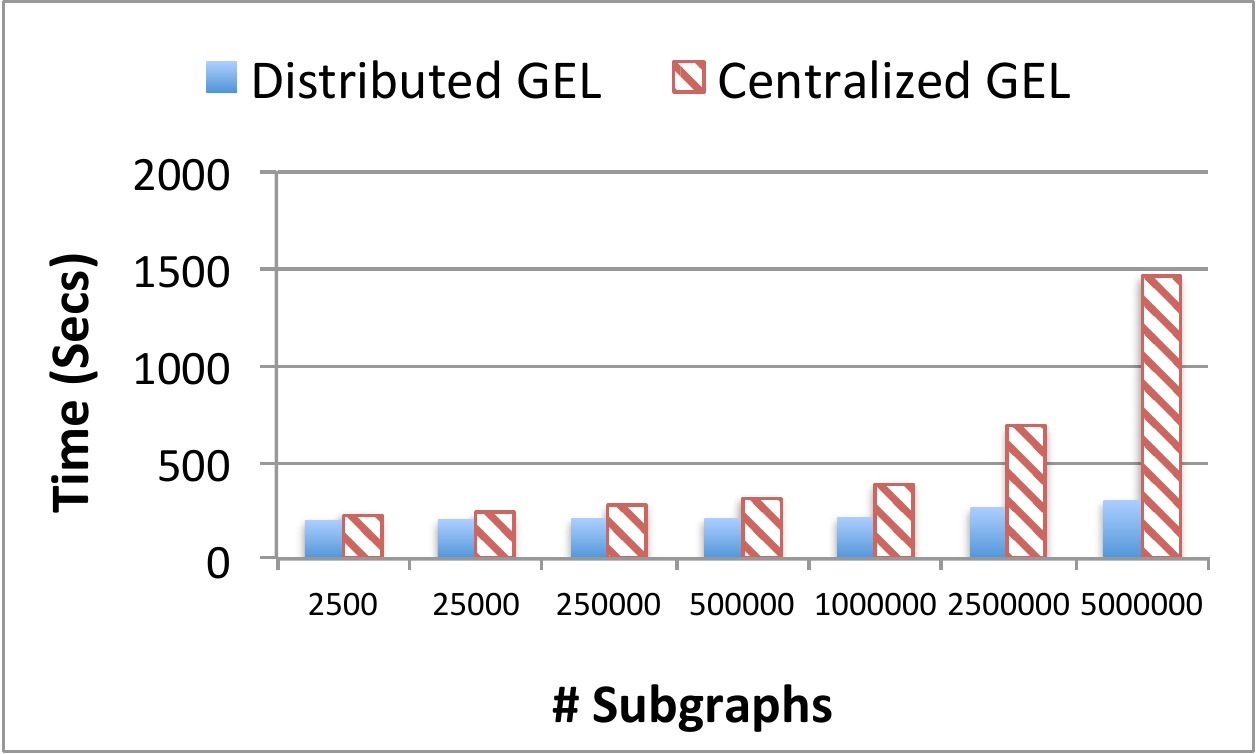}}
\subfigure[]{\label{fig:Distr_GEL_2}\includegraphics[scale=0.44]{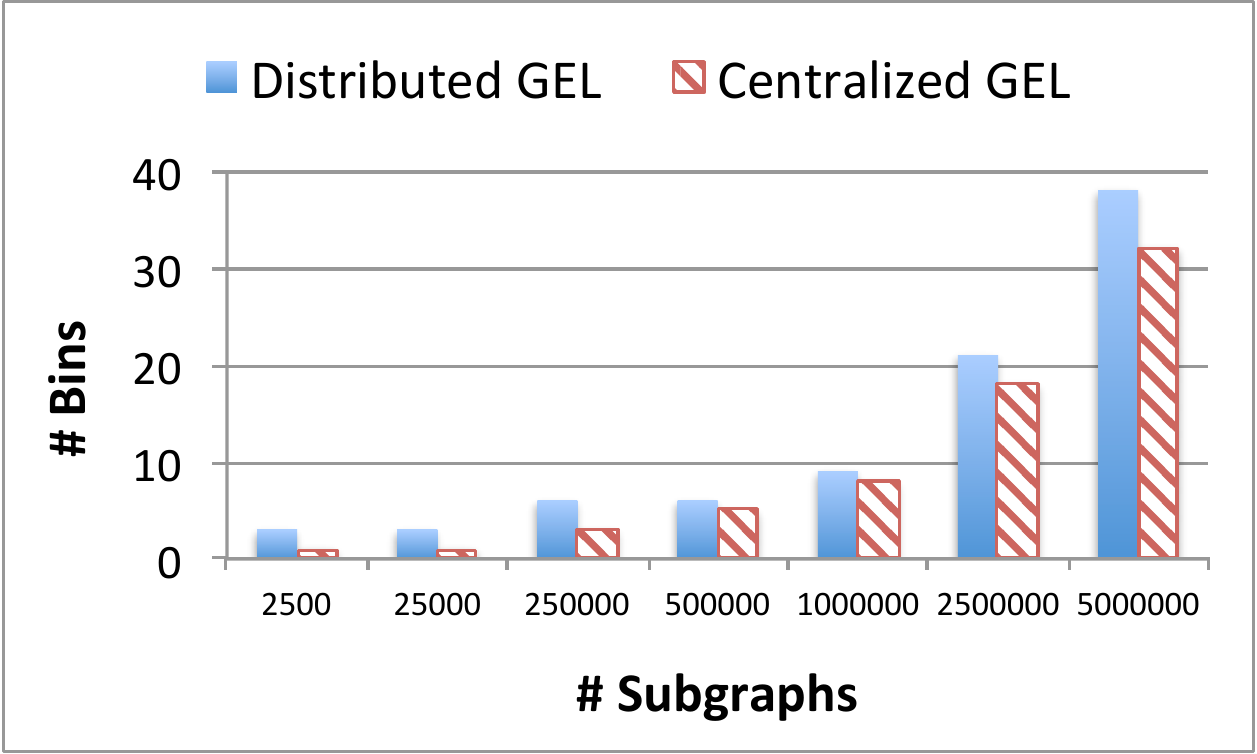}}
\subfigure[]{\label{fig:Distr_GEL_3}\includegraphics[scale=0.44]{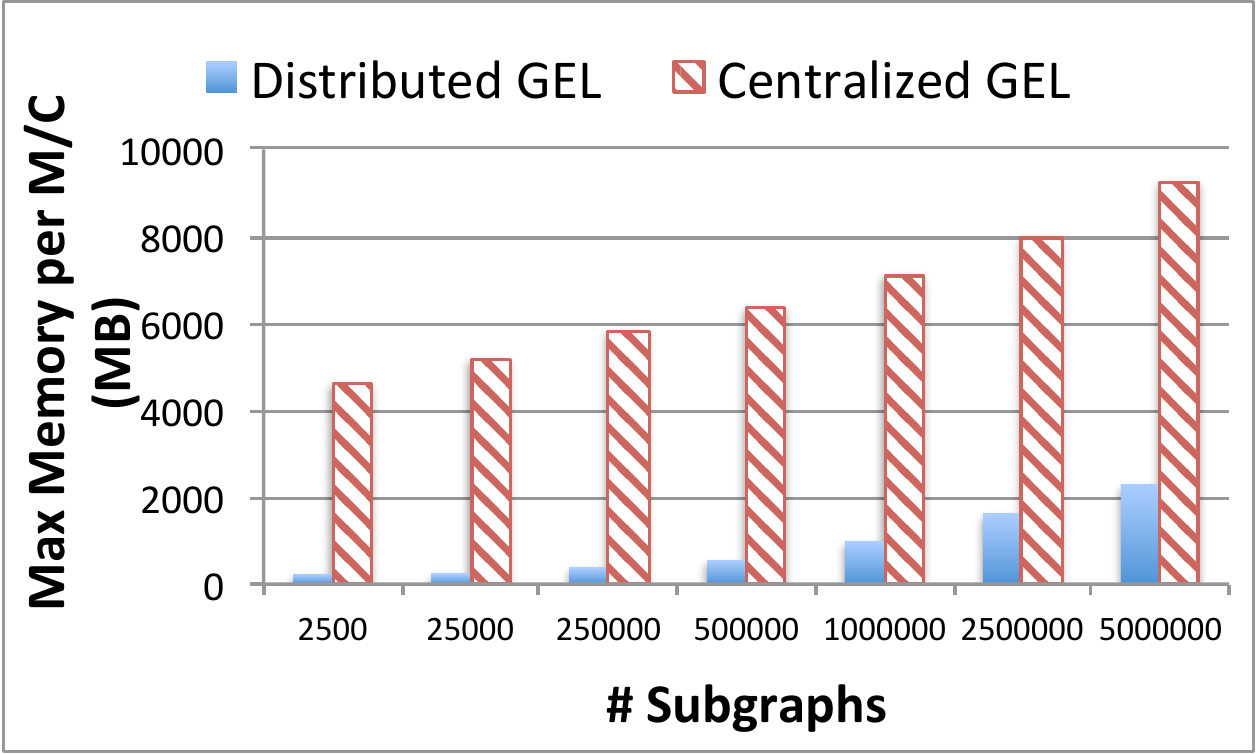}}\\
\vspace{-5pt}
\subfigure[]{\label{fig:Distr_GEL_4}\includegraphics[scale=0.44]{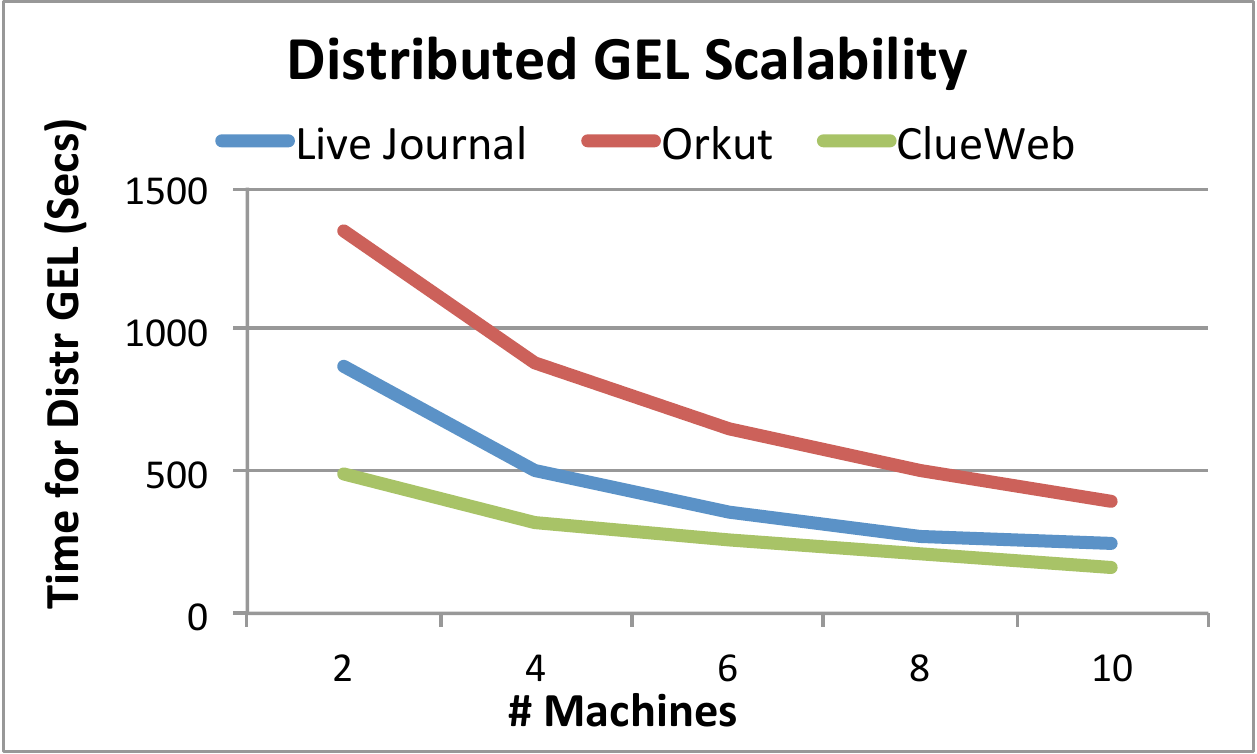}}
\subfigure[]{\label{fig:Distr_GEL_5}\includegraphics[scale=0.44]{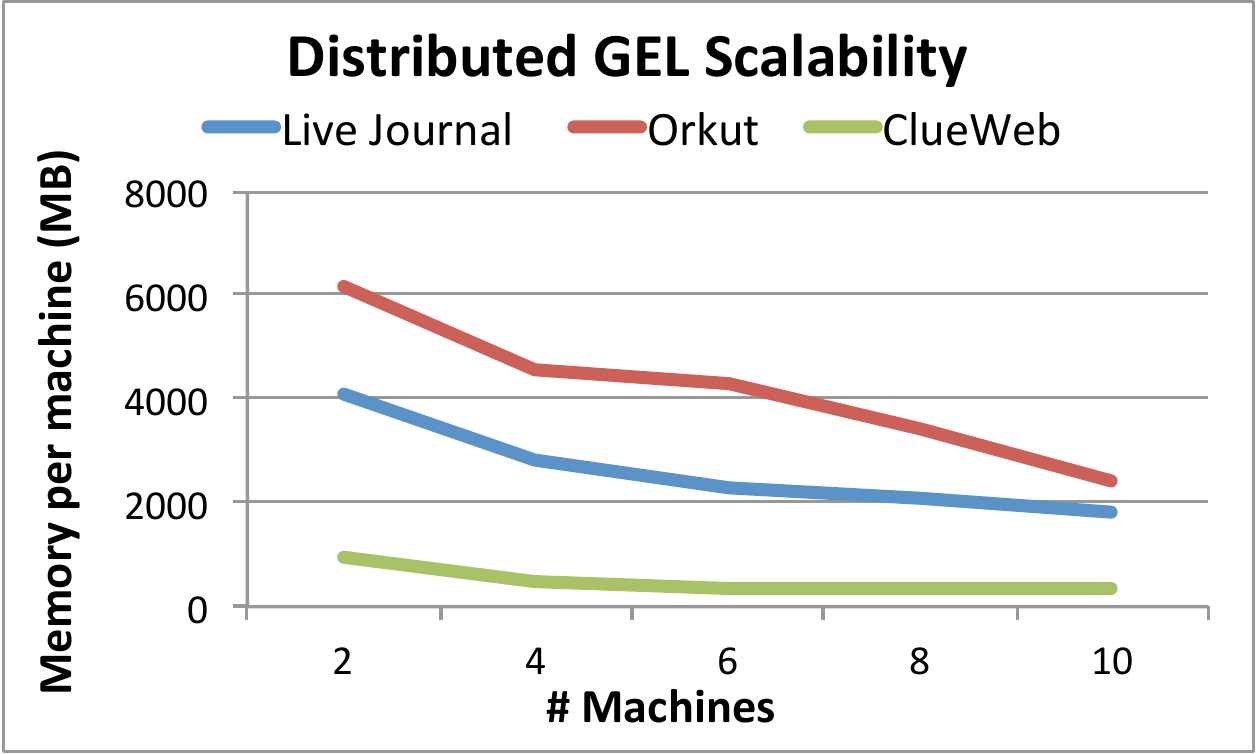}}
\subfigure[]{\label{fig:Distr_GEL_6}\includegraphics[scale=0.44]{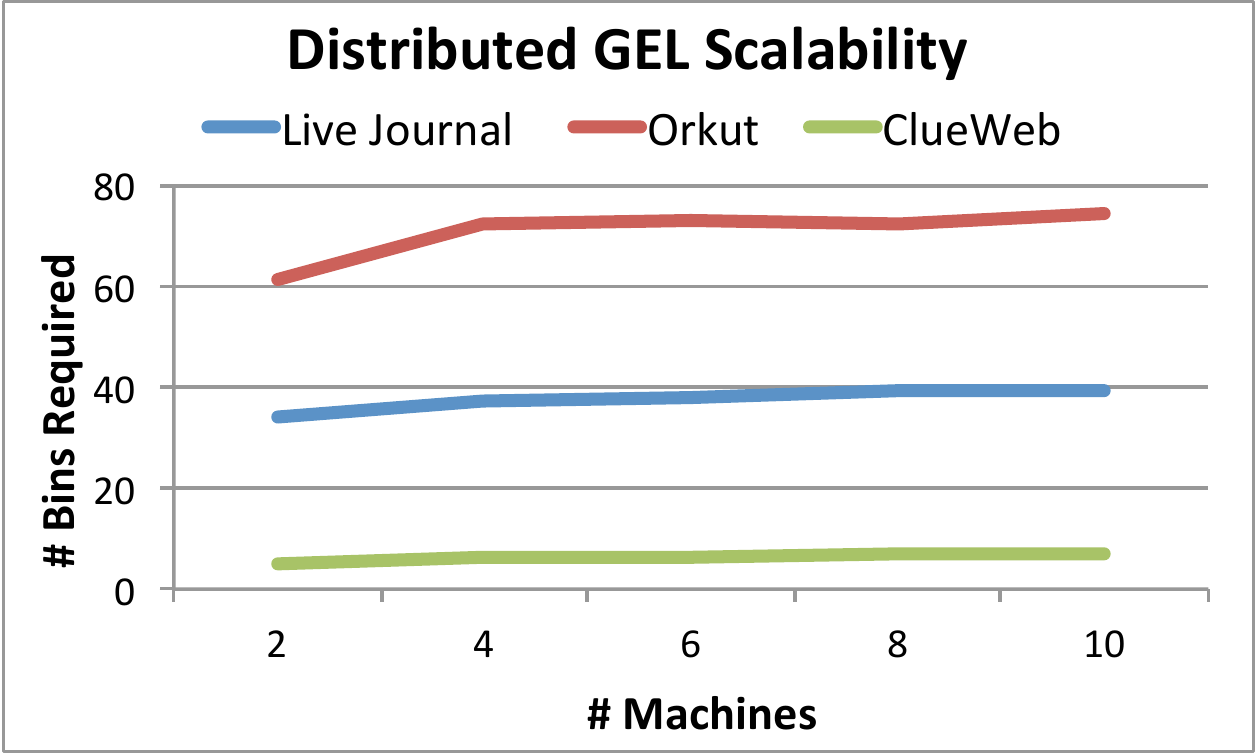}}\\
\vspace{-5pt}
\caption{GEP architecture: (a)-(c) Comparison of centralized and distributed GEP architectures;
        (d)-(f) Distributed GEP architecture: Impact on graph extraction and packing time,  max memory required per bin, and \#bins required for packing with increase in number of machines.}
\label{fig:DistrGEL}
\end{figure*}

\begin{figure*}[t]
  \centering
 
\subfigure[]{\label{fig:execModes}\includegraphics[scale=0.44]{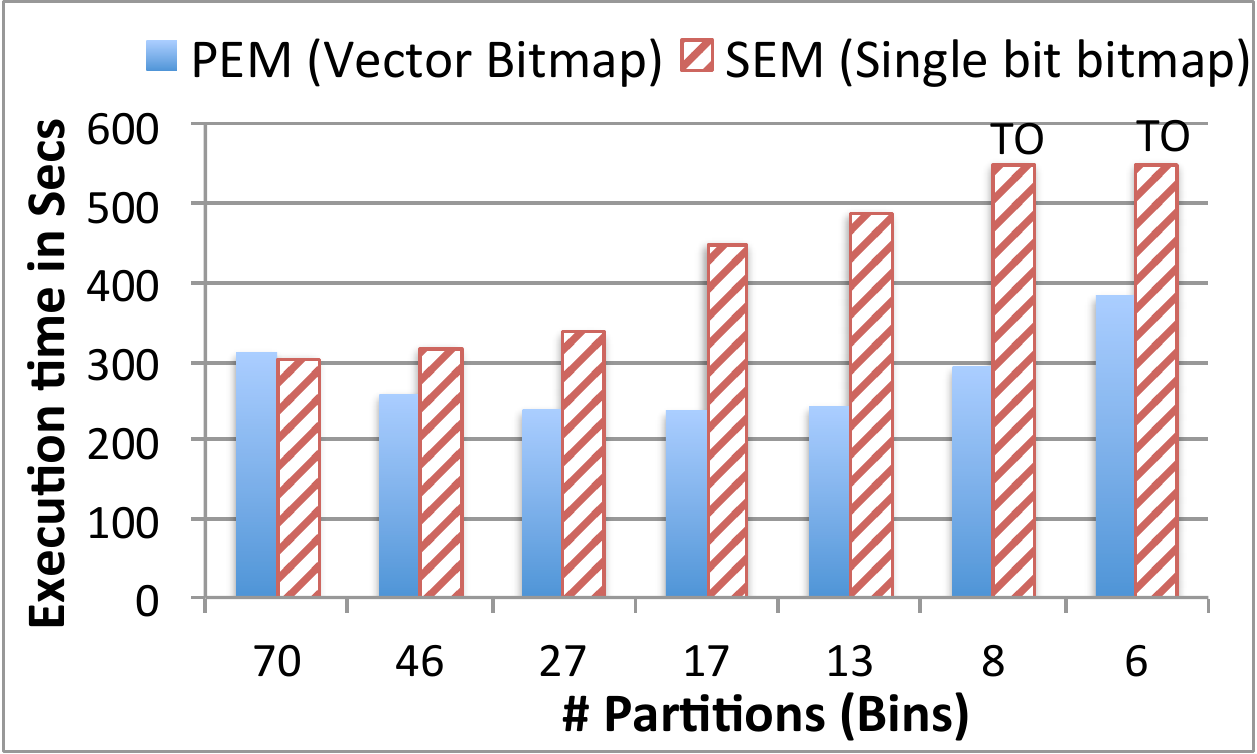}}
\subfigure[]{\label{fig:BitMap1}\includegraphics[scale=0.44]{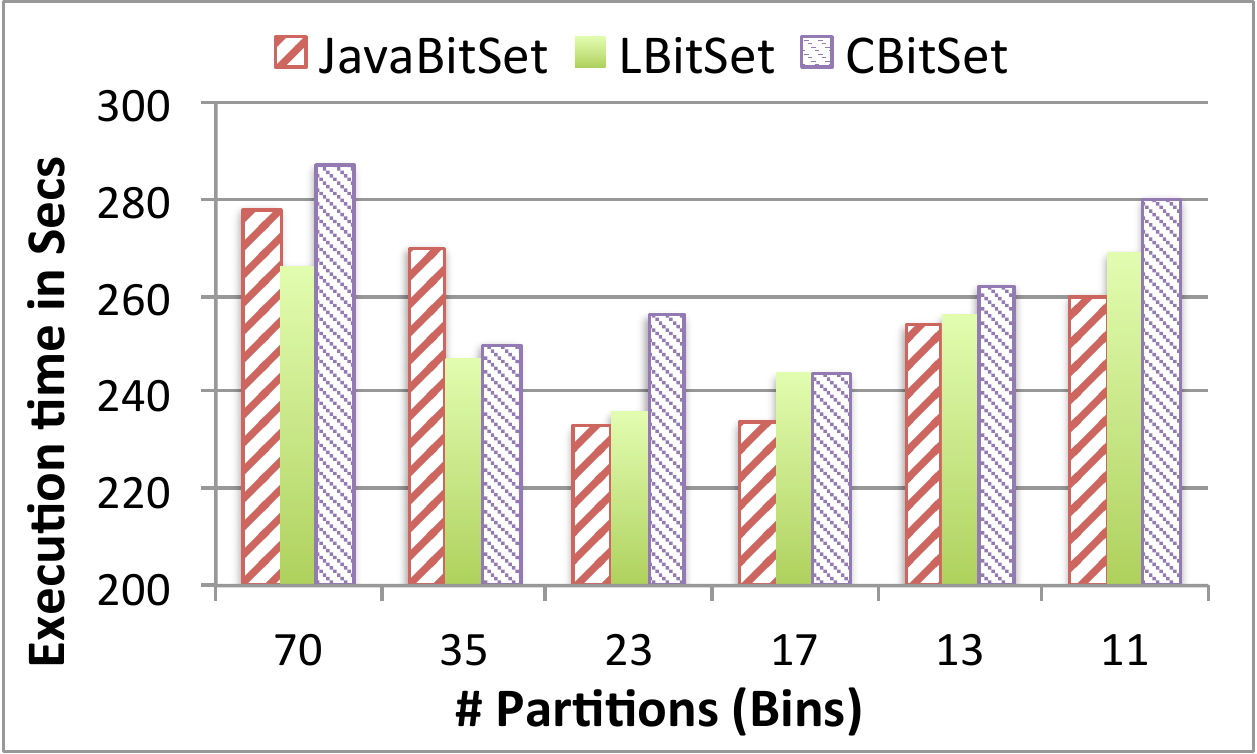}}
\subfigure[]{\label{fig:BitMap2}\includegraphics[scale=0.44]{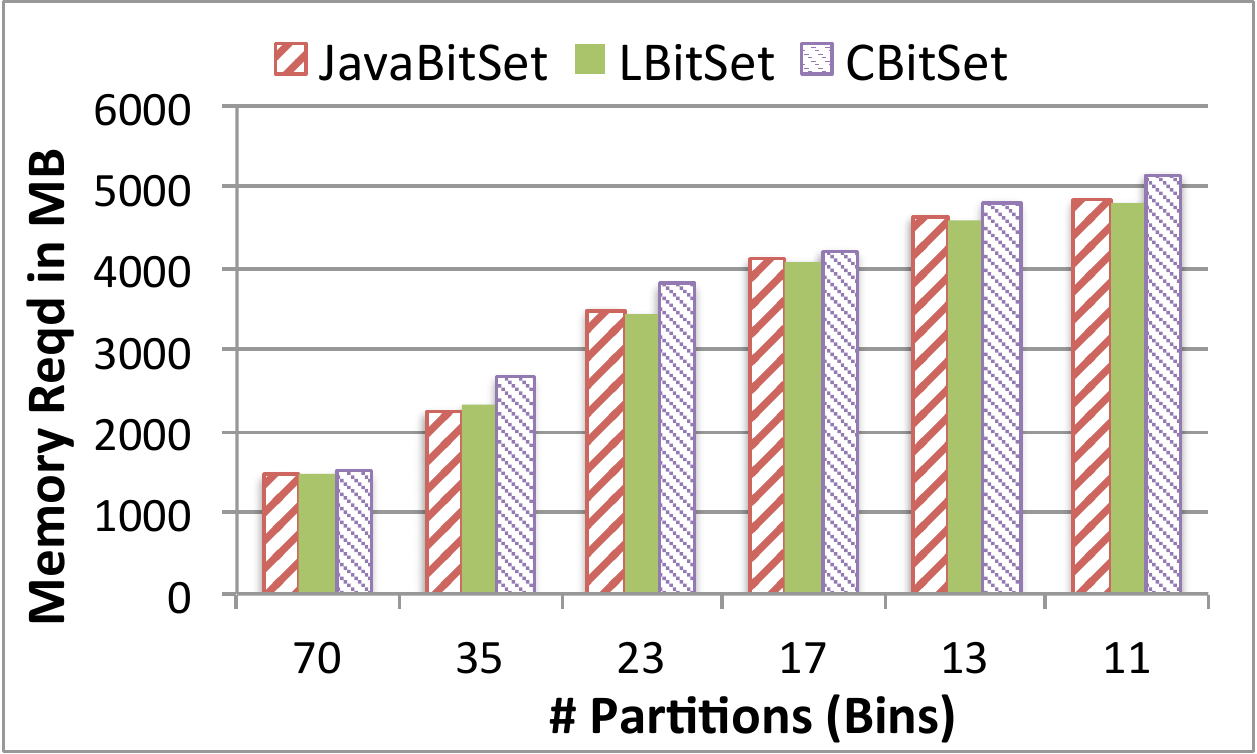}}\\
\vspace{-2pt}
\subfigure[]{\label{fig:End-toEnd}\includegraphics[scale=0.44]{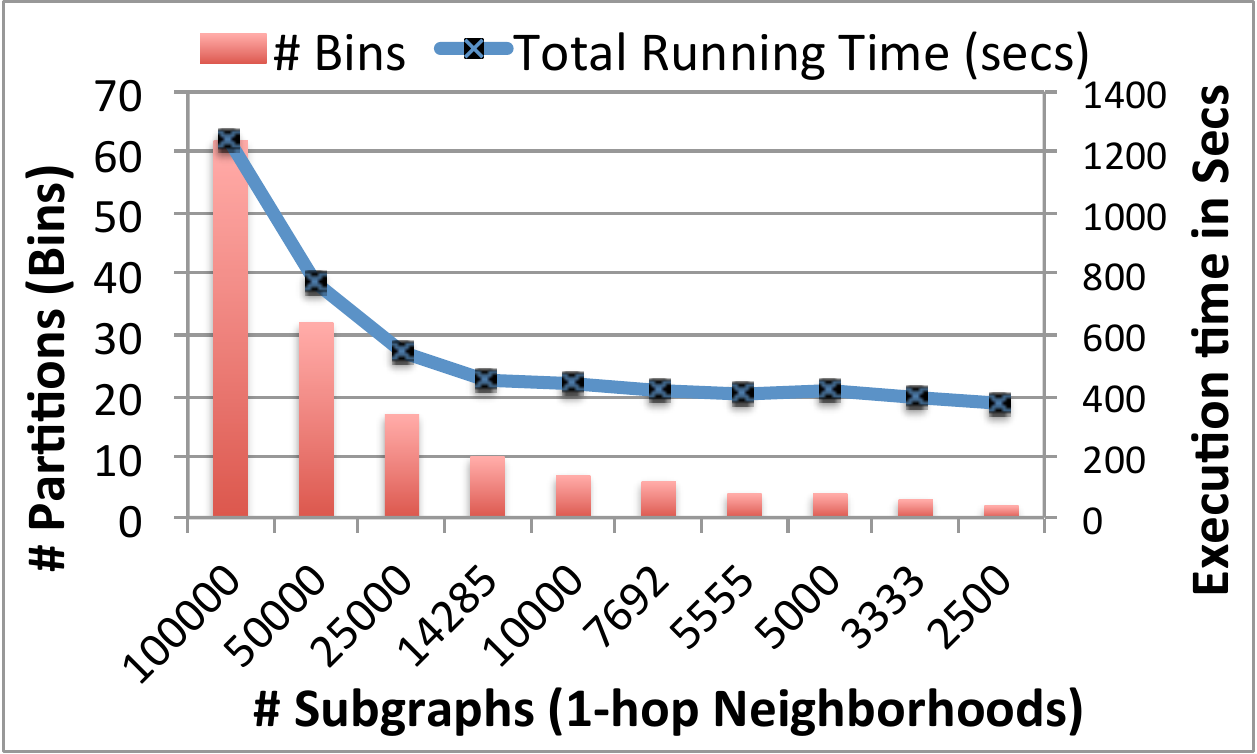}}
\subfigure[]{\label{fig:perfBD1}\includegraphics[scale=0.44]{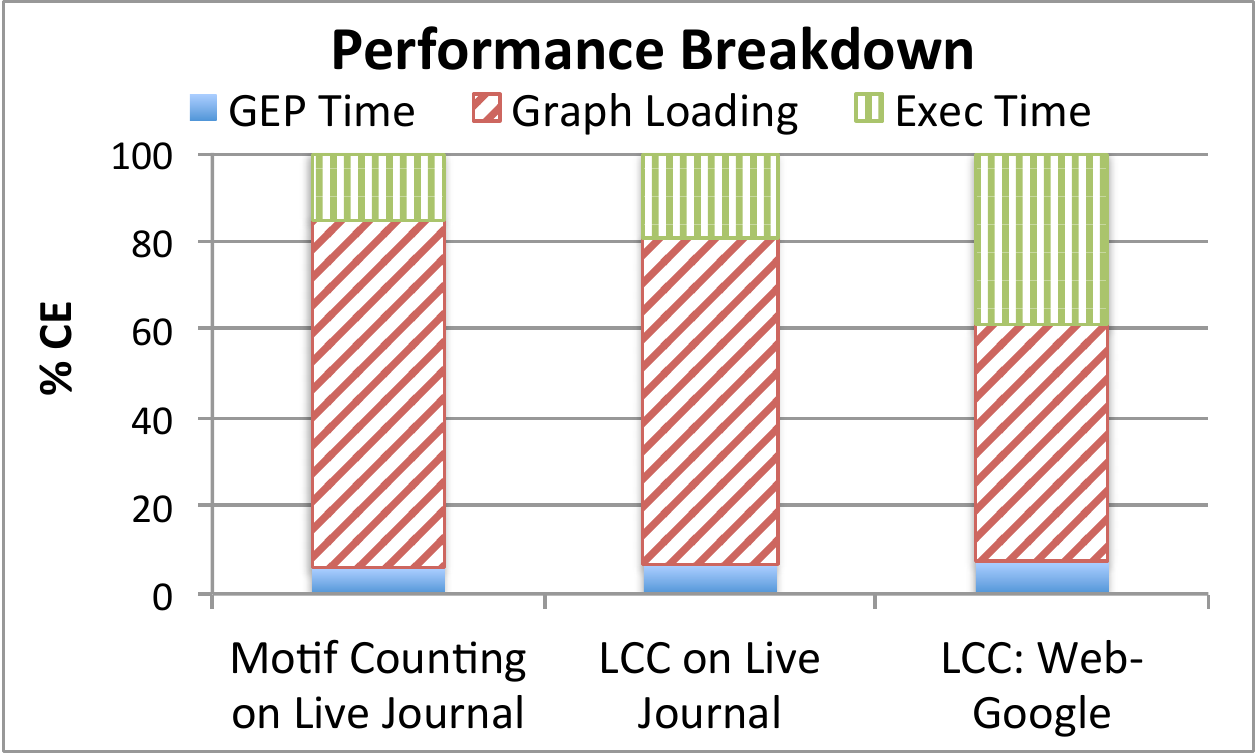}}
\subfigure[]{\label{fig:scalability}\includegraphics[scale=0.44]{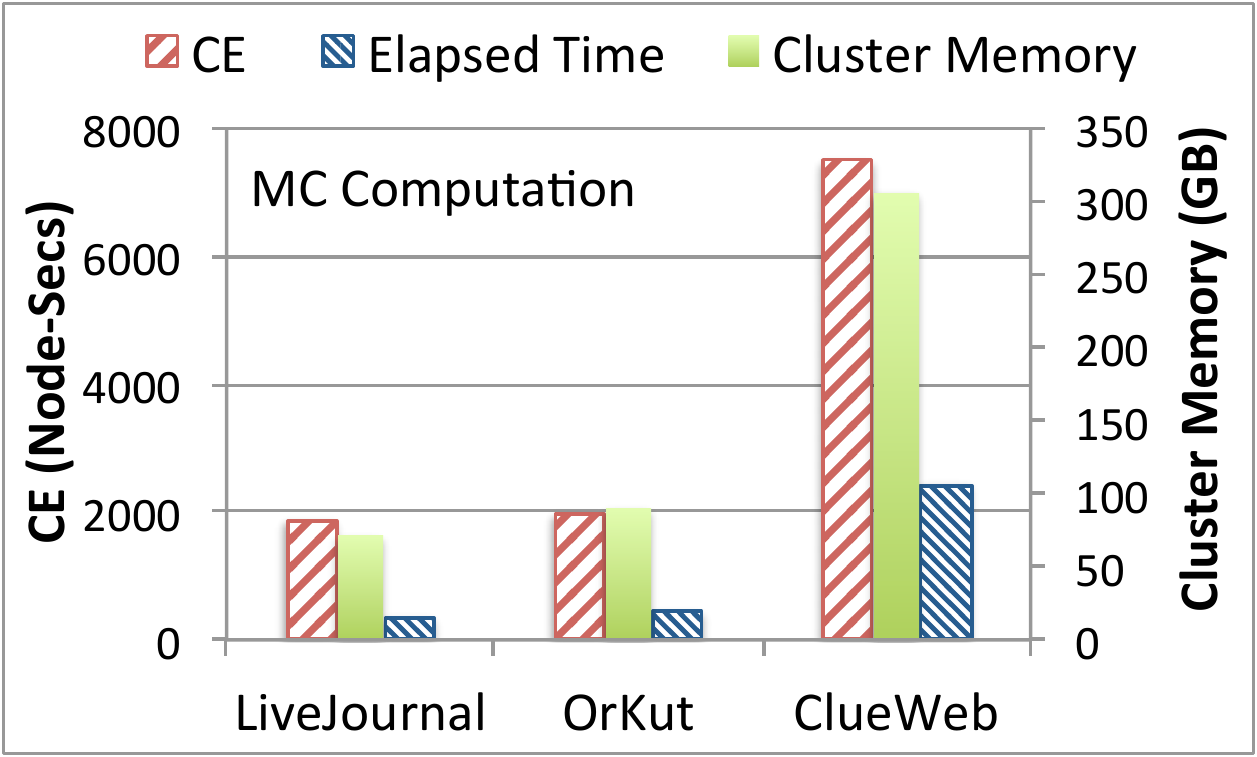}}\\
 \vspace{-12pt} 
\caption{(a) Effect of different execution modes on the running time; 
    (b)-(c) Effect of different bitmap implementations on the
        memory footprints and the running times of the execution engine; 
    (d) End-to-End running time and \#partitions required for different numbers of subgraphs;
    (e) Performance breakdown of different stages of \nframe~for graphs of different sizes and different applications; 
    (f) Scalability: \nframe~performance over large graphs.}
 
  \label{fig:Graph2}
\vspace{-13pt} 
  \end{figure*}

%


\Paragraph{Distributed GEP evaluation.} 
Figures~\ref{fig:Distr_GEL_1},~\ref{fig:Distr_GEL_2}, and~\ref{fig:Distr_GEL_3} compare the
distributed implementation of the GEP module with the centralized version (we use LiveJournal dataset for
this purpose, which is small enough for a centralized solution, and we use 6 machines in the
distributed case).
%
We see that, for a small number of extracted subgraphs, the time taken by the centralized 
solution is comparable to the time taken by
the distributed solution. However as we scale to a large number of subgraphs, the distributed
solution scales much better, and more importantly, the maximum memory required on any 
single machine is much lower, thus removing a key bottleneck of the centralized solution. 
The binning quality of the centralized solution is 
somewhat better, which is to be expected, and hence it would still be preferable to run the GEP 
phase in a centralized fashion. However the gap is not significant, and for large graphs
where running GEP in a centralized fashion is not feasible, distributed GEP generates reasonable
solutions.


Figures~\ref{fig:Distr_GEL_4}, \ref{fig:Distr_GEL_5} and~\ref{fig:Distr_GEL_6}   show the effect of
increasing the number of machines used for distributed GEP on the time taken, memory required 
per machine and the quality of binning solution provided in terms of number of bins required,  for three data sets. 
We see that our distributed GEP mechanism exhibits good scaling behavior without compromising much 
on the quality of binning. It can thus handle very large 
graphs quite effectively. Note that the number of query vertices was set to 3M, so the relative performance for the
different graph does not correlate with the original graph sizes (in particular, ClueWeb has low
        average degree, hence requires fewer bins for the same number of neighborhoods).
%


\begin{figure*}[t]
  \centering
 
\subfigure[]{\label{fig:iter_support_1}\includegraphics[scale=0.44]{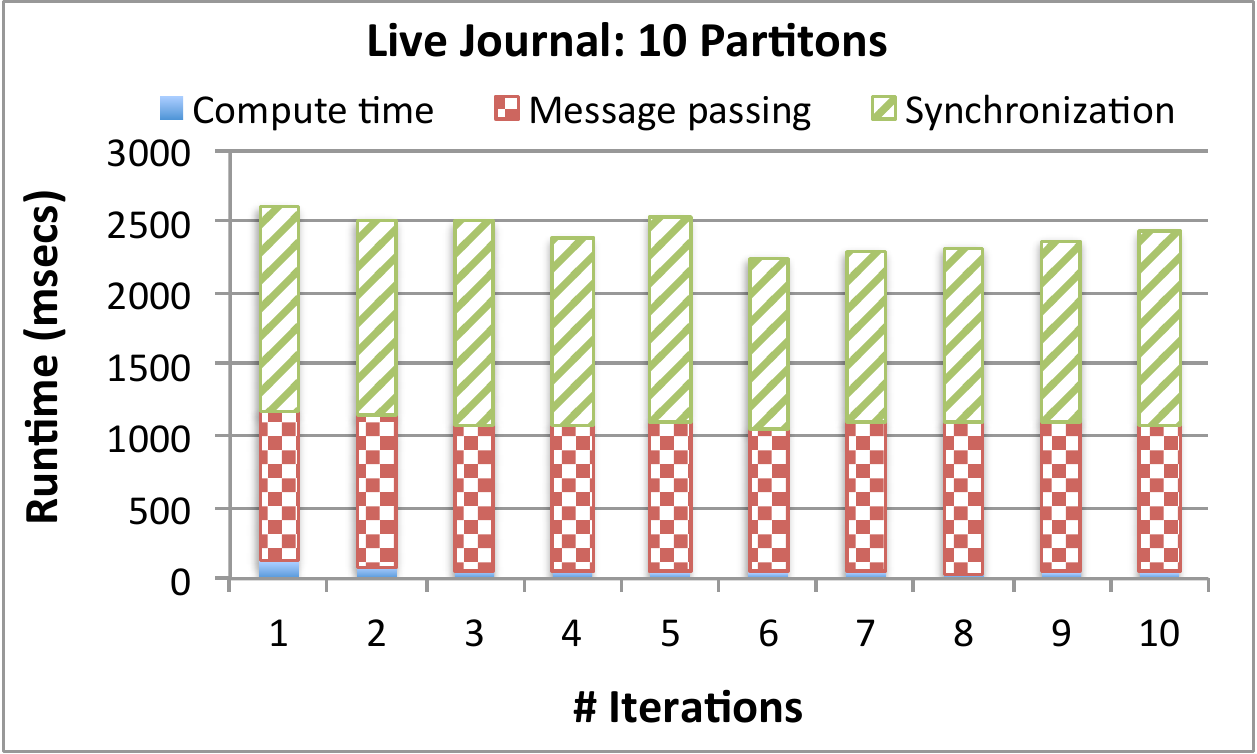}}
\subfigure[]{\label{fig:iter_support_2}\includegraphics[scale=0.44]{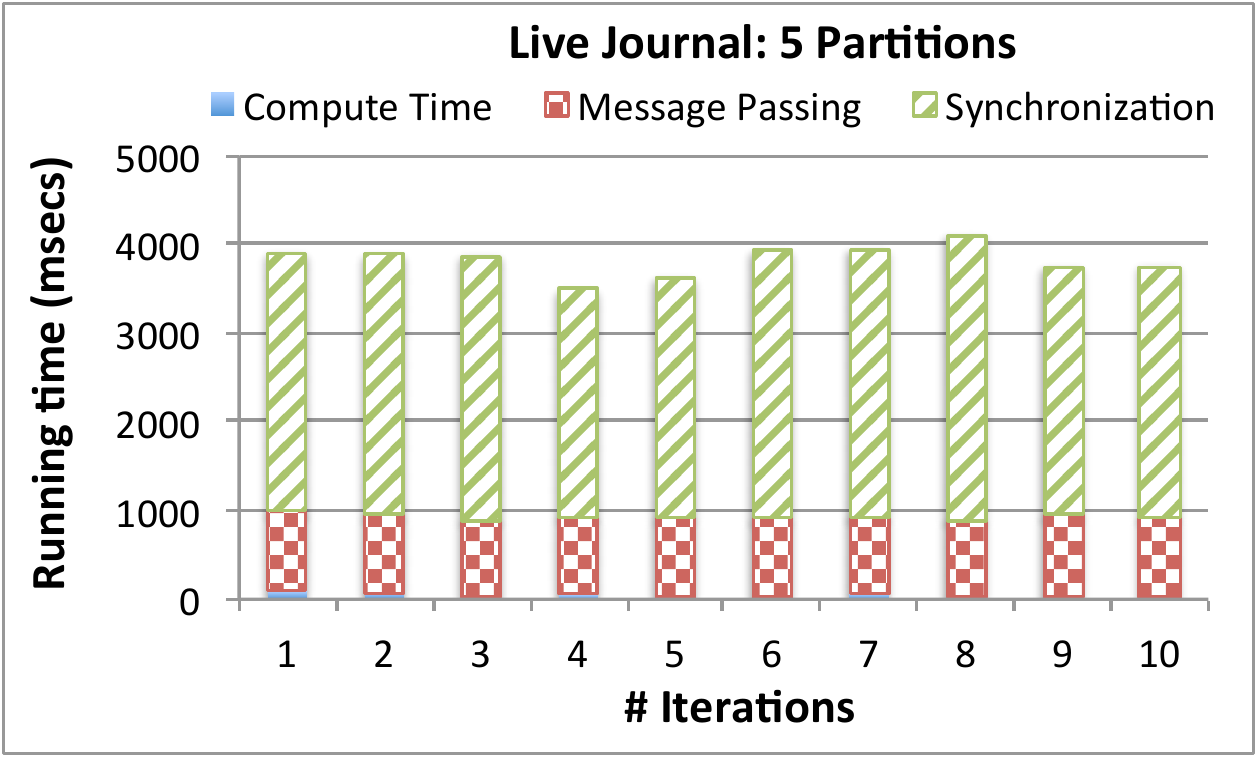}}
\subfigure[]{\label{fig:iter_support_3}\includegraphics[scale=0.44]{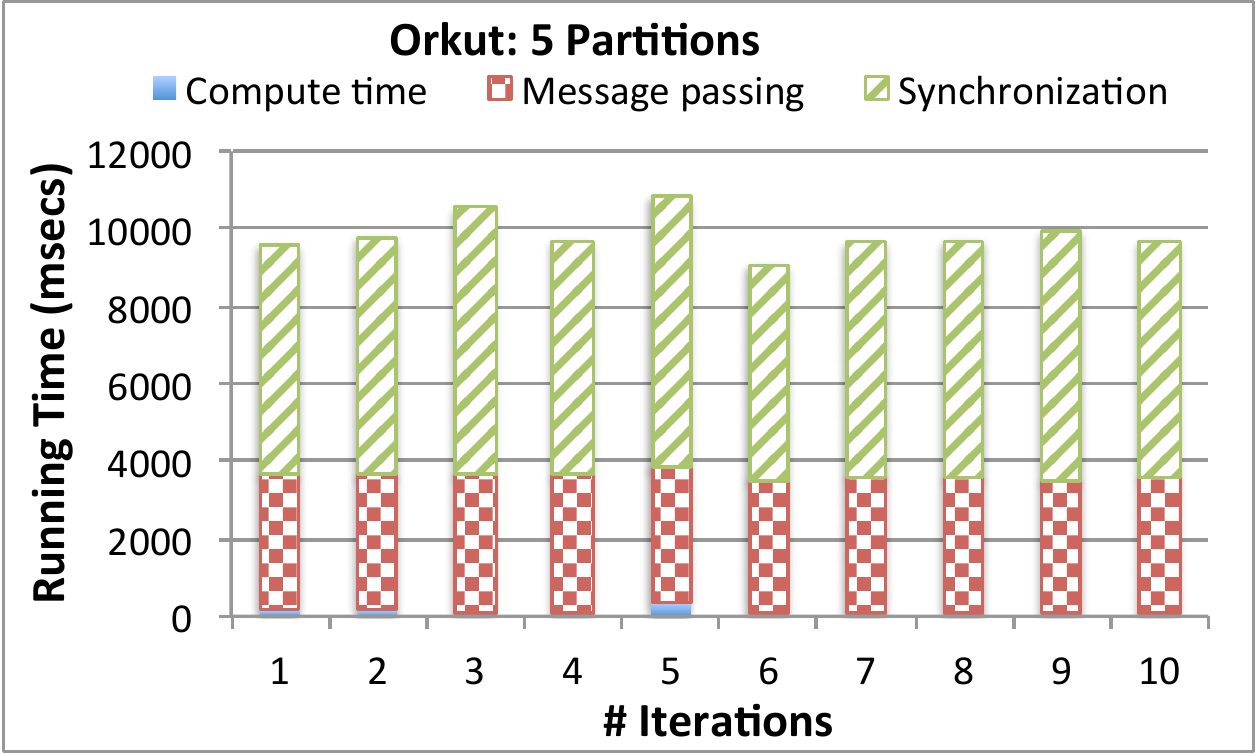}}\\
\vspace{-2pt}
\subfigure[]{\label{fig:iter_support_4}\includegraphics[scale=0.44]{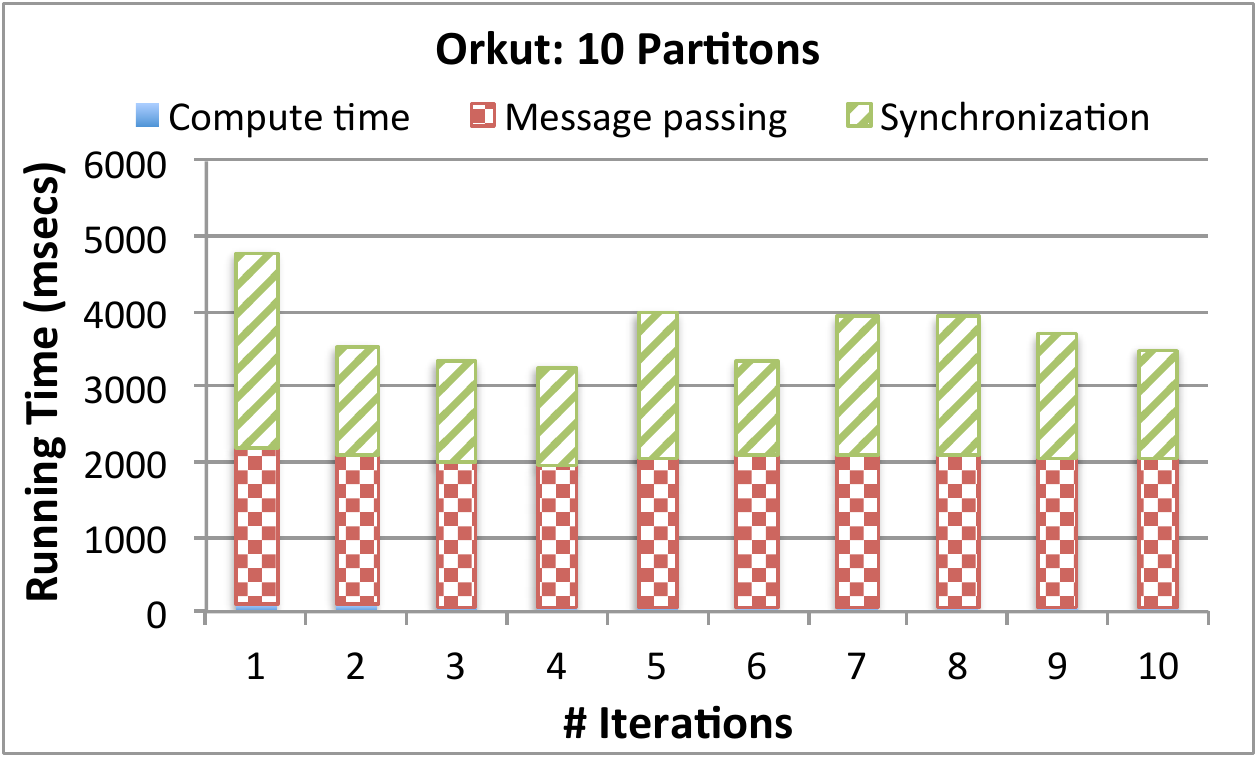}}
\subfigure[]{\label{fig:iter_support_5}\includegraphics[scale=0.44]{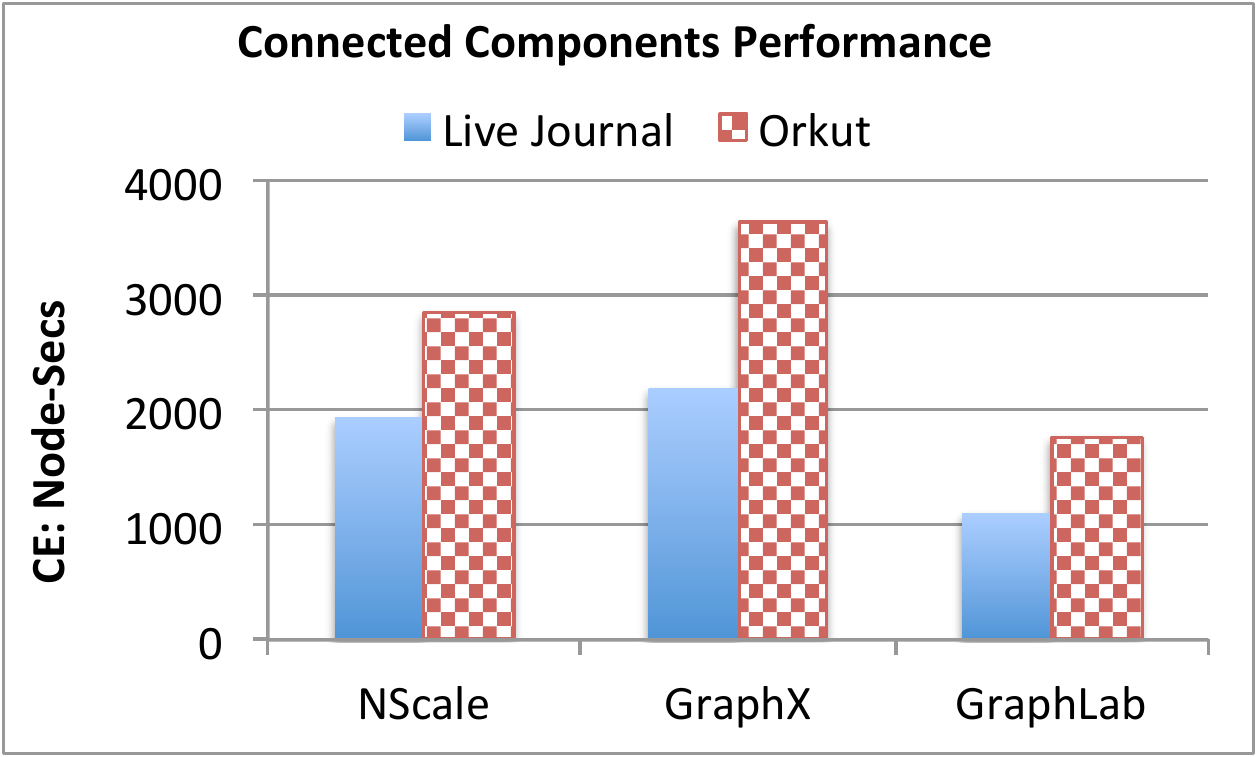}}
\subfigure[]{\label{fig:iter_support_6}\includegraphics[scale=0.44]{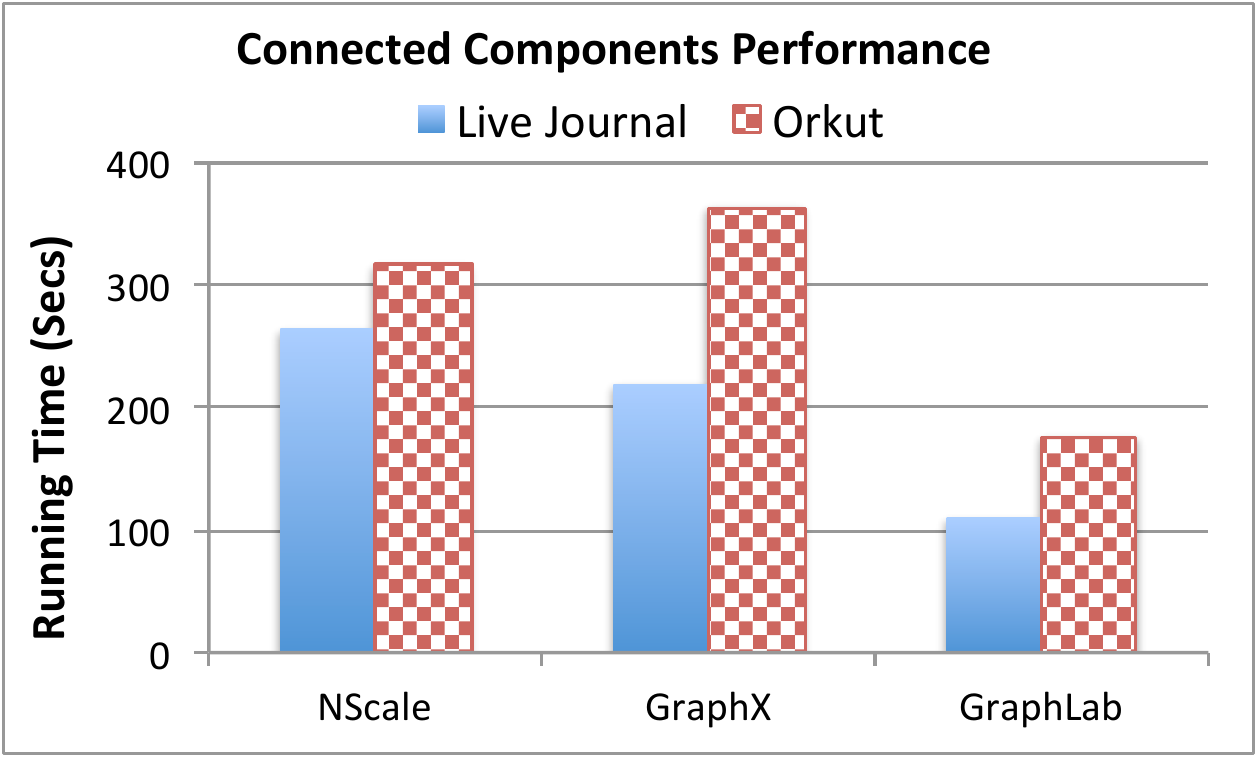}}
 \vspace{-12pt} 
\caption{Connected components: (a-d)  Performance break down for different iterations;
    (e-f) Performance comparison with GraphX and GraphLab in terms of running time and CE (node-secs).}
 
  \label{fig:IterativeSupport}
\vspace{-13pt} 
  \end{figure*}

\vspace{-15pt}
 \subsection{Execution Engine Evaluation}
\vspace{-10pt}
\label{sec:execEngEval}
\Paragraph{Effect of choosing different execution modes.} 
In Figure~\ref{fig:execModes}, we plot the total running times for the single-bit serial (SEM) and vector
bitmap parallel (PEM) execution modes, for the LiveJournal graph, for 25000 extracted subgraphs. 
We see that for 70 partitions, the performance of the two modes is comparable since each partition 
does a small amount of work. However as
the number of partitions decreases, PEM performs much better compared to SEM which times out as the
number of partitions becomes very small. 
On the other hand, SEM uses a single bit bitmap per vertex or edge and hence requires significantly less
memory, and may be useful when we have a large number of low memory machines available for 
graph computation. 

\Paragraph{Bitmap constructions.} 
Figures~\ref{fig:BitMap1},~\ref{fig:BitMap2} compare the different bitmap implementations 
for different numbers of partitions (the setup is the same as above, and hence decreasing number of
partitions implies increasing number of subgraphs per partition). Java BitSet and LBitSet
perform better than CBitSet in terms of execution time, while LBitSet consumes the least amount of
memory as the number of subgraphs in each partition increases. 
As mentioned in Section~\ref{sec:bitmapImpl}, CBitSet is useful
in cases where the overlap between subgraphs is minimum, requiring a small number of bits to be
set. We use LBitSet for most of our experiments. 
\vspace{-5pt}

\vspace{-5pt}
\subsection{System Evaluation}
\vspace{-10pt}

\Paragraph{End-to-End testing.} We evaluate the overall performance of the system for a fixed bin
capacity (8GB) for the LiveJournal graph. We vary the number of subgraphs to be extracted from the
underlying graph and study the effect on the number of bins required to pack them into memory using
the shingle-based bin packing heuristic. We measure the total end-to-end running time of the LCC
computation on each of these subgraphs in PEM mode using LBitSet bitmap construction.
Figure~\ref{fig:End-toEnd} shows that as the number of subgraphs increases, the system distributes
the computation on a larger number of bins and scales well with respect to the increase in the total running time with increase in number of subgraphs
which includes the time required by the GEP phase and the actual graph computation by each instance of the execution engine on each partition.


\Paragraph{Performance breakdown.} Figure~\ref{fig:perfBD1} shows the breakdown in terms of the \%$\mathcal{CE}$ 
required for the different stages of \nframe. The figure shows the performance for two different applications: LCC and  
Motif Counting and two different graphs: LiveJournal and Web-Google. For the smaller graphs like
Web-Google, the \% time taken for
execution is comparable to the graph loading time. For larger graphs like LiveJournal, 
the \% graph loading time dominates all other times
as it includes the time taken to read the disk resident graph and 
associated data, filter and shuffle based on the partitioning obtained from GEP.
In all cases, {\bf GEP constitutes a small fraction of the total time and is the crucial component that  
enables the efficient execution of the graph computation on the materialized
subgraphs in distributed memory using minimal resources. }
As can be seen in the baseline comparisons, without the
GEP phase, other vertex-centric approaches have a very high $\mathcal{CE}$ as compared to
\nframe~for the same underlying datasets and graph computations.

\Paragraph{\nframe~performance for larger graphs.} To ascertain the scalability of \nframe~we
conducted experiments with larger datasets for the Motif Counting application.
Figure~\ref{fig:scalability} shows the results for the scalability experiments on the Social
LiveJournal graph, the Orkut social network graph, and the largest of our datasets, the ClueWeb graph (428M nodes, 1.5B edges). The results show  the
$\mathcal{CE}$ in node-secs and total cluster memory required in GB. 
The results indicate that \nframe~scales well for ego-centric graph computation applications over larger graphs unlike other vertex-centric approaches such as Apache Giraph and GraphLab.    

\vspace{-11pt}
\subsection{Evaluation of Support for Iterative Applications.} 
\vspace{-4pt}
\red{We evaluated the support for iterative applications using the global connected components application. }

\noindent\red{\textbf{Performance breakdown.} We studied the performance
breakdown for the connected components application across different iterations
over two different datasets (LiveJournal and Orkut).
Figures~\ref{fig:iter_support_1}, \ref{fig:iter_support_2},
\ref{fig:iter_support_3}, and \ref{fig:iter_support_4} show the performance
breakdown in terms of compute time, synchronization time (time spent waiting at
the barriers), and message passing time (time spent in updating the key-value
store and fetching updated values from the key-value store). We studied the
performance breakdown for two different scenario, where the graph was partitioned
across 5 or 10 machines (we adjusted the bin capacity parameter to find the setting
which forced \nframe~to use the appropriate number of machines).
As expected, with 5 partitions, the
synchronization overhead is more as compared to the message passing overhead since
the number of ghost vertices that require message passing is smaller. In
comparison, with 10 partitions, the message passing overhead is more as the number of
ghost vertices is relatively higher. The synchronization overhead is less
as each partition does less work and inter-partition skew is smaller.}\\

\noindent\red{\textbf{Performance comparison.} Figures~\ref{fig:iter_support_5}
and \ref{fig:iter_support_6} compare the performance of \nframe~against GraphX
and GraphLab for the connected components application in terms of the running
time (Wall Clock time) and $\mathcal{CE}$ (node-secs) on 10 machines. 
As we can see, our relatively unoptimized implementation compares favorably to both,
and in fact, outperforms GraphX in some of the cases. Overall, GraphLab performs
better in terms of both runtime and $\mathcal{CE}$ for both graph datasets.
This superior performance of GraphLab for iterative computations can be attributed to its highly
optimized MPI-based message passing layer, as well as its implementation in C++.
}

\vspace{-15pt}
\subsection{Discussion.} 
\red{
In summary, our comprehensive experimental evaluation illustrates that \nframe~has comparable performance to the other graph 
processing frameworks for iterative tasks like connected components, while vastly outperforming them 
for more complex analysis tasks. Although \nframe~is able to scale better than the other systems we compared against 
for most of the tasks and it uses fewer resources in general, there is certainly a limit
to the graph sizes that our current implementation can handle given limited resources, and those can be seen or extrapolated 
from our reported numbers (e.g., \nframe~wouldn't be able to do LCC on a graph with 250M edges without at least 62 GB of cluster 
memory). However, \nframe~can process the partitions in sequence on a single machine (for such one-pass analytics tasks) by
loading them one by one, thus the maximum memory needed at any specific time point can be lower (at the expense of increased
wall-clock time). On the other hand, for iterative tasks, \nframe's limits mirror those of Giraph or GraphLab in that, there
must be enough cluster memory to load all the partitions. Some of the recent graph processing systems like X-Stream and
GraphChi do not have this restriction because of their use of disk-based processing; in future work, we plan to investigate
how the \nframe~programming model may be adapted to such settings.
}


\vspace{-11pt}
\section{Conclusion}
\vspace{-4pt}
\label{sec:conclusion}
Increasing interest in performing graph analytics over very large volumes of graph data has led to 
much work on developing distributed graph processing frameworks in recent years, with the
vertex-centric frameworks being the most popular. Those frameworks are, however, severely 
limited in their ability to express and/or efficiently execute complex and rich graph analytics 
tasks that network analysts want to pose. We argue that both for ease-of-use and efficiency, a more
natural abstraction is a {\em subgraph-centric framework}, where the users can write computations
against entire subgraphs or multi-hop neighborhoods in the graph. We show how this abstraction
generalized the vertex-centric programming framework, how it is a natural fit for many commonly used graph analytics tasks, and how it leads to more efficient
execution by reducing the communication and memory overheads.
We also argue that the graph
extraction and loading phase should be carefully optimized to reduce the number of machines required
to execute a graph analytics task, because of the non-linear relationship between that parameter and
the total execution cost; we developed a novel framework for solving this problem, and we show that
it can lead to significant savings in total execution time. Our comprehensive experimental
evaluation illustrates the ability of our framework to execute a variety of graph analytics tasks on
very large graphs, when Apache Giraph,  GraphLab \red{and GraphX} fail to execute them on relatively small graphs.

\newpage
\bibliographystyle{plain}
\bibliography{NScale_VLDBJ}

\begin{thebibliography}{10}

\bibitem{Giraph:Online}
{Apache Giraph {http://giraph.apache.org}}.

\bibitem{Blueprints:Online}
{BluePrints API: {https://github.com/tinkerpop/blueprints/wiki}}.

\bibitem{Furnace:Online}
Furnace: {https://github.com/tinkerpop/furnace/wiki}.

\bibitem{Gremlin:Online}
Gremlin: {http://github.com/tinkerpop/gremlin/wiki}.

\bibitem{Metis:Online}
Metis: {http://glaros.dtc.umn.edu/gkhome/metis}.

\bibitem{Redis:Online}
{Redis {http://redis.io/}}.

\bibitem{snap:online}
{Stanford Network Analysis Project: {https://snap.stanford.edu}}.

\bibitem{oddball}
Leman Akoglu, Mary McGlohon, and Christos Faloutsos.
\newblock {OddBall: spotting anomalies in weighted graphs}.
\newblock In {\em PAKDD}, 2010.

\bibitem{Backstrom:2011:SRW:1935826.1935914}
L.~Backstrom and J.~Leskovec.
\newblock Supervised random walks: Predicting and recommending links in social
  networks.
\newblock In {\em WSDM}, 2011.

\bibitem{burt2007secondhand}
Ronald~S Burt.
\newblock Secondhand brokerage: Evidence on the importance of local structure
  for managers, bankers, and analysts.
\newblock {\em Academy of Management Journal}, 50(1):119--148, 2007.

\bibitem{burt2009structural}
Ronald~S Burt.
\newblock {\em Structural holes: The social structure of competition}.
\newblock Harvard university press, 2009.

\bibitem{Cheng:fg:index:towards}
James Cheng, Yiping Ke, Wilfred Ng, and An~Lu.
\newblock {Fg-index: towards verification-free query processing on graph
  databases}.
\newblock In {\em {SIGMOD}}, 2007.

\bibitem{Cheng:2012:KTP:2168836.2168846}
R.~Cheng, J.~Hong, A.~Kyrola, Y.~Miao, X.~Weng, M.~Wu, F.~Yang, L.~Zhou,
  F.~Zhao, and E.~Chen.
\newblock Kineograph: Taking pulse of a fast-changing and connected world.
\newblock In {\em EuroSys}, 2012.

\bibitem{Unicorn}
Michael Curtiss, Iain Becker, Tudor Bosman, Sergey Doroshenko, Lucian Grijincu,
  Tom Jackson, Sandhya Kunnatur, Soren Lassen, Philip Pronin, Sriram Sankar,
  Guanghao Shen, Gintaras Woss, Chao Yang, and Ning Zhang.
\newblock {Unicorn: A System for Searching the Social Graph}.
\newblock {\em Proc. VLDB Endow.}, 2013.

\bibitem{everett2005ego}
Martin Everett and Stephen~P Borgatti.
\newblock Ego network betweenness.
\newblock {\em Social networks}, 27(1):31--38, 2005.

\bibitem{GraphX}
Joseph~E. Gonzalez, Reynold~S. Xin, Ankur Dave, Daniel Crankshaw, Michael~J.
  Franklin, and Ion Stoica.
\newblock {GraphX:} graph processing in a distributed dataflow framework.
\newblock In {\em OSDI}, 2014.

\bibitem{granovetter2010strentgh}
Mark~S. Granovetter.
\newblock The strength of weak ties.
\newblock {\em American Journal of Sociology}, 1973.

\bibitem{Gupta:2013:WFS:2488388.2488433}
P.~Gupta, A.~Goel, J.~Lin, A.~Sharma, D.~Wang, and R.~Zadeh.
\newblock {WTF}: The who to follow service at twitter.
\newblock In {\em WWW}, 2013.

\bibitem{He:2008:GQL:1376616.1376660}
Huahai He and Ambuj~K. Singh.
\newblock {Graphs-at-a-time: Query Language and Access Methods for Graph
  Databases}.
\newblock In {\em {SIGMOD}}, 2008.

\bibitem{lfgraph}
I.~Hoque and I.~Gupta.
\newblock Lfgraph: Simple and fast distributed graph analytics.
\newblock In {\em TRIOS}, 2013.

\bibitem{scalable-subgraphs}
Jiewen Huang., D.~J. Abadi, and Kun Ren.
\newblock {Scalable SPARQL Querying of Large RDF Graphs}.
\newblock In {\em PVLDB}, 2011.

\bibitem{izumi1998computational}
T.~Izumi, T.~Yokomaru, A.~Takahashi, and Y.~Kajitani.
\newblock Computational complexity analysis of set-bin-packing problem.
\newblock {\em IEICE TRANSACTIONS on Fundamentals of Electronics,
  Communications and Computer Sciences}, 81(5):842--849, 1998.

\bibitem{kashtan2004efficient}
N.~Kashtan, S.~Itzkovitz, R.~Milo, and U.~Alon.
\newblock Efficient sampling algorithm for estimating subgraph concentrations
  and detecting network motifs.
\newblock {\em Bioinformatics}, 2004.

\bibitem{DBLP:journals/im/KolountzakisMPT12}
M.~N. Kolountzakis, G.~L. Miller, R.~Peng, and C.~E. Tsourakakis.
\newblock Efficient triangle counting in large graphs via degree-based vertex
  partitioning.
\newblock {\em Internet Mathematics}, 2012.

\bibitem{Cassandra}
Avinash Lakshman and Prashant Malik.
\newblock {Cassandra: A Decentralized Structured Storage System}.
\newblock {\em {SIGOPS Oper. Syst. Rev.}}

\bibitem{Lee:2012:ICS:2448936.2448946}
Jinsoo Lee, Wook-Shin Han, Romans Kasperovics, and Jeong-Hoon Lee.
\newblock {An in-depth comparison of subgraph isomorphism algorithms in graph
  databases}.
\newblock In {\em {PVLDB}}, 2013.

\bibitem{Leskovec:2006:SLG:1150402.1150479}
J.~Leskovec and C.~Faloutsos.
\newblock Sampling from large graphs.
\newblock In {\em SIGKDD}, 2006.

\bibitem{DBLP:journals/corr/abs-1204-6078}
Y.~Low, J.~Gonzalez, A.~Kyrola, D.~Bickson, C.~Guestrin, and J.~M. Hellerstein.
\newblock {Distributed GraphLab: A Framework for Machine Learning in the
  Cloud}.
\newblock {\em PVLDB}, 2012.

\bibitem{DBLP:journals/pvldb/LowGKBGH12}
Y.~Low, J.~Gonzalez, A.~Kyrola, D.~Bickson, C.~Guestrin, and J.~M. Hellerstein.
\newblock Distributed graphlab: A framework for machine learning in the cloud.
\newblock {\em PVLDB}, 2012.

\bibitem{Malewicz:2010:PSL:1807167.1807184}
G.~Malewicz, M.~H. Austern, A.~J.C Bik, J.~C. Dehnert, I.~Horn, N.~Leiser, and
  G.~Czajkowski.
\newblock Pregel: a system for large-scale graph processing.
\newblock In {\em SIGMOD}, 2010.

\bibitem{NIPS2012_0272}
J.~McAuley and J.~Leskovec.
\newblock {Learning to Discover Social Circles in Ego Networks}.
\newblock In {\em NIPS}, 2012.

\bibitem{MilEtAl02}
R.~Milo, S.~Shen-Orr, S.~Itzkovitz, N.~Kashtan, D.~Chklovskii, and U.~Alon.
\newblock Network motifs: Simple building blocks of complex networks.
\newblock {\em Science}, 2002.

\bibitem{journals/jbcb/MongioviNGPFS10}
Misael Mongiovì, Raffaele~Di Natale, Rosalba Giugno, Alfredo Pulvirenti,
  Alfredo Ferro, and Roded Sharan.
\newblock Sigma: a set-cover-based inexact graph matching algorithm.
\newblock {\em {J. Bioinformatics and Computational Biology}}, 2010.

\bibitem{DBLP:conf/icde/MoustafaNDG11}
W.~E. Moustafa, G.~Namata, A.~Deshpande, and L.~Getoor.
\newblock Declarative analysis of noisy information networks.
\newblock In {\em ICDE Workshops}, 2011.

\bibitem{nguyen13}
D.~Nguyen, A.~Lenharth, and K.~Pingali.
\newblock A lightweight infrastructure for graph analytics.
\newblock In {\em SOSP}, 2013.

\bibitem{P.Cordella:2004:GIA:1018035.1018377}
Luigi P.~Cordella, Pasquale Foggia, Carlo Sansone, and Mario Vento.
\newblock {A (Sub)Graph Isomorphism Algorithm for Matching Large Graphs}.
\newblock {\em {IEEE Trans. Pattern Anal. Mach. Intell.}}, 2004.

\bibitem{Popescu:2013:PTP:2556549.2556553}
Adrian~Daniel Popescu, Andrey Balmin, Vuk Ercegovac, and Anastasia Ailamaki.
\newblock {PREDIcT: Towards Predicting the Runtime of Large Scale Iterative
  Analytics}.
\newblock {\em Proc. VLDB Endow.}, 2013.

\bibitem{journals/ton/PujolESYLCR12}
J.~M. Pujol, V.~Erramilli, G.~Siganos, Xiaoyuan~Y. 0001, N.~Laoutaris,
  P.~Chhabra, and P.~Rodriguez.
\newblock {The Little Engine(s) That Could: Scaling Online Social Networks.}
\newblock In {\em SIGCOMM}, 2010.

\bibitem{mmds}
A~Rajaraman and J.D. Ullman.
\newblock {\em {Mining of Massive Datasets}}.
\newblock Cambridge University Press, 2011.

\bibitem{Rajaraman:2011:MMD:2124405}
Anand Rajaraman and Jeffrey~David Ullman.
\newblock {\em Mining of Massive Datasets}.
\newblock Cambridge University Press, 2011.

\bibitem{xstream}
A.~Roy, I.~Mihailovic, and W.~Zwaenepoel.
\newblock X-stream: Edge-centric graph processing using streaming partitions.
\newblock In {\em SOSP}, 2013.

\bibitem{Salihoglu:2013:GGP:2484838.2484843}
S.~Salihoglu and J.~Widom.
\newblock {GPS:} a graph processing system.
\newblock In {\em SSDBM}, 2013.

\bibitem{DBLP:conf/icde/SeoGL13}
J.~Seo, S.~Guo, and M.~S. Lam.
\newblock Socialite: Datalog extensions for efficient social network analysis.
\newblock In {\em ICDE}, 2013.

\bibitem{DBLP:journals/pvldb/SeoPSL13}
J.~Seo, J.~Park, J.~Shin, and M.~S. Lam.
\newblock Distributed socialite: A datalog-based language for large-scale graph
  analysis.
\newblock {\em PVLDB}, 2013.

\bibitem{Shang:2008:TVH:1453856.1453899}
Haichuan Shang, Ying Zhang, Xuemin Lin, and Jeffrey~Xu Yu.
\newblock {Taming Verification Hardness: An Efficient Algorithm for Testing
  Subgraph Isomorphism}.
\newblock {\em {VLDB}}, 2008.

\bibitem{Shasha:2002:AAT:543613.543620}
Dennis Shasha, Jason T.~L. Wang, and Rosalba Giugno.
\newblock Algorithmics and applications of tree and graph searching.
\newblock In {\em {PODS}}, 2002.

\bibitem{DBLP:journals/corr/SimmhanKWNRRP13}
Y.~Simmhan, A.~G. Kumbhare, C.~Wickramaarachchi, S.~Nagarkar, S.~Ravi, C.~S.
  Raghavendra, and V.~K. Prasanna.
\newblock Goffish: A sub-graph centric framework for large-scale graph
  analytics.
\newblock {\em CoRR}, 2013.

\bibitem{DBLP:journals/pvldb/TianBCTM13}
Y.~Tian, A.~Balmin, S.~A. Corsten, S.~Tatikonda, and J.~McPherson.
\newblock {From "Think Like a Vertex" to "Think Like a Graph"}.
\newblock {\em PVLDB}, 2013.

\bibitem{Tian:2008:TTA:1546682.1547209}
Yuanyuan Tian and Jignesh~M. Patel.
\newblock {TALE: A Tool for Approximate Large Graph Matching}.
\newblock In {\em {ICDE}}, 2008.

\bibitem{Ullmann:1976:ASI:321921.321925}
J.~R. Ullmann.
\newblock {An Algorithm for Subgraph Isomorphism}.
\newblock {\em {J. ACM}}, 1976.

\bibitem{conf/cidr/WangXDG13}
G.~Wang, W.~Xie, A.~J. Demers, and J.~Gehrke.
\newblock {Asynchronous Large-Scale Graph Processing Made Easy.}
\newblock In {\em CIDR}, 2013.

\bibitem{Yan:2004:GIF:1007568.1007607}
Xifeng Yan, Philip~S. Yu, and Jiawei Han.
\newblock {Graph Indexing: A Frequent Structure-based Approach}.
\newblock In {\em {SIGMOD}}, 2004.

\bibitem{Zhao:2007:GIT:1325851.1325957}
Peixiang Zhao, Jeffrey~Xu Yu, and Philip~S. Yu.
\newblock {Graph Indexing: Tree + Delta less than equal to Graph}.
\newblock In {\em VLDB}, 2007.

\bibitem{Zou:2008:NSC:1353343.1353369}
Lei Zou, Lei Chen, Jeffrey~Xu Yu, and Yansheng Lu.
\newblock {A Novel Spectral Coding in a Large Graph Database}.
\newblock In {\em {EDBT}}, 2008.

\end{thebibliography}

\end{document}